\documentclass[pdflatex,sn-mathphys-num]{sn-jnl}

\usepackage{graphicx}%
\usepackage{multirow}%
\usepackage{amsmath,amssymb,amsfonts}%
\usepackage{amsthm}%
\usepackage{mathrsfs}%
\usepackage[title]{appendix}%
\usepackage{textcomp}%
\usepackage{manyfoot}%
\usepackage{booktabs}%
\usepackage{algorithm}%
\usepackage{algorithmicx}%
\usepackage{algpseudocode}%
\usepackage{blindtext}%
\usepackage{adjustbox}%
\usepackage{natbib}%
\usepackage{bm}%
\usepackage{tabularx}%
\usepackage{float}%
\usepackage{rotating}%
\usepackage{xcolor}%
\usepackage{pdflscape}%
\usepackage{longtable}%

\usepackage[font=footnotesize,labelfont=bf]{caption}
\usepackage{subcaption}

\usepackage{listings}
\usepackage{hyperref}%

\theoremstyle{thmstyleone}%
\theoremstyle{thmstyletwo}%

\theoremstyle{thmstylethree}%

\begin{document}

\title[Asset Pricing in Pre-trained Transformers]{Asset Pricing in Pre-trained Transformers}

\author*[1]{\fnm{Shanyan} \sur{Lai}}\email{shanyan.lai@york.ac.uk, annieyanyan125@gmail.com}



\affil*[1]{\orgdiv{Department of Economics and Related Studies}, \orgname{Univiersity of York}, \orgaddress{\street{Heslington}, \city{York}, \postcode{YO10 5DD}, \country{UK}}}



\abstract{This paper proposes an innovative Transformer model, Single-directional representative from Transformer (SERT), for US large capital stock pricing. It also innovatively applies the pre-trained Transformer models under the stock pricing and factor investment context. They are compared with standard Transformer models and encoder-only Transformer models in three periods covering the entire COVID-19 pandemic to examine the model adaptivity and suitability during the extreme market fluctuations. Namely, pre-COVID-19 period, COVID-19 period and 1-year post-COVID-19. The best proposed SERT model achieves the highest out-of-sample $R^2$, 11.94\% and 11.47\% respectively, when extreme market fluctuation takes place, followed by pre-trained Transformer models (11.13\% and 9.72\%). Their Trend-following-based strategy's performance also proves their excellent capability for hedging downside risks during market shocks. The proposed SERT model achieves a Sortino ratio 47\% higher than the buy-and-hold benchmark in the equal-weighted portfolio and 28\% higher in the value-weighted portfolio in the static transaction cost scenario when the pandemic period is considered. It proves that Transformer models have a strong ability to capture patterns of temporal sparsity in asset pricing factor models, especially with high volatility. I also find the softmax signal filter as the common configuration of Transformer models in alternative contexts, which only eliminates differences between models, but does not improve strategy-wise performance, while increasing attention heads improves the model performance insignificantly and applying the 'layer normalization first' method does not boost the model performance in our case.\\

Note: An earlier version is available at Zenodo (DOI: 10.5281/zenodo.15348807)}

\keywords{Stock pricing, Transformer, Attention mechanism, temporal dependency}



\maketitle

\vspace{0.5em}
\noindent
\textbf{Current Version:} June 15, 2026\\
Following insightful comments from Dr Mark Hallam and Dr Chrysovalantis Vaslakis, this version introduces several improvements. First, I correct an error in the out-of-sample $R^2$ calculation caused by the `r2\_score()' function in scikit-learn, which incorrectly aligns predicted returns when computing the denominator. This led to a systematic underestimation of the average OOS $R^2$. I address this by manually computing the mean squared error (MSE) and OOS $R^2$. The denominator now correctly uses the historical mean based on both training and validation samples, consistent with standard asset pricing literature (e.g., Campbell and Thompson, 2008). The updated code is provided in the Appendix. Second, I incorporate a heteroskedasticity and autocorrelation consistent (HAC) estimator into the Diebold-Mariano (DM) test to account for serial correlation in the difference between the two models’ forecast errors. Third, the abbreviations
of the model names are also updated to improve readability, which is exhibited in the Appendix. I am grateful to Dr Mark Hallam and Dr Chrysovalantis Vaslakis for their valuable feedback. All remaining errors are my own.

\vspace{1em}

\section{Introduction}\label{sec:intro_ch4}
Previous studies on machine learning (ML) models for asset pricing or factor investing tasks, such as \citet{Gu2020EmpiricalLearning,Gu2021AutoencoderModels} have already proved that some traditional ML algorithms, for example, the multi-perception (MLP) and random forest (RF), outperform the traditional econometric models when they are applied as model frameworks of asset pricing models. The shallow MLP models show advantages in coping with extreme market fluctuations, such as those caused by the COVID-19 pandemic, with better out-of-sample performance and universal stock adaptivity among large-cap U.S. stocks. Nevertheless, the MLP model still has drawbacks such as overfitting, overlapping (under the method of GKX2020), high sensitivity towards missing values and data noises, as well as low computational efficiency, as it is the simplest neural network (NN) structure with no natural design for detecting the temporal dependency of the time series data. Specifically, inappropriate methods for handling missing values during data preprocessing may introduce significant bias. For example, simply removing the stocks or factors may cause the ‘survival bias’ and filling identical numbers, such as mean or median values, also affects the reliability of the data and the accuracy of prediction if the missing values occupy a considerable fraction of the data. On the other hand, the added complexity of MLP models compared with traditional statistical models multiplies the computational cost. Chapter 3 employed the MLP autoencoder pre-training method together with a recurrent neural network structure to address the disadvantages of MLP models for empirical asset pricing. Moreover, it employs different attention mechanisms to capture longer-term temporal dependency for improving the stock return forecasting accuracy.\\

However, the single-output MLP method individually calculates the stocks, which multiplies the computational cost, especially if the dataset contains thousands of stocks or is extended in length. Moreover, it does not consider cross-sectional effects among individual stock returns, which reduces the model's economic or financial meaning. Thus, the research question remains ‘is there an improved NN structure that satisfies the conditions, including further moderating the overfitting issue, higher computational efficiency and balancing the tradeoff between model interpretability and predictability?’. \citet{Aydogan-Kilic2023ModificationCases,Gu2021AutoencoderModels,Rukmana2024ApplicationAnalysis} attempt to utilize the alternative neural network (NN) structures, including recurrent neural networks and their variations or MLP autoencoders, to solve these issues for stock price forecasting, but the aforementioned issues still have not improved significantly. \\

Fortunately, the recent boom in large language models (LLMs) provides a potential solution. The standard Transformer model \citep{Vaswani2017AttentionNeed}, as the foundation of most commercial LLMs such as openAI’s ChatGPT, Google’s Gemini, X’s Grok and DeepSeek, is built upon an autoencoder NN structure, which is assembled with encoder and decoder blocks. Each encoder or decoder block of a standard Transformer contains an MLP autoencoder module with an auxiliary self-attention layer for capturing and processing the inputs. In between the encoder and decoder blocks, a cross-attention layer is configured to capture the spatial-temporal relationship between input and output data (observable factors and stock excess returns) to improve forecasting accuracy. Since the fundamental structure of the Transformer is the autoencoder structure, which has multiple neurons for multiple outputs, together with the dot-product-based self-attention mechanism, which highly reduces the parameters’ size. They parallel the computation process of multi-stock return series through the autoencoder structure and reduce the parameter estimation workload. Thus, the design substantially increases computational efficiency. Moreover, computing stocks’ returns simultaneously and jointly optimizing multiple stocks’ loss functions also enables the scaling of cross-sectional effects between stocks. Both the self-attention mechanism and the cross-attention mechanism are originally designed for textual information, as the content vectors have different lengths. Therefore, with proper embedding settings, linear in this case, they perfectly fill the missing value by projecting the original input factors. Although it was originally invented for natural language processing (NLP) tasks such as machine translation and context prediction, as a sequence-to-sequence (Seq2Seq) model, it is found highly suited for time series prediction as well \citep{Gezici2024DeepPrediction,Paivarinta2022Transformer-basedEvidence,Zhang2022Transformer-basedPrediction}.\\

Furthermore, researchers have developed various Transformer-based architectures, such as BERT, Realformer, and GPT, some of which have been successfully applied to financial time series forecasting \citet{Cheng2024GPTsFactors}. However, adapting the standard NLP-focused Transformer for time series tasks requires specific architectural modifications. For instance, strict causal masking must be incorporated into both the encoder and decoder to prevent look-ahead bias (data leakage), even when employing a rolling window approach. Additionally, the standard word embedding layer must be replaced with a linear embedding layer to process continuous numerical data effectively. \citet{Cong2021AlphaPortfolio:AI, Ma2023AttentionApproach,Zhang2022AssetLearning} further attempt the Transformer in the context of factor investing and stock pricing on different datasets, but the results are variable according to the stock markets and data methods they are employing. Although it is a formidable challenge applying the LLMs to temporal sparsity data, which is a typical characteristic of financial and economic data that has long-term dependency but fewer observations, for instance, monthly or quarterly economic time series, their work sheds light on advanced machine learning (ML) for empirical asset pricing topics. However, compared with the financial time series, the temporal sparsity characteristic of economic data is more restricted since they are commonly low frequency and have longer time-dependency, hence fewer ML algorithms are explored successfully for economic applications such as empirical asset pricing and factor investing research, not to mention the state-of-the-art (SOTA) large language model (LLM) as Transformer. It is worth putting more effort into exploring the upgraded structure for higher suitability of economic data. In this research, the innovative Transformer model, multi-head single-directional representative Transformers (SERTs), for asset pricing and factor investing tasks, are developed and examined on large-cap U.S. stocks. Also, the innovative pre-trained Transformer models are examined in the context of empirical asset pricing and factor investing. SERT alters the random mask pre-train module of BERT to an MLP autoencoder pre-train module for numerical data, and it enforces the causal masks for the encoder self-attention layer to prevent future data leakage. The innovative pre-trained Transformer simplifies the inside MLP structure and adds the MLP pre-training module for improving the input factors’ quality. \\

Therefore, the contributions of this study are from four aspects: 
\begin{itemize}
  \item Developing the innovative financial economic context LLMs, SERTs and MLP autoencoder pre-trained Transformer, which enriches the scarce literature in ML for asset pricing and factor investing.\\
  
  \item Verifying whether general model improving methods such as multi-head attention, pre-training and layer normalization first (LNF) apply to the specific LLMs for temporal sparsity data such as stock pricing data.\\
  
  \item Offering an anatomy of the ‘black-box’ Transformer models for boosting the transparency and interpretability of the complicated LLMs in the context of stock pricing and factor investing. This increases the financial and economic meanings of the models to a degree and corrects the bias of overthinking the positive relationship between parameter size, model complexity and overfitting \citep{Didisheim2024APTModels}. \\
  
  \item It is the first attempt for the SERT and pre-trained Transformer models to directly work on stock pricing and factor investing. Also, it is the first comprehensive research on fundamental LLM in the context of extreme temporal sparsity data.\\
  
\end{itemize}
This study contains 6 main sections. Section 2 presents the literature review. Section 3 exhibits the data that have been used, while Section 4 discusses the Models employed in this study. Section 5 shows the empirical experiment and results, and Section 6 gives the conclusion and further discussion.

\section{Related work}\label{sec:related_work_ch4}
Although RNN models have highly improved the forecasting performance compared with the MLP models, as proved by \citet{Chen2024DeepPricing,zhou2024learning}, the defects of the gradient vanishing and explosion and low computational efficiency due to the sequential gradatim calculation as a substitute for parallel computation. Thus, the core of the Transformer model, the attention mechanism, was developed for efficiently capturing variable dependency. The classic Transformer models are developed for sentence translation and sentence prediction; thus, they are potentially suitable for short-length data such as monthly returns.\\
 
In the early stage, people developed soft attention to add to the RNN structure for improving the prediction accuracy in NLP tasks. The seminal work of the soft attention mechanism comes from \citet{Bahdanau2015NeuralTranslate}, which also named as additive attention or Bahdanau's attention. The first soft-attention mechanism utilizes the attention-weighted sum of all input states to generate a new hidden state at each time step via a feed-forward network. This mechanism allows the model to dynamically focus on relevant features of the input sequence at each time step. Bahdanau's attention is also named additive attention. Since the attention score was calculated at every time step, it has a drawback of parameter oversizing which latently causes overfitting. It also lowers computational efficiency. Additionally, it has more hyperparameters that require manual tunning, which increases the model training time and reduces the stability of the training results. Also, since it relies on the soft alignment mechanism, it is more suitable for short-term forecasting instead of long-term trend capture, especially in sparse economic data. It was used in financial applications such as \citet{Li2018StockLSTM,Zhang2019AtLSTM}. Concretely, \citet{Li2018StockLSTM} proposed an Attention-based Multi-Input LSTM model for stock price prediction. They used the dual-stage attention mechanism \citep{Qin2017APrediction} to improve the LSTM capability for capturing the patterns of stock pricing and denoise. To handle the noise introduced by naively concatenating auxiliary factors, such as prices of correlated stocks and market indices, with the target stock’s historical series, they separated a mainstream factor from multiple auxiliary factors. Their MI-LSTM processed each factor stream independently and employed mainstream-controlled input gates to regulate auxiliary influence, with a dual-stage additive attention mechanism: the first stage dynamically fused candidate cell-state updates across the different factors via additive attention scores, while the second stage applied the attention to measure the importance of different time steps.
The model was evaluated on historical opening price data from 261 stocks in the Chinese stock market (components of the CSI-300 index) from April 25, 2013 to May 15, 2017. They found their attention-based MI-LSTM significantly outperforms standard LSTM, simple concatenated multi-input LSTM, and alternatives in mean squared error (MSE). The additive attention-based model is validated for the effectiveness of its hierarchical factor modelling and attention-driven selective combination. This work highlighted the effectiveness of the attention mechanisms in managing noisy multivariate financial time series for the forecasting tasks. However, it has limitations such as focusing on short-term (next-day) opening price prediction in a single market, configuration on a short time window (T=10), and latent challenges in generalizing to alternative markets and assets, longer windows, and periods with extreme fluctuations, such as the COVID-19 period. \citet{Zhang2019AtLSTM} proposed AT-LSTM, a two-stage attention-based Long Short-Term Memory (LSTM) model for financial time series forecasting. In the first stage, the additive attention mechanism is applied to distribute the attention weights to input features (e.g. opening price, closing price, trading volume, and related market variables) at each time step. This mechanism computes attention weights based on the previous LSTM hidden state and a linear transformation of the input feature sequence. With the attention layer, the most relevant features are emphasized and the less informative features are suppressed. In the second stage, these attention-weighted representations are fed into a two-layer LSTM network to capture long-term temporal dependencies and predict the next day’s closing price. The model was evaluated on daily data from three major U.S. stock indices: the Russell 2000, the Dow Jones Industrial Average (DJIA) and the Nasdaq from January 2, 1991, to April 30, 2018. Their results demonstrated that AT-LSTM consistently outperformed both the traditional ARIMA model and vanilla LSTM across all three datasets. Nonetheless, it acknowledged limitations, including noticeable time lags in predictions and low sensitivity to extreme market conditions. As demonstrated in Chapter 3, the modern attention mechanisms, such as self-attention and sparse attention, potentially perform better than the additive attention in time series forecasting tasks. This work provides an early example of integrating attention with recurrent networks to enhance predictive accuracy in financial forecasting.\\

Later, \citet{Luong2015EffectiveTranslation} simplifies the Bahdanau attention via three different functions. They proposed three methods for computing the attention score: dot product, generalizing, and concatenation. The asset pricing fitted form is introduced in Chapter 3. Luong's attentions are also known as the multiplication attention mechanism. It is a big step forward in approaching the self-attention mechanism in the Transformer model since it adopts the dot-product function proposed for computing the attention scores. The mathematical equations evidently show that Luong's attention score calculation has freed itself from the dependence of the MLP structure, while the decoder hidden status updating never requires an RNN structure. In that sense, Luong's attention highly reduces the parameter size and promotes computational efficiency. Thus, the design improves training speed significantly and extends the long-term dependency capture. Few economic or financial applications directly based on Luong's attention are found in the existing literature, but it is applied as a built-in module, for example, in \citet{Kim2019OptimizingBoundaries}. However, it is the foundation of the self-attention mechanism, which is built in the classic Transformer models. Although additive attention offers strong expressive power and excels at capturing complex, non-linear alignments, making it effective for tasks with complicated dependencies, it is computationally heavier due to the extra feedforward neural network layer and performs unstably with long sequences. In contrast, Luong's dot-product and general attention significantly improves training speed and parameter efficiency, better supporting long-term dependency modeling, though it may exhibit reduced flexibility when dealing with highly non-linear relationships.\\

Subsequently, the development of attention mechanisms has ushered in a period of prosperity. Numerous attention variations were invented for different contexts. The most well-known attention variations are memory-augmented attention \citep{Weston2014MemoryNetworks}, hierarchical attention \citep{Yang2016HierarchicalClassification}, dual-stage attention \citep{Qin2017APrediction} and self-attention \citep{Vaswani2017AttentionNeed}. Apart from the NLP tasks as they were originally designed, they are used in different time series tasks as well, for example, \citet{Qin2017APrediction} designed the Dual-Stage Attention-based Recurrent Neural Network (DA-RNN), an encoder-decoder framework that integrates two complementary attention mechanisms to enhance multivariate time series forecasting, which is inspired by humans' dual-stage attention of Feature Integration Theory (Preattentive Stage and Focused Attention Stage) \citep{treisman1980feature}. Specifically, the model employs an input attention mechanism in the encoder to dynamically select and distribute the attention weights to the most relevant driving (exogenous) series at each time step, guided by the previous encoder hidden state. In its decoder, a temporal attention mechanism selectively aggregates hidden states across time steps of the encoder according to the current decoder hidden state. By combining the dual-stage attention with the LSTM foundation, it can effectively capture the longer-term temporal dependencies. The model is tested on the SML 2010 smart home sensor dataset (daily) and the NASDAQ 100 stock dataset (daily), and it consistently outperforms traditional statistical models, standard LSTM, and other attention-based RNN variants in terms of prediction error. The paper represents a widely referenced early contribution to attention-augmented recurrent networks for multivariate time series,which are verified by \citep{Liu2020DSTP-RNN:Prediction,Peng2021Dual-stagePrediction,Yang2022AMechanism}, but due to the large parameter size, it may cause overfitting on monthly asset pricing data latently. In that sense, it cannot compete with the universally recognized one in the Transformer model: self-attention, in the asset pricing context. A detailed description of the self-attention mechanism is presented in Section~\ref{subsubsec:Attention mechanism414}.\\

A profound revolution occurred in 2017 when the first Transformer model was proposed. Transformer models are the core and the foundation of most LLM industrialized AI robots, for instance, OpenAI’s ChatGPT, DeepSeek, X’s Grok 3 and Google’s Gemini. With the increasing popularity of AI robots, Transformer models have attracted great attention from researchers. People explore Transformer applications across different contexts, as well as structural variations in Transformer models. For example, \citet{Zhou2021Informer:Forecasting} developed the Informer model, which extends the length of the information that the Transformer can capture, and a review by \citet{Zhao2024ApplicationReview} describes the time series applications of the Informer. For handling sparse time series, \citep{Wu2021Autoformer:Forecasting} also proposed the Autoformer model for longer dependencies. However, economic time series have higher sparsity and longer dependencies than financial time series, which means there is a trade-off between the overfitting caused by oversized parameters and dependency length. These suggest that, for stock pricing factor models, simplified Transformer models may perform better than more complex variants. Wherefore, the single-directional representative from the Transformer (SERT) and pre-trained Transformer with a low-dimensional structure may be suitable for the stock pricing context. \\

Most existing financial or economic applications of Transformers, or more generalized, LLMs, concentrate on textual information modelling, such as sentiment analysis \citep{QianSensitivity2022,Rehman2022TransferAnalysis,Rizinski2024SentimentXLex}, behavioral factor exploration \citep{Ferdus2024TheFactors}, financial and economic news processing \citep{Dolphin2024ExtractingApproach,Miori2024NarrativesDislocations,Sharkey2024BERTEngineering} or social media monitoring \citep{Li2022IncorporatingPrediction}. For example, \citet{Kaplan2023CrudeBERT:Market} designed CrudeBERT as a Transformer model for sentiment analysis on the crude oil future market. It identifies and classifies the news headlines that affect the supply and demand of crude oil to predict its price movement. They proved their model was superior to the prevailing FinBERT \citep{Araci2019Finbert:Models} model on the worldwide crude oil market. \citet{Ferdus2024TheFactors} offer another example for the Transformer model application in financial textual information capture. They employ the Transformer model for modelling sentimental information from X (Twitter) as external factors and prove that their model excels the LSTM, random forest, and LightGBM, which derive the highest stock price forecasts among the S\&P 500 components. In addition to textual data capture, Transformer models are used for speech recognition tasks in finance and economics as well. \citet{Gado2024TransformersReturns} innovatively applies the Transformer model (BERT) to extract the sentimental effect from Canadian and U.S. central bankers’ speeches. They find that negative Federal speech increases the tail stock returns of both the U.S. and Canada, while negative BoC speech decreases the Canadian tail stock returns but has no effect on U.S. tail stocks. Federal sentimental variation causes changes in BoC sentiment but not ‘vice versa’, which indicates the single-directional spillover. Their long-only trading strategy based on negative sentimental Federal speeches achieves 47\% higher cumulative returns than the S\&P 500 and 0.61 in the Sharpe ratio. They also found that the BERT model encompasses cross-sentence context and performs better in stock returns forecast and strategy-wise investing than the individual-sentence context for sentimental analysis. This great finding inspires us to use the sorted portfolios as factors instead of the firm characteristics for cross-sectional consideration. \citet{Li2022IncorporatingPrediction} design a Transformer-based Attention Network (TEA) framework for stock movement prediction. It utilizes textual social media information (Twitter, now X) together with the historical stock prices as predictors to forecast the direction of stock price movement, which excels the traditional benchmark models such as ARIMA, LSTM, StockNet and CNN-BiLSTM-AM, both in accuracy and Matthews Correlation Coefficient (MCC) scores. They find that both sentimental information from X and the price data contribute significantly to facilitating prediction accuracy. Also, the model’s performance is highly improved when the price data works together with sentiment information. \citet{Miori2024NarrativesDislocations} applied GPT-3.5 and economic network theory in the Wall Street Journal to investigate the relationship between news topic structures and financial market dislocations. They find the key economic topics (e.g., inflation, interest rates, Federal Reserve decisions) dominate market narratives, network fragmentation correlates with financial market shocks, and GPT-based narrative structures provide a higher accuracy view for economic discourse than traditional topic modelling techniques.\\

Recently, \citet{Nie2024AChallenges} conducted a survey for the financial application of large language models (LLM), which exhibits examples of LLMs in different tasks of finance. It assists researchers in finding directions in which they are interested. Their introduction emphasizes more on specific financial Transformer variation models for individual tasks rather than the generalized models in different contexts of the financial domain. For example, FinBERT \citep{Araci2019Finbert:Models} is a model for financial sentiment analysis, and FinVIS-GPT \citep{Wang2023Finvis-gpt:Analysis} is for financial chart analysis. Although their work encourages future researchers to seek models for their assignments as well as demonstrates the challenges of applying these models to a degree, it is insufficiently comprehensive for researchers who are working on it. Fortunately, \citet{Zhao2024ApplicationReview} provided a more comprehensive literature review of LLMs for financial applications. Similarly, they classify the LLM for finance into five main categories, namely, sentiment analysis, question answering, named entity recognition, time series forecasting and mathematical reasoning, but they concentrate on the mechanism of models for distinctive tasks, which furnishes readers with the blueprint of groups of models’ capability to achieve. Moreover, they highlight the performance of GPT-4 for different financial tasks and indicate that a textual Financial LLM with the fundamental structure of GPT-4 has better performance than otherwise. Nevertheless, both LLM review papers agree that it is a significant challenge to apply LLMs directly to numerical data, especially time series data with pronounced temporal sparsity. Therefore, developing and adjusting LLM models and making them optimized for numerical data could be a new direction for future researchers to endeavour with.\\

Compared to research on processing financial or economic textual data, the direct application of Transformer models to numerical financial data is relatively immature. \citet{Korangi2023AMarkets} use the Transformer model for mid-cap companies’ short-term and medium-term default probability prediction. Their encoder-only Transformer model achieves the highest predictive accuracy among the benchmarks of logistic regression, XGBoost, LSTM, and Temporal Convolutional Networks models. \citet{Sun2022DeepMechanism} applies a convolutional neural network (CNN) assembled Transformer (named Conv-Transformer) with a graph attention mechanism (GAT) unit on the portfolio optimization task of Chinese CSI 300 stocks. The model optimizes the return-risk ratio directly and improves the asset allocation decisions. Specifically, the Conv-Transformer is deployed to capture temporal dependencies in individual stocks, while the GAT reflects the cross-sectional relationship between stocks. The model dynamically learns the correlation between stocks and overcomes the limitations of static stock relationships in existing methods. \citet{Tevare2023ForecastingTransformer} employ the Stack Transformer model for New York Stock Exchange (NYSE) stock price forecasting. They substituted the Sinusoidal Positional Encoding (SPE) positional encoding method with the Time2Vec learnable positional encoding method for a better fit for the time series data and applied their model to daily and weekly stock prices and technical analysis indicators. Their model derives the lowest mean square error (MSE), root mean square error (RMSE) and mean absolute error (MAE) compared with the benchmark LSTM models. \citet{Zhang2024Deep20202022} develop a novel model named Quantformer and examine it on the Chinese stock market. It tailors the standard Transformer model for time series data via a simplified decoder block of the standard Transformer model and adopts it for building financial trading strategies based on technical analysis indicators (emphasized on the price-volume factors). They notice the effect of different temporal distances of financial time series, but removing masks of the attention layer may cause future data leakage even under the rolling window method. They also verify the Transformer models’ capability for handling market fluctuations. \citet{Wheeler2024MarketGPT:Series} employs the innovative MarketGPT model based on the most prevailing commercial LLM, generative pre-trained Transformer (GPT), for modelling the order flow of the NASDAQ stock market. They invent an order generation engine within a discrete event simulator to enable the realistic replication of limit order book dynamics by building a high-fidelity market simulation environment. The model offsets the drawbacks of traditional time series models (ARIMA, GARCH, LSTM, and GANs), such as missing market microstructure, divergence between generative and real data, and neglecting long-term temporal dependencies. The empirical results of their work show that MarketGPT successfully simulates stylized facts such as heavy-tailed distribution, volatility clustering and long-range dependence. Thus, their model provides an innovative proposition of depicting the unobservable factors for asset pricing factor models. However, MarketGPT has imperfections of high computational cost (requirements of high computational power GPU), flash crash during the extreme market fluctuation period and no direct market trend predictions. This means the model possibly can solve the omitted factor problem to some degree after positive modifications. \citet{Lezmi2023TimeManagement} provide another encoder-only Transformer model application on multi-step time series forecasting and asset management of the global markets. They examined their model on the daily and weekly main global stock indexes, such as the S\&P500 and Euro Stoxx 50, as well as 10-year global futures markets, such as the U.S., Germany and UK. They construct a trading system based on a trend-following strategy in which trading signals are generated through the model’s binary trend prediction and a portfolio optimization process based on the model’s volatility forecasting. It proved the effectiveness of the model for long temporal dependency capturing in financial data among the traditional models such as ARIMA, GARCH and LSTM. They also confirm that the self-attention mechanism allows the model to efficiently analyze both temporal and spatial relationships. Nonetheless, their model struggles to handle extreme market fluctuations during the COVID-19 period. Moreover, the short step size configuration of the rolling window method may cause overlapping problems. \\

Less literature on LLM model applications on numerical economic data is found, particularly ‘Transformer for asset pricing’. The most relevant works are \citep{Cong2021AlphaPortfolio:AI,Gabaix2024AssetEmbeddings,Ma2023AttentionApproach,Zhang2022AssetLearning}. \citet{Ma2023AttentionApproach} utilize a CNN-assembled stacked encoder-only Transformer model as a factor model framework for stock return and volatility forecasting in the Chinese stock market. Concretely, they deploy 72 firm characteristics and 8 macroeconomic indicators as the monthly observable factors to model and predict 2466 Chinese A-share stocks. In their trading system, the returns forecast is for trading signal generation, and the volatility forecast is for asset allocation (portfolio optimization). Their model with four encoder stacks achieved both the best OOS $R^2$ and investing performance among the benchmarks of linear regression, principal component analysis (PCA), random forest and LSTM. \citet{Cong2021AlphaPortfolio:AI} conduct a comparative study on commonly applied NN models, which include RNN, LSTM, bi-directional LSTM, LSTM with attention mechanism, GRU and encoder-only Transformer. They applied models on the monthly returns of over 20000 stocks from 1970 to 2016 and deployed 54 firm characteristics as the factors. However, all the models they applied encountered surprisingly low OOS fitness. The highest OOS $R^2$ appears in the LSTM model, which is 0.45\%, and the lowest one is the encoder-only Transformer, which only has 0.12\%, followed by the RNN model (0.13\%). This paper does not mention how the trading signals are generated, hence the investment performance indicators, such as Sharpe ratio and annualized returns have less reference significance. This work is possibly a warning for researchers who misunderstand the relationship between data size and model complexity, and evidence that a simplified Transformer structure may not successfully handle an overly large output dimension. In the factor model case, the stock number N cannot be far beyond the time length. However, \citet{Zhang2022AssetLearning} conducts a similar large-scale comparative study of the U.S. stock market and reports more promising results. Some differences from \citet{Cong2021AlphaPortfolio:AI}’s work, \citet{Zhang2022AssetLearning} organizes the input factors via \citet{Gu2020EmpiricalLearning} method and adjusts the fixed time span to a dynamic time span from 5 to 40 years. They find that all their deep learning models outperform the traditional statistical models, and the RNN with attention mechanism model, and the Transformer model have the highest OOS model fitness. For investment performance, the RNN with attention mechanism achieves the highest Sharpe ratio, which is superior to all alternatives. \citet{Didisheim2024APTModels} present a clue for this phenomenon from their Artificial Intelligence Pricing Theory (AIPT), which challenges the argument that ‘sparse factors work better’ believed by the traditional Arbitrage Pricing Theory (APT). They enlarge the dimension of the input non-linear factors up to 360000 and find that it significantly reduces the pricing errors and promotes the Sharpe ratio. Consequently, they prove that large complex factor models surpass the traditional sparse models from statistical theory and further prove that even if the factor numbers are far beyond the observation numbers (time length), their conclusion still holds since large factor models considered the complex interaction between factors and have a better approximation to the true Stochastic Discount Factor (SDF). They also refute the alternative linear dimension reduction methods, such as PCA, which have better performance than the large factor models. This inspires us for the proposed pre-trained transformer models and SERT models. However, no matter the firm characteristics as factors or factors organized by the method of GKX2020, it increased the completeness of the input information but highly sacrificed the economic meanings. Thus, it is worth attempting to utilize firm characteristic-sorted portfolios as factors to improve the economic meanings. \citet{Ye2024FromPricing} provided a comprehensive review of ML for empirical asset pricing. They introduce ML methods as solutions from different angles of empirical asset pricing, for example, return prediction, price direction predictions, ranking of assets and portfolio optimization. Their work discloses substantially interesting future research directions for researchers to work on. Their introduction encompasses task-specific ML models without a detailed explanation, which may struggle audiences without a computer science or machine learning background. \citet{Giesecke2024AIResearch} provides a detailed overview of ML applications in frontier economic research. They focus on five categories of ML applications for economics: data curation, sentiment analysis, clustering, classification, and function approximation. Among the five categories, data curation is the revolutionary application according to other literature. They find the ML algorithm can capture textual information from government reports, which enables extracting new factors from these textual reports for economic prediction tasks, for example, \citet{Giesecke2023StateStructure} applied the LayoutLMv3 and LLM models to collect and restructure debt maturity data from more than 400,000 pages of financial disclosures, then utilizes them for predicting government bond rates. They illustrate the challenges in AI-driven economics research, such as computational cost, model interpretability and data quality, hence a hybrid approach with the deterministic (economic rules) and stochastic (ML insights) methods should be considered for addressing these challenges. \citet{gezici2024deep} applied the vision Transformer models (Vision Transformer, Data Efficient Image Transformers) and vision neural network models (ConvMixer) on nine volatile ETFs' prices and technical analysis indicators from January 1st, 2002, to January 1st, 2022. They visualized the ETFs' data into a two-dimensional $65 \times 65$ pixel figures, then utilized the advanced vision neural network models for generating trading signals such as `buy', `sell' and `hold'. They confirmed that the ConvMixer has the highest prediction accuracy (87.94\%), but since the high occupation of the `hold' signals, the model's profitability is inferior to the vision Transformer models (achieved the annualized return of 17.19\%). Although it advances the application of advanced neural network models research on alternative financial data (images), it still has limitations. For instance, high unnecessary computational cost due to the images being transformed from the numerical data. Additionally, they simplified one dollar per transaction as the static transaction cost, which can not reflect the real trading environment and how the turnover rate could affect the profitability of these advanced models. And they did not show any generalised examination results of their models on alternative ETFs or different assets, which means the sample size of nine ETFs is relatively small. Thus, compared with their work, the proposed SERT models and pre-trained Transformer models highly improve the computational efficiency and are more realistic in the transaction cost configurations. Also, the proposed models are tested on 420 large-cap U.S. stocks, which covers a larger sample. \citet{wang2025machine} verified the existence of short or long-term dependencies (autocorrelation) and seasonality (e.g. January effect) effects in U.S. stock returns, and cast a doubt on \citet{Gu2020EmpiricalLearning}'s work that shallow MLP models work better on stock return prediction. They employ the Autoformer model, which was designed specifically for time series data on GKX2020's data set and data method from 1957 to 2021. They followed the rule-of-thumb ML data split, which uses 70\% of the data for training, 20\% for validation, and 10\% for testing. The Autoformer model, which replaces the traditional self-attention mechanism in Transformer models with a Fast Fourier-based autocorrelation detection mechanism, and embeds series decomposition blocks within the encoder and decoder to isolate data into `seasonal' and `trend-cyclical' components. Their model also extended the neural network depth to a dual-block autoencoder structure with two attention heads. They examined their models on the step-size of 1 month, 3 months and 12 months, and found that the predictive performance is significant in the longer term due to the low-frequency structures, trend components, and seasonality being stronger over longer horizons. From a profitability perspective, their model achieves an excellent Sharpe ratio of 3.40 (peaking at 4.41). Nonetheless, they did not consider the transaction cost, turnover rate and the short-side transaction borrowing rate, as well as evaluate how the turnover rate influences the profitability. Moreover, due to the extensive factor dimension and the deep neural network structure, the large parameter size (over 40,000) requires massive computational costs, which is not quite realistic in industrial implementation. Additionally, they did not implement the vanilla Transformer as the benchmark to show the advantage of the innovative Autoformer model. Therefore, the proposed SERT and pre-trained Transformer could be a better solution for balancing predictive performance and computational cost. Furthermore, in this Chapter, the static transaction cost method is employed to evaluate the models' profitability in a conservative penalty environment, and the dynamic transaction cost (Turnover rate with 20bps for large-cap stocks) to evaluate how turnover rate would affect the profitability. That assists the practitioners in real-time market trading.\\

\section{Data description}\label{sec:Data description}
The data approach used in this chapter mainly follows the one in Chapter 3. The entire data length covers from January 1957 to December 2022. The separation of the training period, validation period and testing period is listed in Table~\ref{tab:data_des_ch3}. I use the rolling window method instead of the prevailing extending window method in previous literature, such as \citet{Gu2020EmpiricalLearning}, and this is different from Chapter 2. The validation window size is 30\% of the training size. The 182 factors provided by \citet{AndrewY.Zimmermann2020OpenPricing}, which excludes the factor with missing values larger than 40\% for maintaining a reasonable data quality. Although the MLP autoencoder pre-training approach can promote data quality, it is crucial to exclude series containing an inordinate number of missing values to reduce the bias or the potential `hallucination' of LLMs. \\

The number of latent-layer neurons in the feedforward autoencoder in the main body of the Transformer model is configured to 70\% of the input dimension to balance better representation of the input factors and prevent latent-layer overfitting. After examining the training window of 102, 204, 306 and 408 with the validation window of 30, 60, 90 and 120, respectively, the window size of training is configured to 102 (validation size 30). This aligns with the relatively short sight of standard Transformer model design. In other words, the standard Transformer model is originally designed for short-context prediction, for example, one phrase to predict another phrase, or one sentence to forecast another sentence. Parameters are re-estimated every 12 months (step size 12) for the balance of computational efficiency, training stability and forecasting accuracy. Therefore, the `in-sample total' observation number in Table~\ref{tab:data_des_ch3} includes 102 training, 30 validation and 45 iterations with 12 observations as step size moving forward ($102+30+45\times12=672$). 
\begin{table}[htbp!]
\centering
\begin{tabular}{cccc}
\hline
\textbf{Name} & \textbf{Start date} & \textbf{End date} & \textbf{Observation No.} \\
\hline
In-sample total & 1/1957 & 12/2012 & 672 \\
Testing (OOS) for ‘1911’ & 1/2013 & 11/2019 & 83 \\
Testing (OOS) for ‘2112’ & 1/2013 & 12/2021 & 108 \\
Testing (OOS) for ‘2212’ & 1/2013 & 12/2022 & 120 \\
\hline
\end{tabular}
\caption[Data splits for in-sample and out-of-sample data.]{Data splits for in-sample and out-of-sample data. ‘1911’ means the testing period ends before the pandemic happened, `2112’ means the testing period contains the pandemic period and ‘2212’ means the testing period contains not only the pandemic period but also one year after the pandemic.}
\label{tab:data_des_ch3}
\end{table}

\section{Models}\label{sec:Models}
\subsection{Overview of Transformer}\label{subsec：overview_Transformer}
\begin{figure}[htbp!]
\centering
\includegraphics[width=0.9\columnwidth]{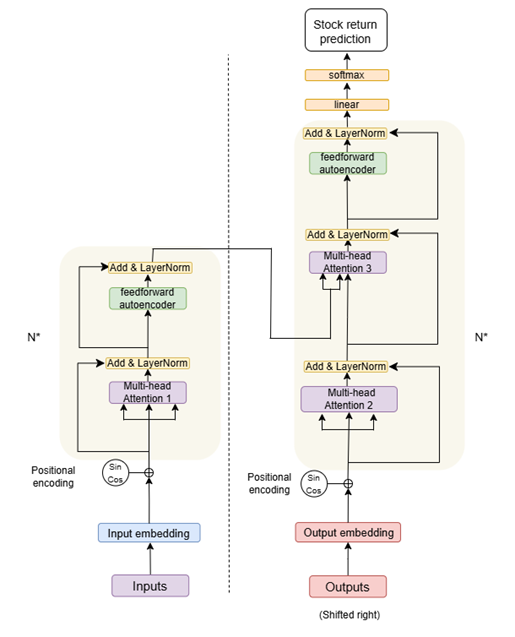}
\caption[The structure of standard Transformer]{The structure of standard Transformer,  source from Vaswani, A., Shazeer, N., Parmar, N., Uszkoreit, J., Jones, L., Gomez, A. N., Kaiser, Ł., \& Polosukhin, I. (2017). Attention is all you need. Advances in Neural Information Processing Systems, 30. The left side is the encoder, and the right side is the decoder. Both the inputs and outputs experienced the word embedding process which turned the word sequences into machine-readable vectors. N× means how many blocks are on each side. N is a hyperparameter. The structures in a rounded rectangle are defined as one block.}
\label{fig:sd_trans_411_ch3}
\end{figure}

The underlying structure of the Transformer is the autoencoder neural network structure, which was first proposed by \citet{Rumelhart1986LearningErrors}. Figure~\ref{fig:sd_trans_411_ch3} shows the earliest version of the Transformer from \citet{Vaswani2017AttentionNeed}. The left-side panel with ‘input’ are the encoder blocks, and the right-side panel with ‘output’ are the decoder blocks. This is a complex autoencoder structure developed from a simple autoencoder framework. $N^*$ is denoted as the number of encoder or decoder blocks. To simplify the structure and fit it to the context of asset pricing, the $N$ is configured to $1$. For the encoder block, the standard Transformer has one multi-head attention layer assembled with a one-hidden-layer feedforward (FFN) MLP autoencoder structure. Inputs, after embedding and positional encoding processes, turn into sequences with identical dimensions of output embedding results and are fed into the encoder block as the features. These features have two forms of utilization, one goes into the add and layer normalization (Add \& Norm) process through a bypass, while another form goes into the multi-head attention layer, then arrives at the Add \& Norm process and is jointly calculated with the previous form. Likewise, the sum of the attention layer output and the features is either passed through the bypass and fed into the next Add \& Norm process, or fed into the FFN structure after layer normalization. The output from the second Add \& Norm passes through the cross-sectional attention layer, the multi-head attention layer links the encoder and decoder block, which is also known as the latent layer or latent space of an autoencoder.\\

In the decoder block, the original output experiences the output embedding and positional encoding, then either travels directly to the Add \& Norm process or is fed through the ‘multi-head attention2’ layer and acts as the output of that layer to the Add \& Norm process. Repetitively, the outcome of the first Add \& Norm process either flows to the next Add \& Norm process or generates the attention ‘Query’ for the cross-sectional attention layer. The ‘Key’ and ‘Value’ of the cross-attention layer (‘multi-head attention3’) are derived from the output of the connected encoder. The output of the cross-attention layer is processed directly by the Add \& Norm process or passed through the final feedforward MLP autoencoder, then rejoins the ‘Add \& Norm’ process. The entire procedure, after estimation and validation, yields the optimal weights for the value forecasts (linear layer) or the probability forecasts (softmax layer). In this chapter, the standard Transformer with an MLP autoencoder pre-training module (Pre-trained Transformer) and the encoder-only (with no cross-attention and decoder block) with an MLP autoencoder pre-training module (SERT) are examined for asset pricing tasks. They are benchmarked with the layer-normalization first (LNF) transformation models and standard Transformer and encoder-only Transformer models.\\

\subsubsection{Positional encoding}\label{subsubsec：Positional encoding}
\begin{figure}[htbp]
    \centering
    \begin{subfigure}[c]{0.95\textwidth}
        \centering
        \includegraphics[width=\linewidth, height=0.25\textheight, keepaspectratio]{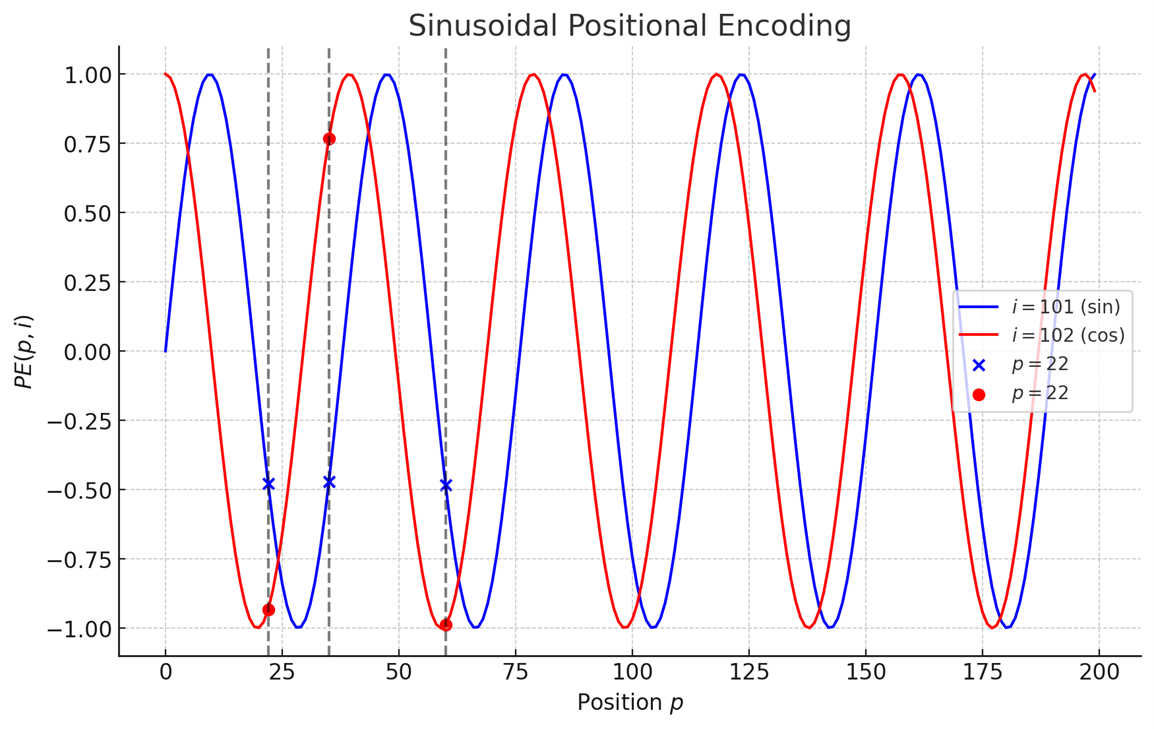}
        \caption{Up}
    \end{subfigure}
    
    \vspace{0.5em}
    \begin{subfigure}[c]{0.95\textwidth}
        \centering
        \includegraphics[width=\linewidth, height=0.25\textheight, keepaspectratio]{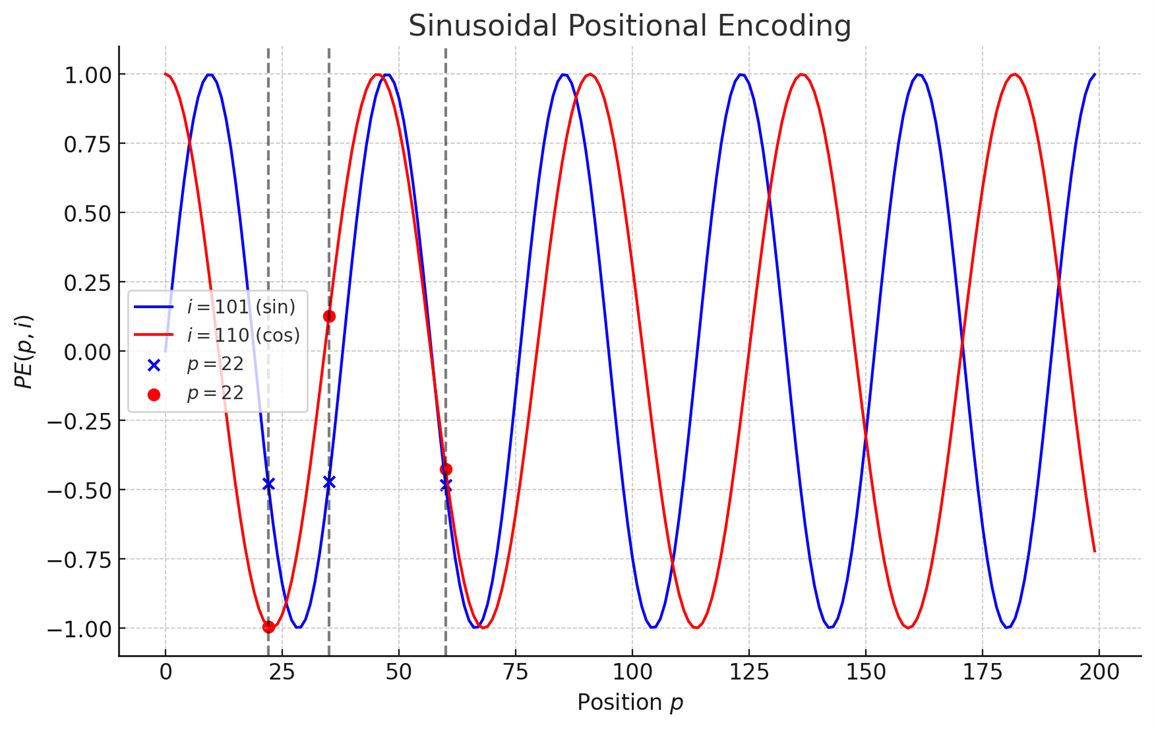}
        \caption{Middle}
    \end{subfigure}
    
    \vspace{0.5em}
    \begin{subfigure}[c]{0.95\textwidth}
        \centering
        \includegraphics[width=\linewidth, height=0.25\textheight, keepaspectratio]{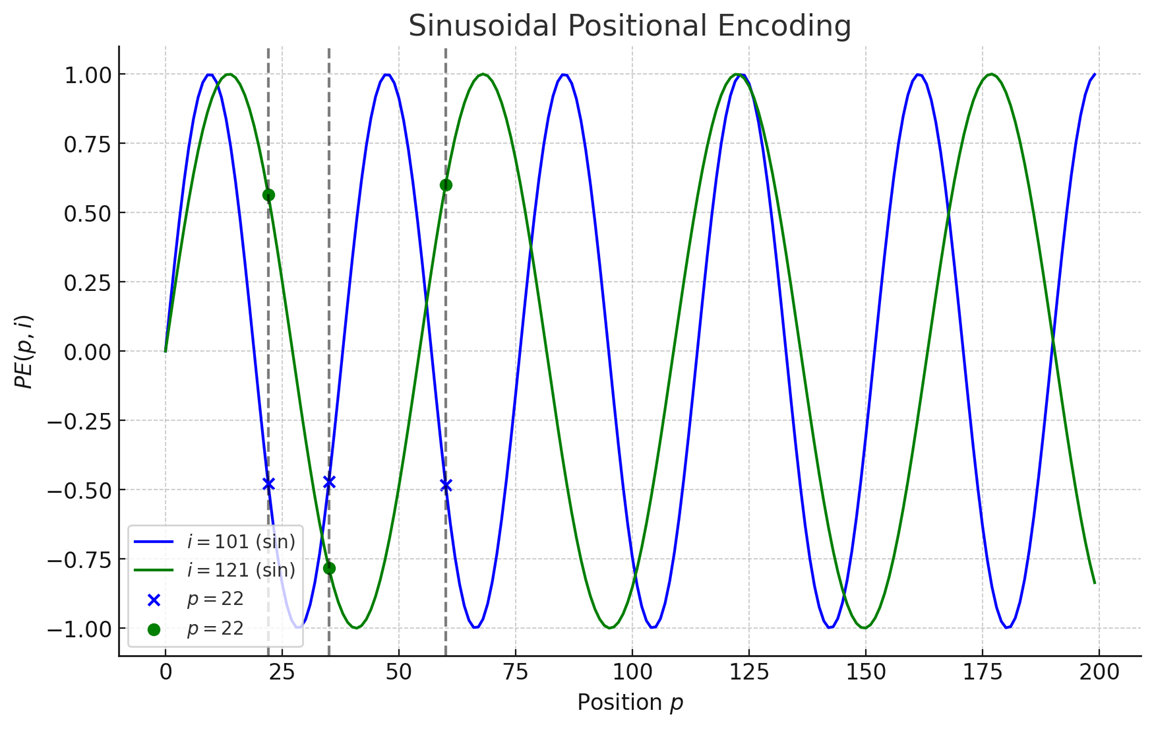}
        \caption{Bottom}
    \end{subfigure}

    \caption[Sinusoidal positional encoding plots]{Sinusoidal positional encoding plots. The X-axis shows the positions (time steps), and the Y-axis shows the positional codes. The upper panel shows the encoding values of the 101\textsuperscript{st} and 102\textsuperscript{nd} features; the middle panel shows the 101\textsuperscript{st} and 110\textsuperscript{th} features; and the bottom panel shows the 101\textsuperscript{st} and 121\textsuperscript{st} features. $p=22$ means the 22\textsuperscript{nd} time step.}
    \label{fig:pos_ch3}
\end{figure}

Like textual data, time series data has sequences, but unlike the recurrent neural network (RNN) models, the parallel structure improves computational efficiency, but the self-attention mechanism cannot identify the sequence unless an added function provides a guide. Also, the MLP autoencoder structure (feedforward neural network) inside the Transformer model has no mechanism designed for temporal sequence capturing. A talented positional encoding design was proposed in Transformer’s seminal paper \citep{Vaswani2017AttentionNeed}, named Sinusoidal Positional Encoding (SPE). It can be mathematically presented as:
\begin{align}
PE_{(pos, 2i)} &= \sin\left(\frac{pos}{10000^{\frac{2i}{d_{\text{model}}}}} \right)\label{eq:pos_enc1_ch3} \\
PE_{(pos, 2i+1)} &= \cos\left(\frac{pos}{10000^{\frac{2i}{d_{\text{model}}}}} \right)\label{eq:pos_enc2_ch3}
\end{align}
Where $PE$ is the positional encoding, $pos$ is the position, $i$ is the index of the dimension (features), $d_{model}$ is the dimension of the input. These functions imply that features with odd sequential numbers are encoded with the sine function, and the ones with even sequential numbers are encoded as cosine functions. Figure~\ref{fig:pos_ch3} exhibits an example of how positional encoding marks the positions and notes periodic information for individual features. The sine and cosine functions originally have directionality for indicating the time steps, while the difference between the sine and cosine functions indicates the difference between two features at the identical time step. 
We can see the mechanism of SPE from Figure~\ref{fig:pos_ch3} that in the individual time step, positional codes identify different features via different wavelengths, and the distance between the round point and cross point on the identical grey dashed line demonstrates the difference between the two features. For individual features, the periodic characteristic in the sine or cosine function assists in marking the periodicity of the input features.\\

Compared with other large data applications of Transformer, parameter efficiency is a non-negligible aspect which should be considered cautiously. That is the main reason that non-parametrized SPE is preferred to Time2Vec positional encoding deployed in the works of \citet{Lezmi2023TimeManagement,Ma2023AttentionApproach,Tevare2023ForecastingTransformer}. Moreover, the summation adding positional information is superior to concatenation adding with a simplified parameter structure. In the stock pricing, if the $X'\in \mathbb{R}^{T\times d_{\text{model}}}$
 is defined as the embedded factor matrix, the positional encoded input $X’$ is equal to $X+PE$. How the positional codes work with the attention mechanism for capturing spatial-temporal variation is introduced in the appendix of Chapter 4. \\

The SPE fulfils all requirements for a good positional encoding solution, for example, unique for individual time steps, deterministic, consistent distance between any two time steps and better generalizability for different data lengths. It not only marks out the absolute positions for time steps but also indicates the relative positions. In addition, \citet{Vaswani2017AttentionNeed} also mention that there is no significant difference between fixed and learnable positional encoding methods in their Transformer model. \\

\subsubsection{Autoencoder}\label{subsubsec：Autoencoder4.1.3}
The fundamental frame of the Transformer is an autoencoder structure, which was first proposed by \citet{Rumelhart1986LearningErrors}. It is developed for dimensional reduction and feature learning assignments. An autoencoder has at least one encoder block and one decoder block. The left panel of Figure~\ref{fig:autoenc2_4141_ch3} shows the framework of the general autoencoder. The encoder connects the input layer, and the decoder connects the output layer. For an autoencoder, the input dimension and output dimension need not be identical if it has suitable dimension embeddings. The autoencoder mechanism is highly specialized in pivotal factor extraction. Thus, it was widely used for tasks such as data denoise, anomaly detection and prediction. Although the asset pricing data size sometimes is ‘a drop in the ocean’ compared with alternative contexts, such as video generation and speech recognition, the simplified autoencoder structure can still be fit for such tasks, which are exemplified by \citet{Gu2021AutoencoderModels}.\\ 

The Transformer employs complex structures upon the autoencoder framework; the simplified Transformer model with a single encoder and single decoder has a single-layer MLP autoencoder module linked to one multi-head attention layer, respectively. It deploys a cross-attention layer (layer in blue colour in Figure~\ref{fig:autoenc2_4141_ch3}) as the latent layer connects the encoder and decoder. For stock pricing or factor investing, linear embedding is a typical substitution for the ‘Word2Vec’ embedding commonly applied in NLP tasks.\\

\subsubsection{Attention mechanism}\label{subsubsec:Attention mechanism414}
\begin{figure}[htbp!]
\centering
\includegraphics[width=0.95\columnwidth, height=0.4\textheight, keepaspectratio]{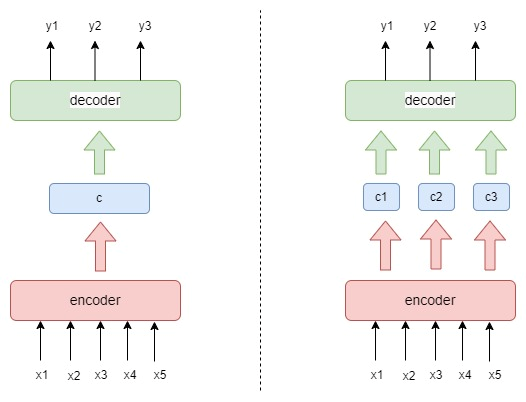}
\caption[General attention mechanism workflow chart.]{General attention mechanism workflow chart. The left panel is the general autoencoder structure, while the right panel is the autoencoder with an attention mechanism. In the left panel, $c$ is the output vector of the encoder, which means the autoencoder without an attention mechanism distributes equal weights to the output of the encoder in each time step. In the right panel, $c1,c2,c3$ indicate attention mechanism adds different weights on different time step $t$}
\label{fig:autoenc2_4141_ch3}
\end{figure}

The self-attention mechanism was first proposed by \citet{Vaswani2017AttentionNeed}. It is developed from \citet{Luong2015EffectiveTranslation}’s dot-product attention, which is introduced in Chapter 3. As discussed in the section on related work of this chapter, the traditional attention mechanisms rely on the output of the encoder, which means they could be highly affected by the quality of the encoder output. In addition, compressing the entire series in a fixed hidden state may cause information losses, especially for long temporal dependency data. In the self-attention mechanism, the calculation of the attention score is independent, which implies no external information, such as lagged labels (stock returns in this case), assistance. In the standard Transformer model, the self-attention layer in the encoder computes attention scores only from the input series, whereas the one in the decoder computes them only from the output series. It can lower the noise from external information. Also, the self-attention mechanism computes attention weights for all positions; it captures dependencies between any two time steps. Moreover, the self-attention in Transformer models enables parallel computing since its calculation is based on a matrix, which significantly increases computational efficiency. The right panel of Figure~\ref{fig:autoenc2_4141_ch3} provides a general picture of the attention mechanism in the autoencoder. \\

\begin{figure}[htbp!]
\centering
\includegraphics[width=0.95\columnwidth, height=0.4\textheight, keepaspectratio]{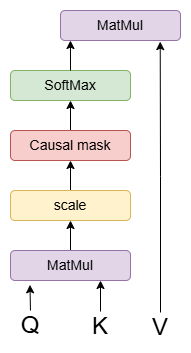}
\caption[Workflow of single-head attention mechanism]{Workflow of single-head attention mechanism. ‘MatMul’ means matrix multiplication. ‘scale’ implies the division by $\sqrt{d_{\text{model}}}$.}
\label{fig:satt1_ch3}
\end{figure}

The cross-attention mechanism is the bridge between the encoder and decoder in the Transformer; it matches the information supported by the encoder and decoder dynamically. In other words, the joint layer of the Transformer cannot only capture the information provided by the extracted factors from the original input factors but also dynamically concentrate on the lagged stock excess returns as factors. Section~\ref{fig:satt1_ch3} further demonstrates the mechanism of cross-attention. \\

Transformer models were originally designed for NLP tasks, it could be single-directional including generative pre-trained Transformer (GPT) or bidirectional such as BERT. The bidirectional Transformer models for NLP predictions predict the missing phrase according to the preceding and succeeding context, which implies that future data leakage may exist if the model is applied directly to time series data. Alternations for Transformer models, such as causal masks, should be taken on for a specific time series compulsorily. Section~\ref{subsubsec:Single-head self-attention4141} also discusses how causal masks prevent future data leakage.\\
\paragraph{Single-head self-attention}\label{subsubsec:Single-head self-attention4141}
Unlike the Single-head self-attention in Chapter 3, the self-attention in the Transformer model is based on the autoencoder architecture. This means it can be computed in parallel, matrix-wise. It computes the attention score according to the entire feature group. One self-attention layer contains three input vectors. They are Query (Q), Key (K) and Value (V). Query determines which information should be focused on. The Key is the match of Query for calculating the attention weights, while the Value represents the actual information, and it multiplies with the attention weights for the layer output. Figure~\ref{eq:sattention1_ch3} shows the workflow of single-headed attention. Concretely, the first step is generating Q,K,V via the transformation of input features X. Equation~\eqref{eq:sattention1_ch3} illustrates the derivation of Q,K,V.\\
\begin{equation}
Q = W^Q X,\quad K = W^K X,\quad V = W^V X
\label{eq:sattention1_ch3}
\end{equation}
Where $W^Q,W^K,W^V$ are the weight matrices derived via estimation like MLP models. The second step is computing stabilized attention scores via dot product Q,K and dimension-wise scaling operation.\\
\begin{equation}
\text{Attention Score} = \frac{QK^\top}{\sqrt{d_{\text{model}}}}
\label{eq:sattention2_ch3}
\end{equation}
For the transformer in a time series context, the third step (causal mask in Figure~\ref{fig:satt1_ch3}) is indispensable and crucial, since it is not only sequential but also temporal. Practically, due to the softmax function nature, it can be represented as '0' for visible and $-\infty$ for non-visible.
In Equation~\eqref{eq:sattention_mask3_ch3}, $M_{ij}$ is the notation of the mask matrix, $i$ is the time step that earlier than time step $j$ initially, and it can be presented as: 
\begin{equation}
M_{ij} = 
\begin{cases}
0, & \text{if } i < j \\
-\infty, & \text{if } i > j
\end{cases}
\label{eq:sattention_mask3_ch3}
\end{equation}
\begin{equation}
M = \begin{bmatrix}
0 & \cdots & -\infty \\
\vdots & \ddots & \vdots \\
0 & \cdots & 0
\end{bmatrix}
\label{eq:sattention_mask4_ch3}
\end{equation}
Where the first column indicates the time step 1, the last column indicates the time step $t$. Thus, the masked attention output can be presented as:
\begin{equation}
\textit{Masked Attention}(Q, K, V) = \text{Softmax} \left( \frac{\text{QK}^\mathrm{T}}{\sqrt{d_{\text{model}}}} + M \right) V
\label{eq:satt_ch3}
\end{equation}
Where the $\frac{\text{QK}^\mathrm{T}}{\sqrt{d_{\text{model}}}} + M$ is defined as the attention score. It is weighted via the softmax function (attention weights). 

\paragraph{Multi-head self-attention}\label{subsubsec:Multi-head self-attention4142}
Multi-head self-attention is designed for the case where the dimension of extracted factors is large. It performs exceptionally well among the cross-sectionally oversized features in the general ML context. It separates the latent factors into subgroups and applies single-head attention within each factor group. It is an ideal approach for increasing the quality of information extraction. Concretely, for $Head_i$, the masked attention for feature group $i$ is:
\begin{align}
\textit{Head}_i(Q_i, K_i, V_i) 
&= \text{Softmax} \left( \frac{Q_i K_i^\mathrm{T}}{\sqrt{d_k}} + M_i \right) V_i, \quad i \in \{1,2,\dots,h\}
\label{eq:att_head_ch3}
\end{align}

After concatenating and linearly connecting the outputs of each head, the final output of the masked multi-head attention can be computed:
\begin{equation}
\textit{Masked MultiHead}(Q, K, V) = \text{Concat}( \text{Head}_1, \text{Head}_2, \dots, \text{Head}_h ) W^O
\end{equation}
Where $W^O \in \mathbb{R}^{h d_k \times d_{\text{model}}}$ is the weight matrix for linear transformation on heads outcomes. $d_k$ is the dimension for head $i$, $h$ is the number of heads, where $d_k = \frac{d_{\text{model}}}{h}$.\\

\begin{figure}[htbp!]
\centering
\includegraphics[width=0.95\columnwidth, height=0.4\textheight, keepaspectratio]{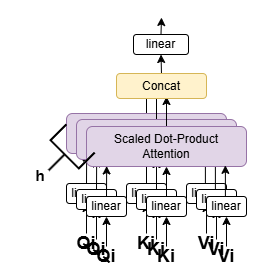}
\caption[workflow of multi-head attention mechanism.]{workflow of multi-head attention mechanism. Source from: Vaswani, A., Shazeer, N., Parmar, N., Uszkoreit, J., Jones, L., Gomez, A. N., Kaiser, Ł., \& Polosukhin, I. (2017). Attention is all you need. Advances in Neural Information Processing Systems, 30. The ‘Qi’, ‘Ki’ and ‘Vi’ are the Query, Key and Value in group i, while h is the number of the heads. ‘linear’ is noted as the linear function.}
\label{fig:multi_h_att_41421_ch3}
\end{figure}

Compared with single-head attention, multi-head attention potentially improves the model's representability by jointly considering the relation of factors intra and inter-factor groups. It also stabilizes the model training by sharing the workload for each head. Figure~\ref{fig:multi_h_att_41421_ch3} shows the workflow of multi-head attention. \\
\paragraph{Cross-attention}\label{subsubsec:Cross-attention4143}
The conjunction of encoder and decoder blocks in a standard Transformer is the cross-attention layer. Unlike the self-attention mechanism, which takes Q,K,V all from the input features, it takes Q from the decoder output and K,V from the encoder output:
\begin{equation}
Q_{de} = H_{de} W^Q, \quad K_{en} = H_{en} W^K, \quad V_{en} = H_{en} W^V
\end{equation}
Where $H_{de},H_{en}$ are the outputs of the decoder and encoder, respectively. Afterwards,
\begin{equation}
\textit{Masked Cross Attention}(Q_{de}, K_{en}, V_{en}) 
= \text{Softmax} \left( \frac{Q_{de} K_{en}^\mathrm{T}}{\sqrt{d_H}} + M_{\text{cross}} \right) V_{en}
\end{equation}
Where $M_{cross}$ is the mask for the cross-attention layer, and $d_H$ is the dimension of the hidden layer output of the encoder and decoder. Similarly, the multi-head cross attention is adding heads to handling the grouped inputs.\\

The feedforward MLP autoencoder in the Transformer is the same as that in the pre-training module in Chapter 3 and in this chapter. A more generalized introduction can be found in the appendix of Chapter 4.\\

\subsubsection{Add and layer normalization}\label{subsubsec:Add and layer normalization416}
As introduced in the ‘overview of Transformer’ section, each outcome from the previous stage is added to the output of this stage. After layer normalization, the summation outcome is passed to the next stage. Using the ‘Add\&Norm’ after single-head self-attention layer as an example, it comes:
\begin{equation}
X' = \text{LayerNorm} \left( X + \textit{Masked Attention}(Q, K, V) \right)
\end{equation}
Where $X$ is the output of the previous layer.
\subsubsection{Estimation}\label{subsubsec:Estimation417}
The estimation of Transformer models in this study minimises the MSE function with an L1 regularization term (to prevent latent overfitting) through the method of Stochastic Gradient Descent (SGD) with the Adaptive Moment Estimation (Adam) optimizer. The detailed derivation of estimation for the standard Transformer model is explained in the appendix of Chapter 4.
\subsection{The proposed pre-trained Transformer}\label{subsec:The proposed pre-trained Transformer42}
Researchers further improve the Transformer models’ performance by incorporating an additional module named ‘pre-training’. \citet{Clark2020Electra:Generators, Raffel2020ExploringTransformer}, verified that pre-trained Transformers can significantly increase the predictability of NLP tasks. In this study, the pre-trained Transformer, as well as the layer normal first (LNF) pre-trained Transformer, are also examined in the context of stock pricing and factor investing. It deploys an MLP autoencoder as the pre-training module to pre-train the inputs, namely, the portfolio-sorted factors. \\

The advantages of the MLP autoencoder structure as a pre-training method mainly lie in denoising and pattern detection via input factor restructuring, thereby assisting the attention mechanism in the main body of the Transformer model. The proposition of structurally increasing the size of economic and financial data is supported by \citet{Didisheim2024APTModels}. In the context of the factor model in this study, the noisy input factors with missing values first pass through the pre-trained autoencoder module to project the dimensionality from 182 to 420, aligning with the output dimension for the cross-attention layer. Slightly different from the MLP autoencoder in Chapter 3, the MLP autoencoder in the pre-train module is employed as the dimensional projection (extending the dimension to align with the decoder) method instead of the information extraction (dimensionality reduction). Explicitly, it can be seen as the generalized principal components analysis (PCA) process \citep{Gu2021AutoencoderModels} or PCA without the restriction of orthogonal and linearity, which extracts the original factors with better representativity. In this case, the MLP autoencoder improves the information captured from noisy original inputs and enlarges the limited data size to some extent via the high dimensionality of the latent space. As a byproduct, it perfectly solves the missing value existing in the input factors as well as minimizes the noise of the original input. \\

\begin{figure}[htbp!]
\centering
\includegraphics[width=0.95\columnwidth, height=0.8\textheight, keepaspectratio]{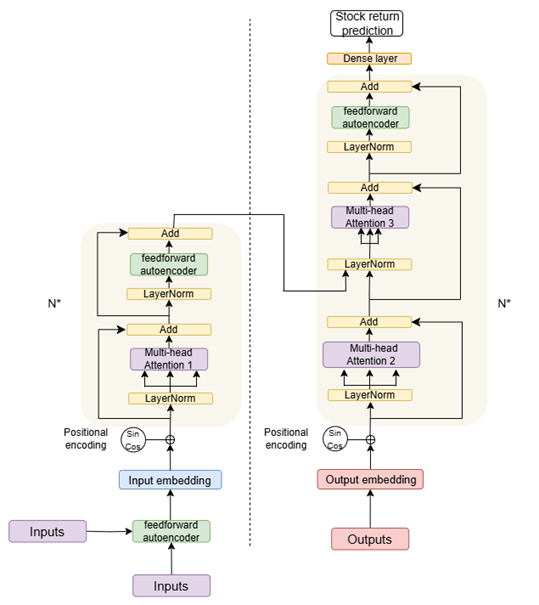}
\caption[The structure of the proposed pre-trained Transformer with layer normalization first.]{The structure of the proposed pre-trained Transformer with layer normalization first.}
\label{fig:PTrans423_LNF_ch3}
\end{figure}

In addition, \citet{Liu2020EnhancingNetworks} suggest altering the ‘Add \& LayerNorm’ process into layer normalization first (LNF), then ‘Add’ could stabilize the training process, moderate the ‘gradient vanish’ and ‘explosion’ issues and improve the training speed. Mathematically, it can alter the Equation~\eqref{eq:LNF_ch3} into: 
\begin{equation}
X' = X + \textit{sublayer} \left( \text{LayerNorm}(X) \right)
\label{eq:LNF_ch3}
\end{equation}
Where $\textit{sublayer}(\cdot)$ is the function of output from the previous layer. Figure~\ref{fig:PTrans423_LNF_ch3} shows the proposed model of the LNF version.\\

\begin{figure}[htbp!]
\centering
\includegraphics[width=0.95\columnwidth, height=0.8\textheight, keepaspectratio]{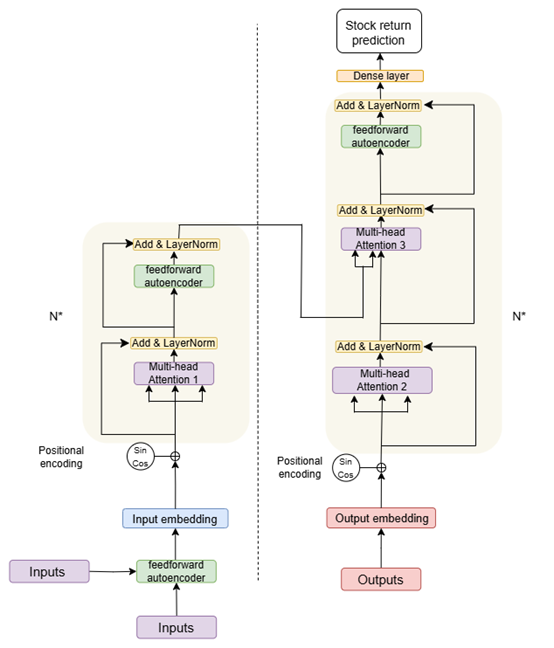}
\caption[The structure of the proposed pre-trained Transformer.]{The structure of the proposed pre-trained Transformer. In the left panel, the first feedforward autoencoder is the pre-trained MLP autoencoder for enlarging the input dimension. And the input embedding is the linear embedding for numerical data. The rest of the two feedforward autoencoder modules are normal MLP autoencoders with one input and one output layer. `Multi-head attention 1' and `Multi-head attention 2' are self-attention modules, while `Multi-head attention 3' is a cross-attention module which connects the encoder and decoder. $N$ is the number of encoder or decoder blocks. And $N^*$ is the number of encoder or decoder blocks.}
\label{fig:PTrans422_ch3}
\end{figure}
\subsection{Single directional Encoder Representations from Transformer (SERT)}\label{subsec:SERT43}
\begin{figure}[htbp!]
    \centering
    \begin{minipage}[c]{0.45\textwidth}
        \centering
        \includegraphics[width=0.9\textwidth]{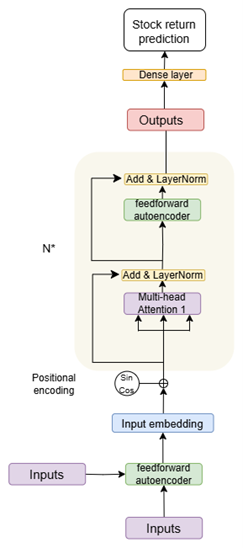}
    \end{minipage}
    \begin{minipage}[c]{0.45\textwidth}
        \centering
        \includegraphics[width=0.9\textwidth]{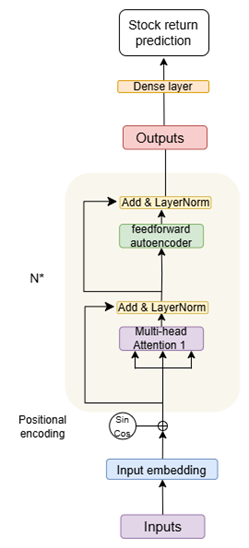}
    \end{minipage}
    \caption[single directional encoder representations from Transformer (SERT) and benchmark model.]{single directional encoder representations from Transformer (SERT) and benchmark model from \citet{Cong2021DeepPricing}. The left panel presents the structure of the proposed SERT model, and the right panel presents the standard encoder-only Transformer model in some FinTech and financial economics literature.}
\label{fig:SERT431_ch3}
\end{figure}

\citet{Kenton2019Bert:Understanding} propose a simplified Transformer structure for NLP tasks based on the standard Transformer. It is named Bidirectional Encoder Representations from Transformer (BERT). Compared with the standard Transformer, BERT removes the decoder structure, which significantly reduces the number of parameters; hence, the design of BERT is more suitable for data with smaller sizes. It moderates the overfitting issue and improves computational efficiency with less sacrifice of model predictability. However, since BERT is designed for language processing, its random masking pre-training process and bidirectional mechanism make it inapplicable for single-directional numerical time series prediction. Therefore, this study also proposes an alternative model named single-directional encoder representations from Transformer (SERT) for the time series in the context of stock pricing. The SERT changes the bidirectional mechanism into a single-directional mechanism via the time series causal mask. Also, it alters the random masking pre-training process for textual prediction to the MLP autoencoder pre-training process for adapting numerical time series data with missing values and noise. In that sense, SERT also belongs to the encoder-only Transformer model family. The left panel of Figure~\ref{fig:SERT431_ch3} presents the structure of the proposed SERT model. As no decoder block is present, the encoder-only Transformer models remove the cross-attention layer and the entire decoder block, which contains a feedforward MLP autoencoder module and a self-attention layer. Instead, it connects the feedforward MLP autoencoder module of the Transformer’s encoder block linearly to the output layer. After estimation and optimization, the optimal weights are derived for stock return prediction.\\

Analogous to the SERT model, standard encoder-only Transformer models are widely used in different contexts. \citet{Cong2021DeepPricing} employed the model as a representation of the Transformer models for comparative study on NN models in the asset pricing context, but it does not have a significant out-of-sample fitness. In this study, their model is also set as a benchmark to see whether the SERT performs better than the encoder-only Transformer model in the stock pricing or factor investing context. The right panel of Figure~\ref{fig:SERT431_ch3} shows the benchmark model employed by \citet{Cong2021DeepPricing}.\\

\subsection{Transformers in stock pricing and factor investing}\label{subsec:Trans_invest}
In the context of stock pricing factor models, the original input is the sorted portfolio factors \citep{AndrewY.Zimmermann2020OpenPricing}, while the original output is the large capital U.S. stock excess returns. Unlike the work in \citep{Gezici2024DeepPrediction,Paivarinta2022Transformer-basedEvidence,Zhang2022AssetLearning}, they deploy the entire Transformer model including the softmax layer for predicting the probability of the trading signals, this study modifies the Transformer models with a linear dense layer for predicting the out-of-sample (OOS) returns directly and using the softmax function as the trading signal filter for a comparative analysis to the sign signal generation (trend following) method. Ten Transformer models and ten encoder-only Transformer models (including SERT) are examined. They are separated into three groups: the LNF group (Group 1), the proposed pre-trained group (Group 2) and the standard group (Group 3), which are shown in Table~\ref{tab:Model_config_ch3}. The difference between the models within the group is the number of heads.\\

The Transformer models were implemented via Python (v3.8) using the PyTorch \citep{paszke2019pytorch} package of `nn.Transformer'. Training was conducted on the NVIDIA GeForce RTX 3070 GPU. Due to the parallel computation method, it is possible to apply it on a laptop without a GPU if the data size is limited.\\

\section{Empirical results}\label{subsec:Empirical results_ch3}
To study how the attention head affects the models’ performance in stock pricing, in this research, the standard Transformer-based models and encoder-only Transformer-based models are separated into 3 groups. ‘P\_Trans\_LNF’ and ‘SERT\_LNF’ are denoted for proposed pre-trained models with LNF. ‘P\_Trans\_H1’ to ‘P\_Trans\_H7’ and ‘SERT\_H1’ to ‘SERT\_H7’ as the proposed pre-trained models with different numbers of heads, while the ‘C\_Trans\_H1’ to ‘C\_Trans\_H4’ and ‘En\_Trans\_H1’ to ‘En\_Trans\_H4’ are the benchmark models without pre-training modules with heads. The module configurations are listed in Table~\ref{tab:Model_config_ch3}.
\begin{table}[htbp!]
\centering
\begin{tabular}{cccc|cccc}
\hline
\textbf{Model} & \textbf{Heads} & \textbf{Group No.} & \textbf{LNF} 
& \textbf{Model} & \textbf{Heads} & \textbf{Group No.} & \textbf{LNF} \\
\hline
SERT\_LNF  & 1 & 1 & Y & P\_Trans\_LNF  & 1 & 1 & Y \\
\hline
SERT\_H1  & 1 & 2 & N & P\_Trans\_H1  & 1 & 2 & N \\
SERT\_H2  & 2 & 2 & N & P\_Trans\_H2  & 2 & 2 & N \\
SERT\_H3  & 3 & 2 & N & P\_Trans\_H3  & 3 & 2 & N \\
SERT\_H4  & 4 & 2 & N & P\_Trans\_H4  & 4 & 2 & N \\
SERT\_H6  & 6 & 2 & N & P\_Trans\_H6  & 6 & 2 & N \\
SERT\_H7  & 7 & 2 & N & P\_Trans\_H7  & 7 & 2 & N \\
\hline
En\_Trans\_H1  & 1 & 3 & N & C\_Trans\_H1  & 1 & 3 & N \\
En\_Trans\_H2  & 2 & 3 & N & C\_Trans\_H2  & 2 & 3 & N \\
En\_Trans\_H4  & 4 & 3 & N & C\_Trans\_H4 & 4 & 3 & N \\
\hline
\end{tabular}
\caption[Model configurations.]{Model configurations. The ‘Heads’ shows the head number of the attention layers, and the ‘Group No.’ is the notation to classify model groups. The ‘LNF’ indicates the option of ‘layer normalization first’ where ‘Y’ means yes, and ‘N’ means no. 'En' indicates encoder-only, and 'H' indicates head followed by the head number. 'C' means classic Transformer models. All models in Group 3 are the benchmark models without the MLP pre-training process.}
\label{tab:Model_config_ch3}
\end{table}
The model performance and backtesting performance evaluation follow the methods of Chapter 2. Due to the projected large-dimension latent factors (from the pre-training process) being diluted from the original factors, the correlation between the latent factors and original factors is insignificant to evaluate the impact of the factors. Therefore, evaluating variable importance (factor impact) is of less significance in this chapter. \\

\subsection{Model performance of Transformer}\label{subsec:Model performance of Transformer_ch3}
Table~\ref{tab:model_perform511_ch3} shows the model performance of the proposed Transformer models and their benchmarks which encompass the three periods denoted as ‘1911’, ‘2112’ and ‘2212’ respectively. From the average OOS $R^2$ and MSE, the proposed Transformer models with different heads (P\_Trans\_H1 to P\_Trans\_H7) outperform the benchmark models (P\_Trans\_LNF and C\_Trans\_H1 to C\_Trans\_H4) in all periods. The advantages of the proposed models are significant, particularly during the extreme market fluctuations (sharp uptrend in ‘2112’ and sharp up-down fluctuation in ‘2212’). This means the proposed models have marked advantages in tolerating extreme market conditions. However, within each group, there are fewer differences shown between average OOS fitness indicators. This means that adding heads to attention layers helps but does not improve models’ fitness significantly. These indicators agree that Model P\_Trans\_H3, the proposed pre-trained Transformer with 3 attention heads, is the best-performing model in all periods. Interestingly, across all periods and within each group, the average OOS $R^2$ increases with head numbers, but it peaks and then declines. For example, in the proposed pre-trained Transformer group, it peaks at 3 heads and in the standard Transformer group, it peaks at 2 heads. This implies that there exists an optimal attention head number for Transformer models in a certain case.\\

From the factor investing perspective, all models in different periods show significant and positive mean $\alpha$ values. The average positive residual $\alpha$ values provide strong evidence of market inefficiencies and excess predictability across the three periods. These anomalies suggest the existence of the traditional $\alpha$ and the uncaptured non-linear risk premiums, which is where factor investors can potentially benefit from.\\

\begin{table}[htbp]
  \centering
  \begin{tabular}{lccccc}
    \toprule
    \textbf{Model} & \textbf{Avg\_R2} & \textbf{Avg\_MSE} & \textbf{avg\_$\alpha$} & \textbf{Ann.$\alpha$} & \textbf{$\alpha$\_t\_statistic} \\
    \midrule
    \multicolumn{6}{c}{\textbf{Pre-COVID-19 Period (1911)}} \\
    \midrule
    P\_Trans\_LNF & 0.0377 & 0.0057 & 0.0082 & 0.0981 & 21.2026*** \\
    P\_Trans\_H1  & 0.0658 & 0.0055 & 0.0116 & 0.1388 & 30.0030*** \\
    P\_Trans\_H2  & 0.0679 & 0.0055 & 0.0120 & 0.1439 & 30.9728*** \\
    P\_Trans\_H3  & 0.0763 & 0.0054 & 0.0094 & 0.1126 & 24.9751*** \\
    P\_Trans\_H4  & 0.0746 & 0.0054 & 0.0111 & 0.1336 & 29.2173*** \\
    P\_Trans\_H6  & 0.0737 & 0.0054 & 0.0107 & 0.1281 & 28.3103*** \\
    P\_Trans\_H7  & 0.0679 & 0.0055 & 0.0120 & 0.1442 & 31.0214*** \\
    C\_Trans\_H1  & 0.0536 & 0.0055 & 0.0118 & 0.1421 & 30.1773*** \\
    C\_Trans\_H2  & 0.0571 & 0.0055 & 0.0071 & 0.0851 & 18.2654*** \\
    C\_Trans\_H4  & 0.0325 & 0.0057 & 0.0088 & 0.1058 & 22.9111*** \\
    \midrule
    \multicolumn{6}{c}{\textbf{COVID-19-Inclusive Period (2112)}} \\
    \midrule
    P\_Trans\_LNF & 0.0224 & 0.0077 & 0.0101 & 0.1207 & 31.3827*** \\
    P\_Trans\_H1  & 0.1029 & 0.0071 & 0.0113 & 0.1362 & 34.8214*** \\
    P\_Trans\_H2  & 0.1036 & 0.0070 & 0.0118 & 0.1420 & 35.9810*** \\
    P\_Trans\_H3  & 0.1113 & 0.0070 & 0.0098 & 0.1174 & 30.7129*** \\
    P\_Trans\_H4  & 0.1095 & 0.0070 & 0.0117 & 0.1399 & 35.6825*** \\
    P\_Trans\_H6  & 0.1052 & 0.0070 & 0.0106 & 0.1274 & 33.0697*** \\
    P\_Trans\_H7  & 0.1040 & 0.0070 & 0.0124 & 0.1487 & 37.1062*** \\
    C\_Trans\_H1  & 0.0273 & 0.0076 & 0.0127 & 0.1526 & 36.9573*** \\
    C\_Trans\_H2  & 0.0276 & 0.0077 & 0.0093 & 0.1115 & 29.2770*** \\
    C\_Trans\_H4  & 0.0123 & 0.0078 & 0.0105 & 0.1262 & 32.5920*** \\
    \midrule
    \multicolumn{6}{c}{\textbf{Including COVID-19 and One-Year After (2212)}} \\
    \midrule
    P\_Trans\_LNF & 0.0234 & 0.0081 & 0.0088 & 0.1053 & 31.3703*** \\
    P\_Trans\_H1  & 0.0907 & 0.0076 & 0.0099 & 0.1187 & 35.9190*** \\
    P\_Trans\_H2  & 0.0911 & 0.0075 & 0.0104 & 0.1245 & 37.4614*** \\
    P\_Trans\_H3  & 0.0972 & 0.0075 & 0.0085 & 0.1017 & 30.9821*** \\
    P\_Trans\_H4  & 0.0960 & 0.0075 & 0.0101 & 0.1217 & 36.9343*** \\
    P\_Trans\_H6  & 0.0928 & 0.0075 & 0.0092 & 0.1108 & 33.7605*** \\
    P\_Trans\_H7  & 0.0917 & 0.0075 & 0.0108 & 0.1299 & 38.6464*** \\
    C\_Trans\_H1  & 0.0267 & 0.0081 & 0.0111 & 0.1332 & 38.6795*** \\
    C\_Trans\_H2  & 0.0266 & 0.0081 & 0.0081 & 0.0967 & 29.2690*** \\
    C\_Trans\_H4  & 0.0144 & 0.0082 & 0.0091 & 0.1095 & 33.4911*** \\
    \bottomrule
  \end{tabular}
\caption[Model performance indicators of Transformers.]{Model performance indicators of Transformers. Model P\_Trans\_LNF is the proposed model with layer normalization first. P\_Trans\_H1 to P\_Trans\_H7 are proposed pre-trained Transformer models and C\_Trans\_H1 to C\_Trans\_H4 are the standard Transformer models as benchmarks. In the ‘alpha\_t\_statistic’ column, values with three stars mean it is significant at the 99\% level, two stars mean 95\% significance level, and one star means 90\% significance level. The notation of 'Ann' means annualized.}
\label{tab:model_perform511_ch3}
\end{table}

The three-stage Diebold-Mariano (DM) test with HAC estimator robust results shown in Table~\ref{tab:dm_trans_ch3} illustrate the significance level that one model outperforms the alternatives. The negative value means the model named with the column label outperforms the model named with the row label, such as the ‘-0.44’ in the first column, which indicates that P\_Trans\_LNF outperforms C\_Trans\_H4. From the panel of ‘1911’, the difference between models is insignificant. However, during the extreme market fluctuations, models in different groups perform significantly differently from each other, although models still perform insignificantly within groups. While the proposed pre-trained Transformer models demonstrate substantial improvements in overall predictive accuracy in Table~\ref{tab:Model_perform521_SERT_ch3}, the Diebold-Mariano (DM) test with HAC standard errors suggests that the superiority is not strictly uniformly distributed across the testing period. The Diebold-Mariano (DM) test, which is based on Mean Absolute Error (MAE), yields limited statistical significance. This divergence highlights the varying sensitivities of evaluation metrics to extreme market fluctuations. Specifically, MSE disproportionately penalizes large errors, revealing that pre-trained Transformer models' primary advantage lies in successfully capturing massive structural breaks during the highly volatile COVID-19 periods. In contrast, the MAE-based DM test is robust to outliers and primarily reflects median daily performance, where the marginal absolute differences between models are less pronounced. Furthermore, the HAC estimator strictly penalizes the strong error autocorrelation typical of volatile periods, further suppressing the test statistic. Consequently, the insignificant DM results do not imply inferior model performance; rather, they demonstrate that pre-trained Transformer models' superiority is asymmetrically concentrated in absorbing extreme tail risks rather than yielding uniform improvements. The pre-trained Transformer models outperform benchmark models in all periods and significantly surpass the benchmarks in the extreme market conditions (‘2112’ and `2212’) and insignificantly in the ‘1911’. This proves that pre-trained Transformer models have a great capability to handle extreme market fluctuations. \\

Additionally, although \citet{Liu2020UnderstandingTransformers} suggest that LNF could increase the training efficiency, stabilize the training process and improve the model performance for Transformer models, especially for deep structure, in this examination, LNF does not assist in increasing model performance in all testing periods. With the sacrifice of the FFN module dimension (20\% of the input dimension) and minimization of the hyperparameters for training stability, the LNF Transformer model demonstrates suboptimal efficiency regarding the performance of standard Transformer models. In the asset pricing context, this means incomplete information is extracted from the original input factors. In LNF cases, adding attention heads, increasing dimensions, and larger hyperparameters cause serious overfitting directly during the examinations. \\
\begin{sidewaystable}[htbp]
\centering
\small
\begin{tabularx}{\textwidth}{l *{9}{X}}
\toprule
\multicolumn{10}{c}{\textbf{Pre-COVID-19 Period (1911)}} \\
\midrule
 & P\_H1 & P\_H2 & P\_H3 & P\_H4 & P\_H6 & P\_H7 & P\_LNF & C\_H1 & C\_H2 \\
\midrule
P\_H1 & & & & & & & & & \\
P\_H2 & -0.44 & & & & & & & & \\
P\_H3 & -1.37 & -0.48 & & & & & & & \\
P\_H4 & -1.44 & -0.28 & 0.17 & & & & & & \\
P\_H6 & -0.66 & -0.25 & 0.05 & -0.06 & & & & & \\
P\_H7 & -0.47 & 0.16 & 0.66 & 0.61 & 0.4 & & & & \\
P\_LNF & 0.83 & 1 & 1.15 & 1.08 & 1.32 & 0.92 & & & \\
C\_H1 & -0.72 & -1.04 & -1.26 & -1.06 & -1.14 & -0.88 & 0.46 & & \\
C\_H2 & 0.13 & -0.34 & -0.59 & -0.5 & -0.65 & -0.28 & 0.97 & -0.59 & \\
C\_H4 & 1.12 & 1.34 & 1.46 & -1.37 & -1.57 & -1.23 & -0.51 & 0.95 & 1.41 \\
\bottomrule
\end{tabularx}

\vspace{0.3em}

\begin{tabularx}{\textwidth}{l *{9}{X}}
\toprule
\multicolumn{10}{c}{\textbf{COVID-19-Inclusive Period (2112)}} \\
\midrule
 & P\_H1 & P\_H2 & P\_H3 & P\_H4 & P\_H6 & P\_H7 & P\_LNF & C\_H1 & C\_H2 \\
\midrule
P\_H1 & & & & & & & & & \\
P\_H2 & -0.53 & & & & & & & & \\
P\_H3 & -1.62 & -0.56 & & & & & & & \\
P\_H4 & -1.75* & -0.45 & 0.03 & & & & & & \\
P\_H6 & -0.48 & -0.05 & 0.36 & 0.41 & & & & & \\
P\_H7 & -0.79 & -0.03 & 0.58 & 0.68 & 0.03 & & & & \\
P\_LNF & 1.83* & 2.03** & 2.11** & 2.07** & 2.24** & 1.94* & & & \\
C\_H1 & -1.88* & -2.14** & -2.20** & -2.11** & -2.09** & -2.02** & 0.2 & & \\
C\_H2 & 1.65* & -1.82* & -1.94* & -1.90* & -1.95* & -1.79* & 0.55 & -0.41 & \\
C\_H4 & 2.14** & 2.37** & 2.40** & -2.35** & -2.45** & -2.26** & -1.05 & 1.05 & 1.33 \\
\bottomrule
\end{tabularx}

\vspace{0.3em}

\begin{tabularx}{\textwidth}{l *{9}{X}}
\toprule
\multicolumn{10}{c}{\textbf{Period Including COVID-19 and One-Year After (2212)}} \\
\midrule
 & P\_H1 & P\_H2 & P\_H3 & P\_H4 & P\_H6 & P\_H7 & P\_LNF & C\_H1 & C\_H2 \\
\midrule
P\_H1 & & & & & & & & & \\
P\_H2 & -0.49 & & & & & & & & \\
P\_H3 & -1.56 & -0.56 & & & & & & & \\
P\_H4 & -1.68* & -0.45 & 0.03 & & & & & & \\
P\_H6 & -0.48 & -0.06 & 0.34 & 0.39 & & & & & \\
P\_H7 & -0.77 & -0.06 & 0.56 & 0.65 & 0.04 & & & & \\
P\_LNF & 1.82* & 2.01** & 2.09** & 2.05** & 2.22** & 1.92* & & & \\
C\_H1 & -1.88* & -2.13** & -2.19** & -2.10** & -2.09** & -2.02** & 0.18 & & \\
C\_H2 & 1.66* & -1.82* & -1.94* & -1.90* & -1.95* & -1.79* & 0.53 & -0.4 & \\
C\_H4 & 2.14** & 2.36** & 2.39** & -2.34** & -2.44** & -2.25** & -1.09 & 1.06 & 1.33 \\
\bottomrule
\end{tabularx}
\caption[DM test results of proposed pre-trained Transformer and benchmark models.]{DM test results of proposed pre-trained Transformer and benchmark models. Negative values demonstrate that the model with column labels as names performs worse than the one with row labels as names. 'P\_H(i)' and 'C\_H(i)' $i= 1,2,\dots,7$ is equivalent to 'P\_Trans\_H(i)' and 'C\_Trans\_H(i)' respectively, indicating the Transformer models. Significance levels: * $p<0.1$, ** $p<0.05$, *** $p<0.01$.}
\label{tab:dm_trans_ch3}
\end{sidewaystable}

\clearpage
\begin{sidewaysfigure}[htbp!]
\centering
\begin{subfigure}{0.32\textwidth}
  \includegraphics[width=\linewidth, height=0.22\textheight, keepaspectratio]{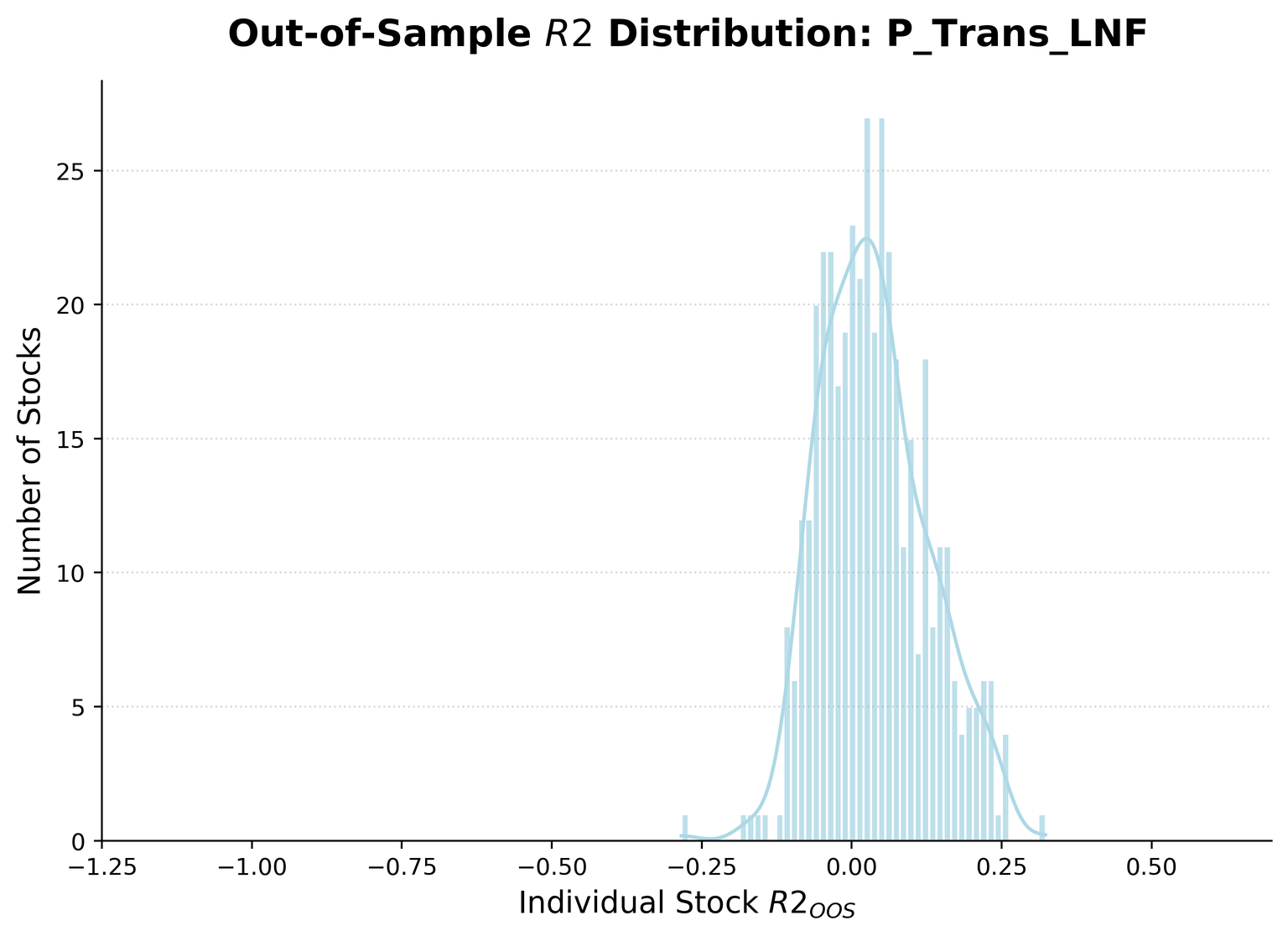}
  \caption{P\_Trans\_LNF(1911)}
\end{subfigure}
\hfill
\begin{subfigure}{0.32\textwidth}
  \includegraphics[width=\linewidth, height=0.22\textheight, keepaspectratio]{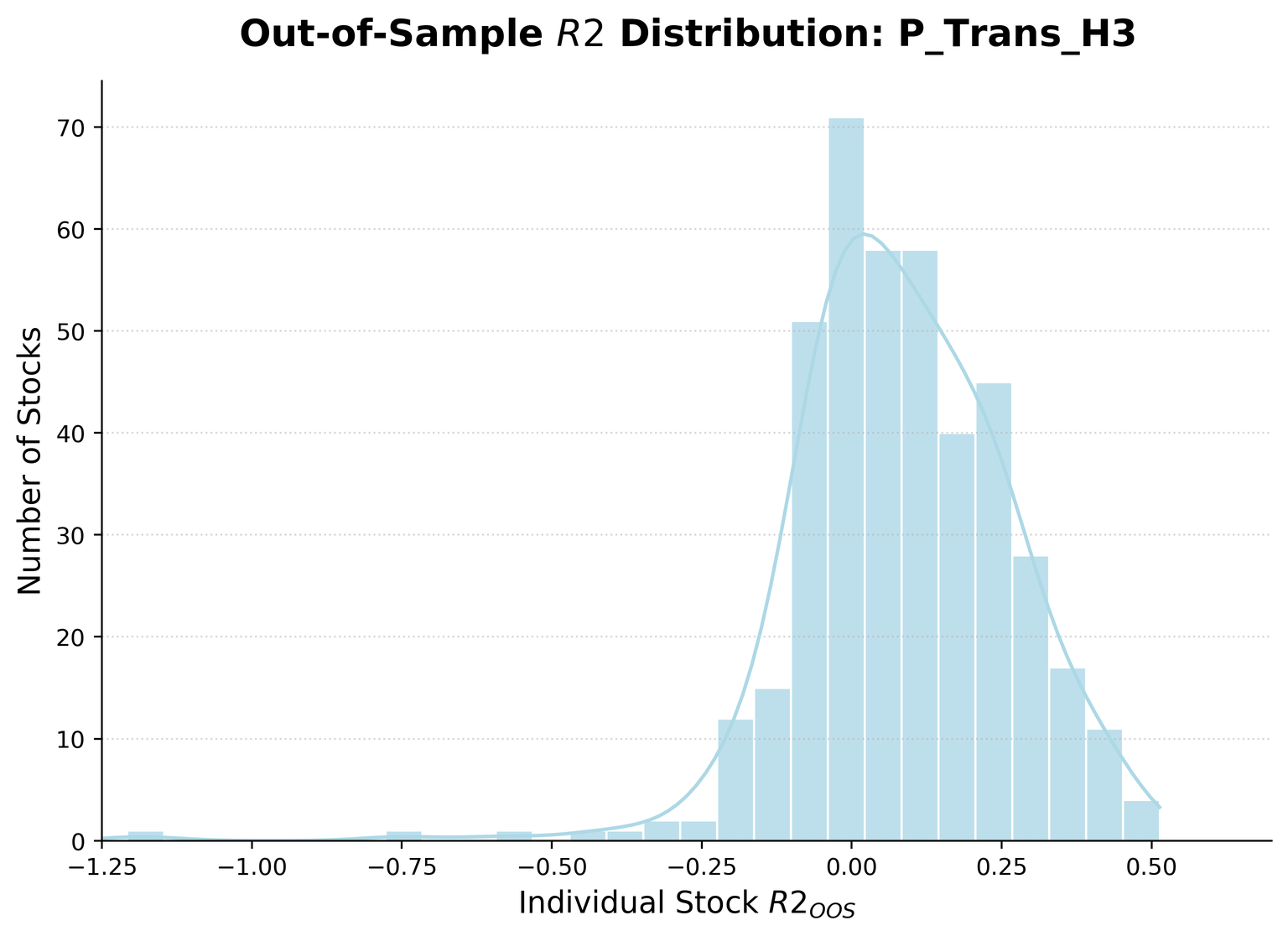}
  \caption{P\_Trans\_H3(1911)}
\end{subfigure}
\hfill
\begin{subfigure}{0.32\textwidth}
  \includegraphics[width=\linewidth, height=0.22\textheight, keepaspectratio]{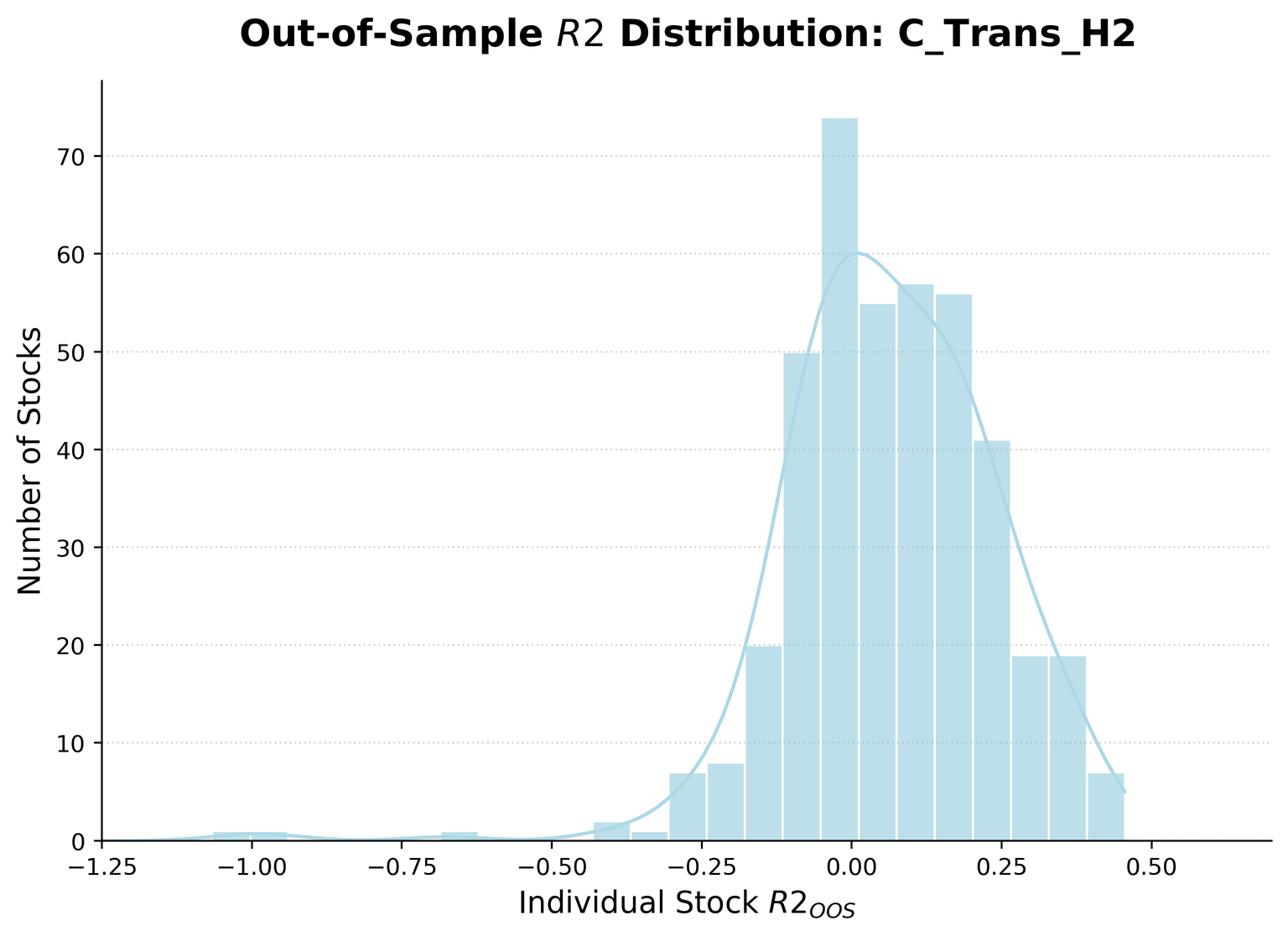}
  \caption{C\_Trans\_H2(1911)}
\end{subfigure}

\vspace{0em} 

\begin{subfigure}{0.32\textwidth}
  \includegraphics[width=\linewidth, height=0.22\textheight, keepaspectratio]{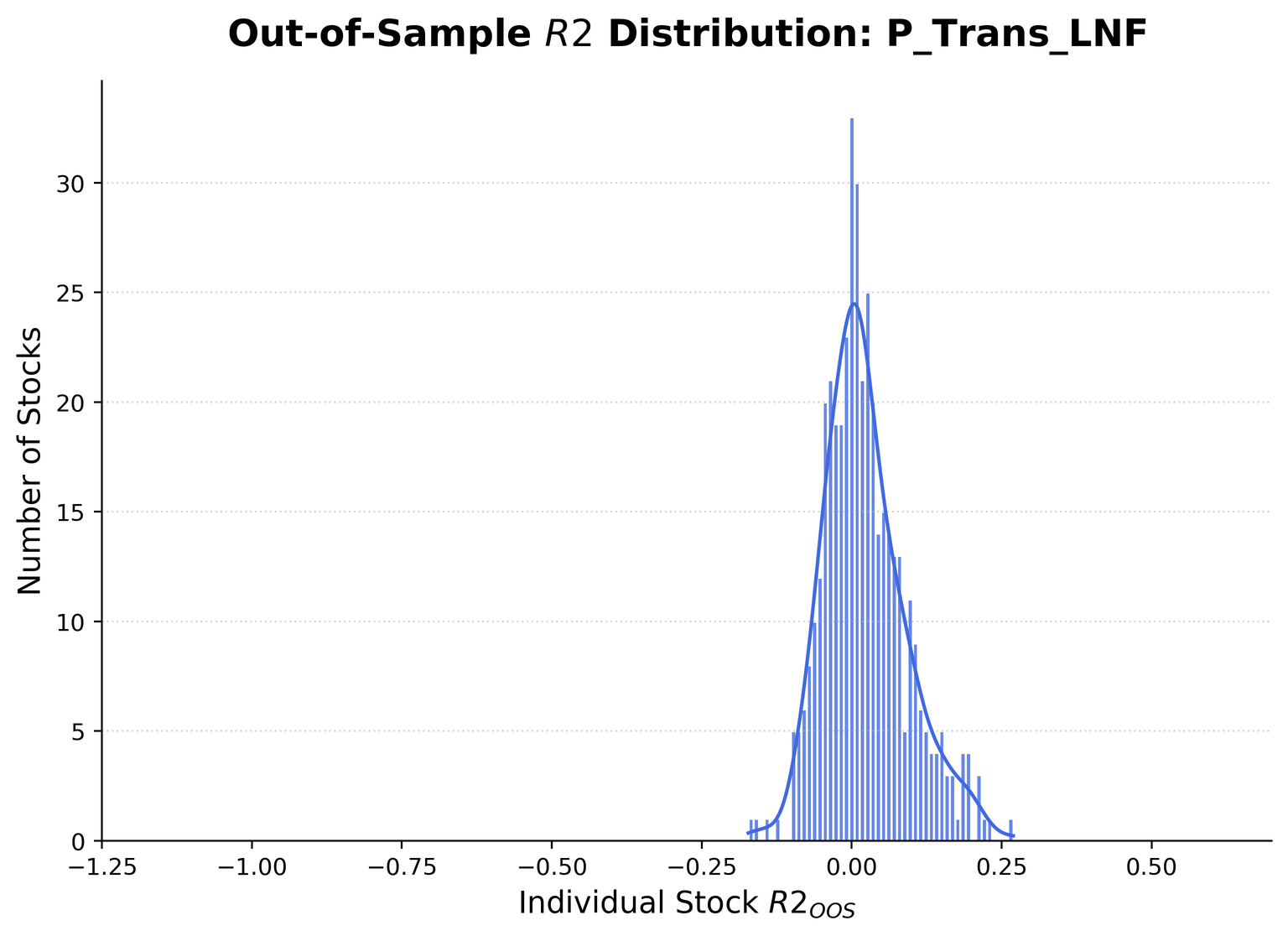}
  \caption{P\_Trans\_LNF(2112)}
\end{subfigure}
\hfill
\begin{subfigure}{0.32\textwidth}
  \includegraphics[width=\linewidth, height=0.22\textheight, keepaspectratio]{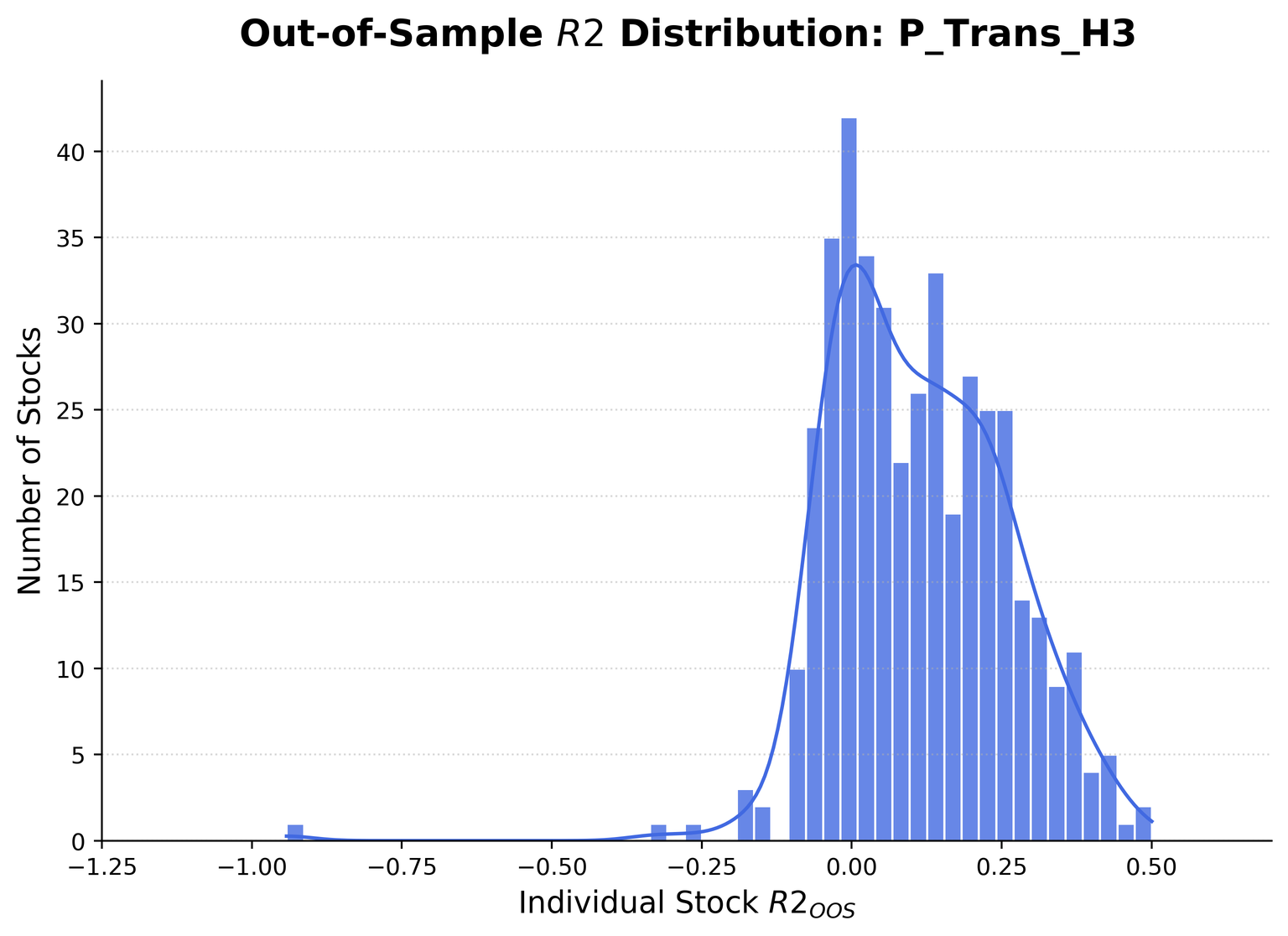}
  \caption{P\_Trans\_H3(2112)}
\end{subfigure}
\hfill
\begin{subfigure}{0.32\textwidth}
  \includegraphics[width=\linewidth, height=0.22\textheight, keepaspectratio]{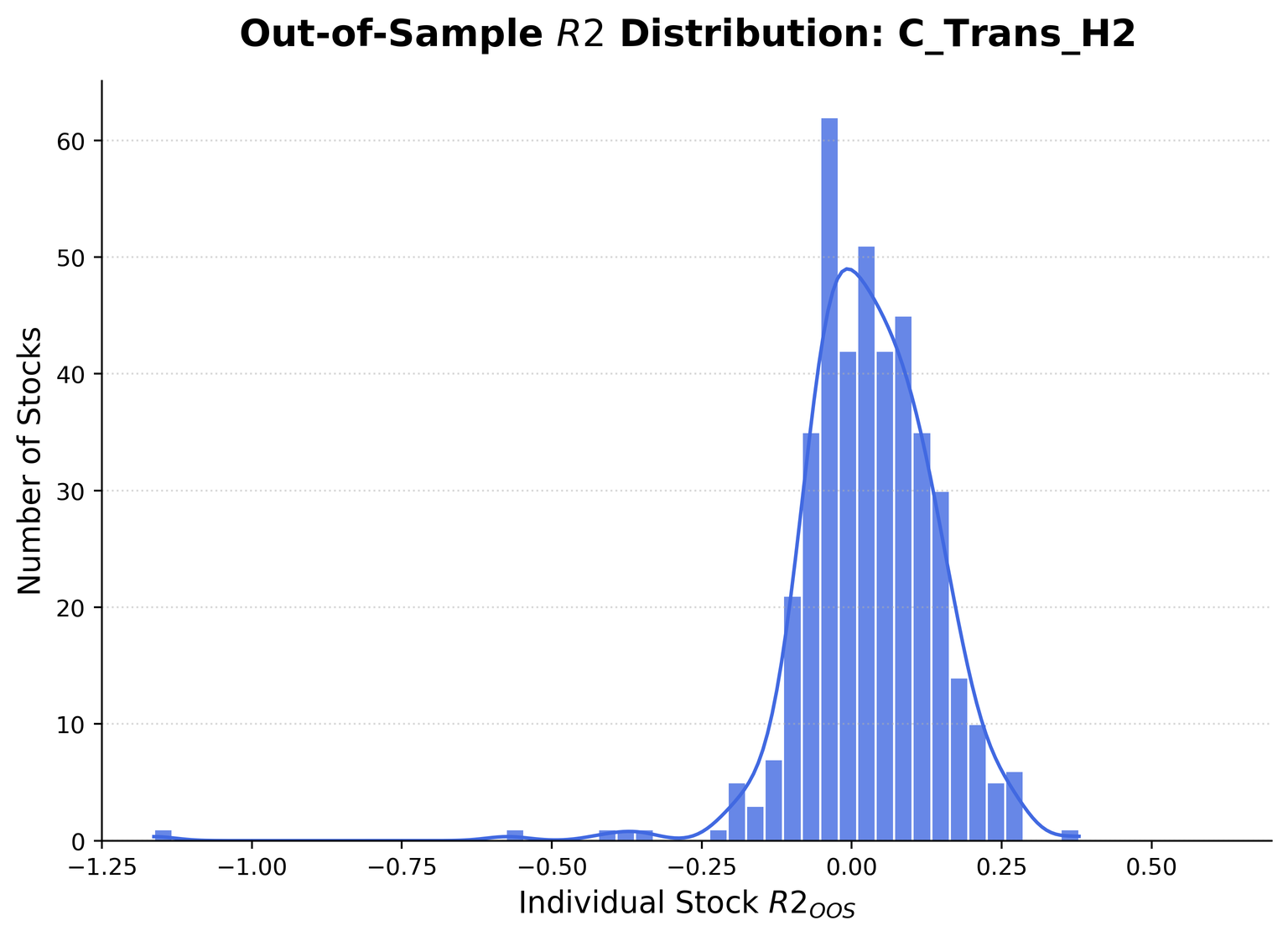}
  \caption{C\_Trans\_H2(2112)}
\end{subfigure}

\vspace{0em} 

\begin{subfigure}{0.32\textwidth}
  \includegraphics[width=\linewidth, height=0.22\textheight, keepaspectratio]{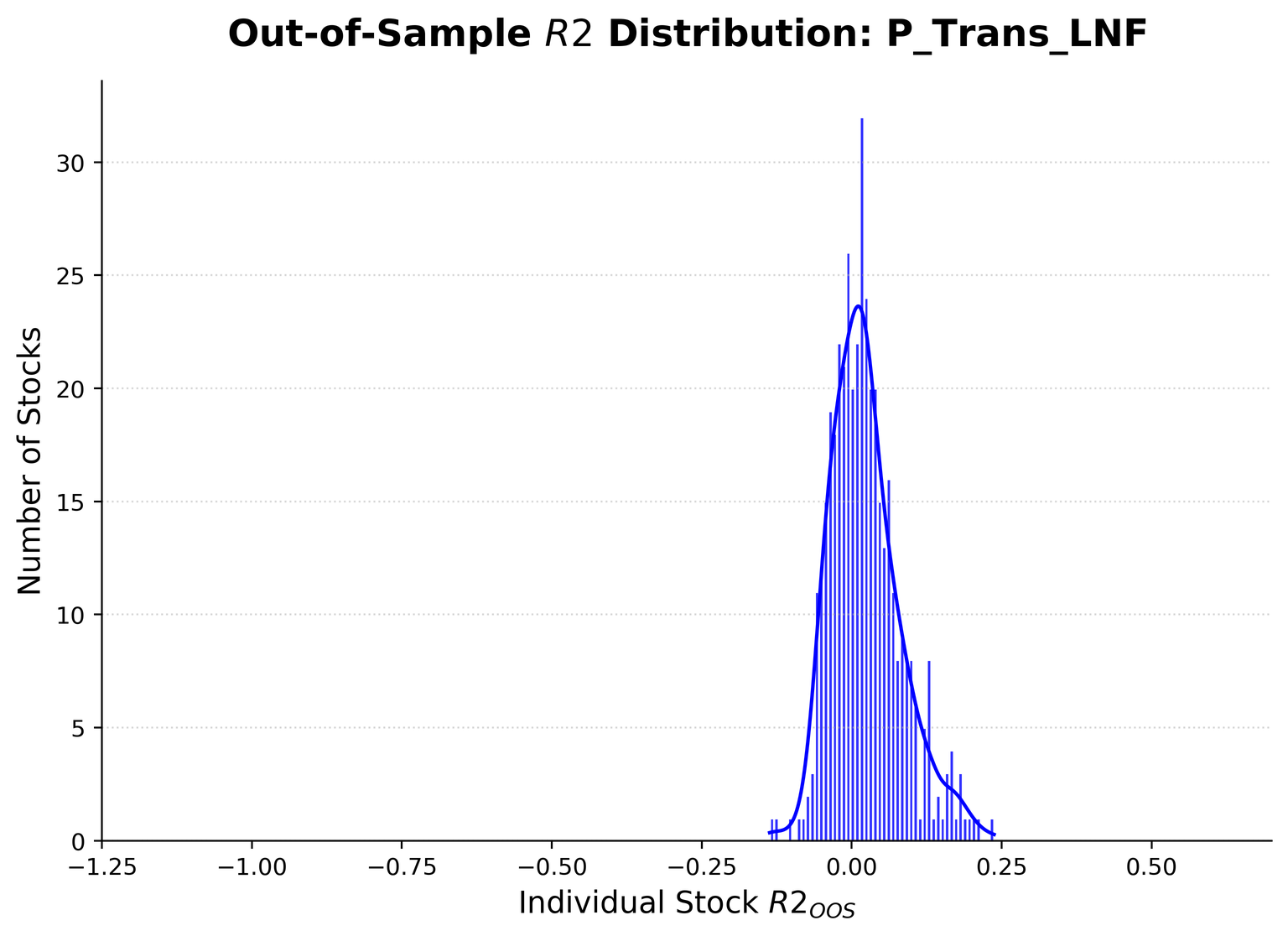}
  \caption{P\_Trans\_LNF(2212)}
\end{subfigure}
\hfill
\begin{subfigure}{0.32\textwidth}
  \includegraphics[width=\linewidth, height=0.22\textheight, keepaspectratio]{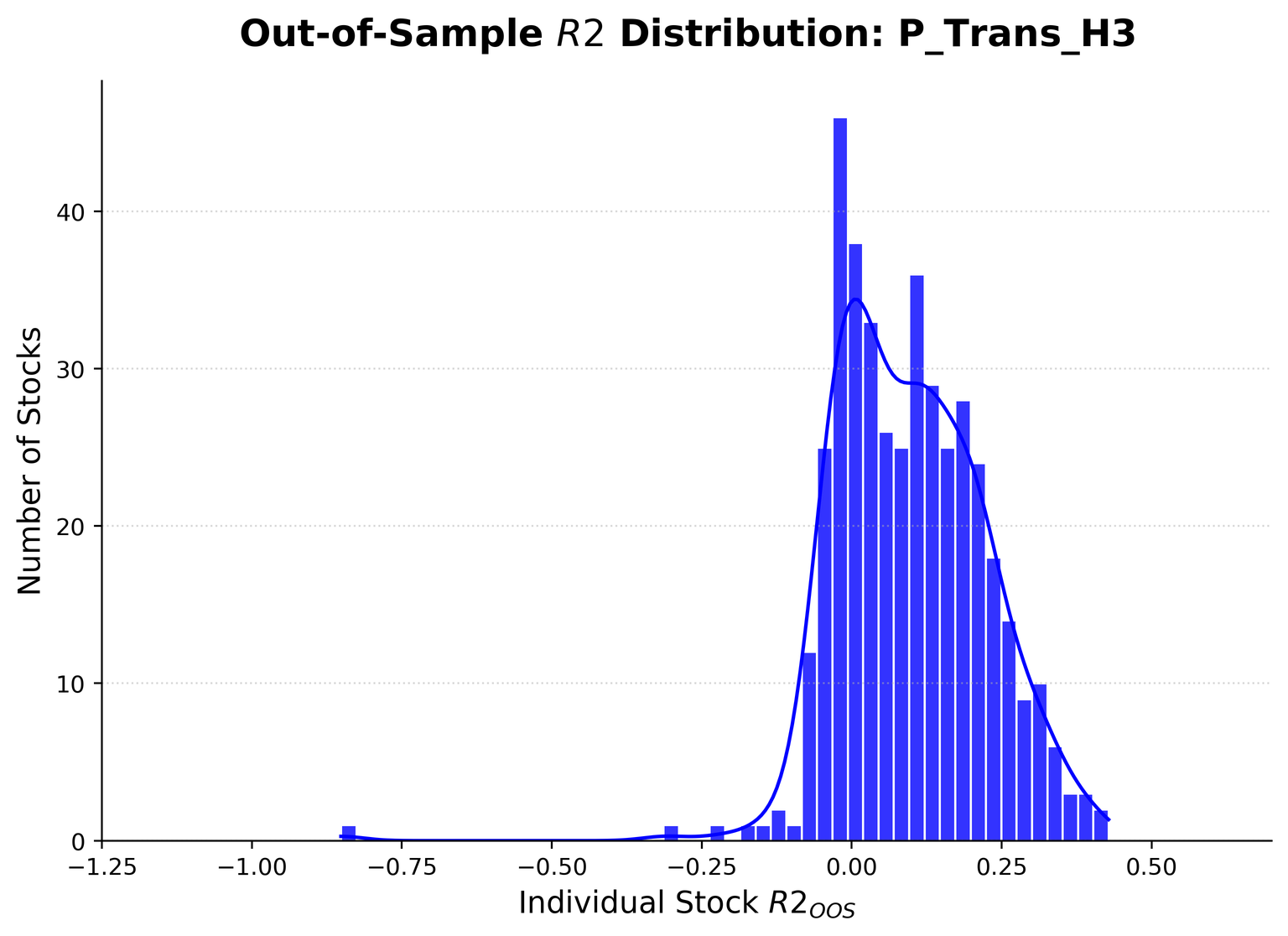}
  \caption{P\_Trans\_H3(2212)}
\end{subfigure}
\hfill
\begin{subfigure}{0.32\textwidth}
  \includegraphics[width=\linewidth, height=0.22\textheight, keepaspectratio]{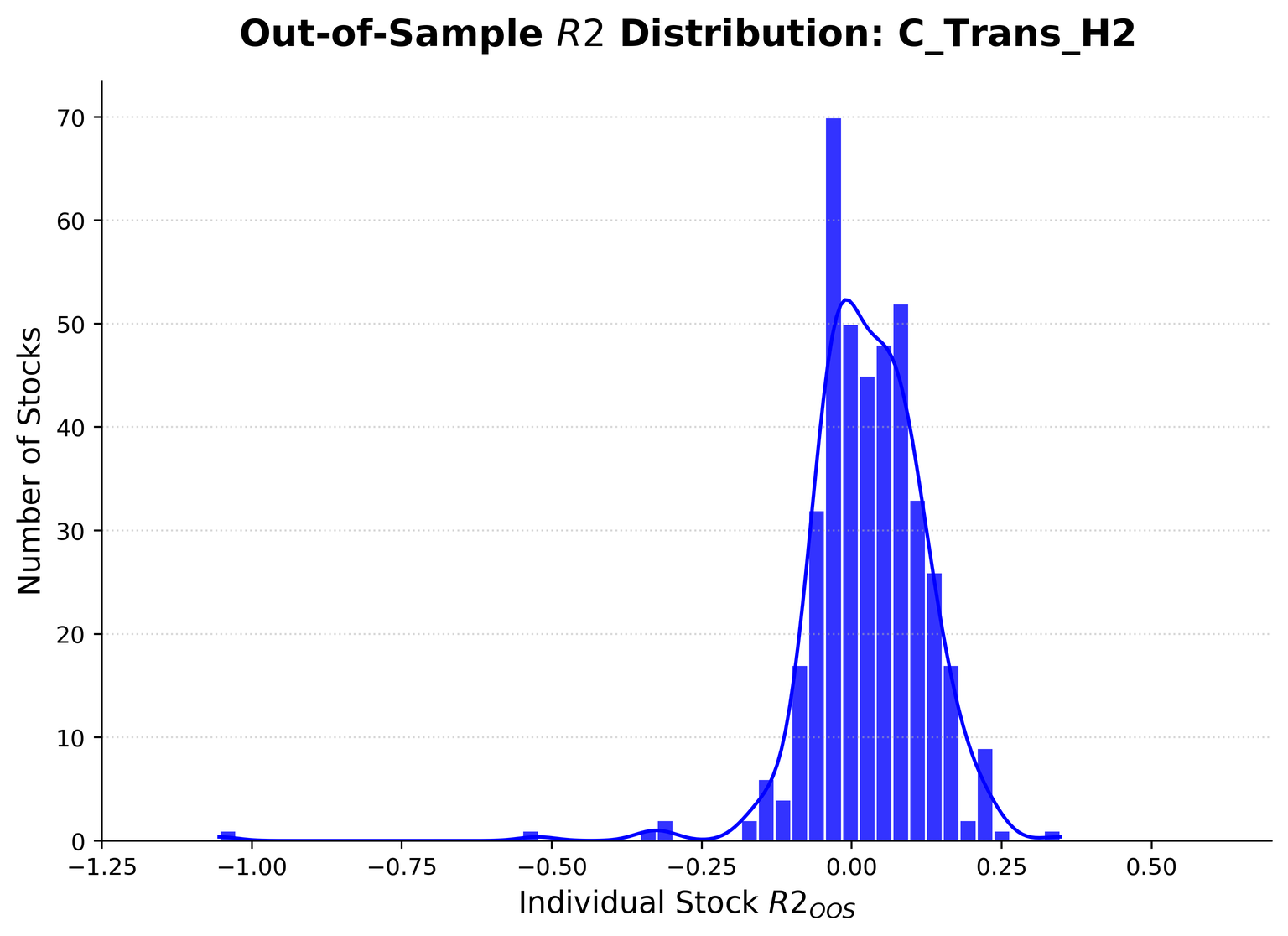}
  \caption{C\_Trans\_H2(2212)}
\end{subfigure}
 \caption[OOS $R^2$ distribution diagrams of best-performed Transformer model in each group for each period.]{OOS $R^2$ distribution diagrams of best-performed Transformer model in each group for each period. The 1911, 2112 and 2212 are short for Pre–COVID-19 Period, COVID-19–Inclusive Period and Period Including COVID-19 and One-Year After, respectively. X-axis shows the OOS $R^2$ value ranges in each subgraph, while the Y-axis shows the stock numbers.}
  \label{fig:r2_distribution512_Trans_ch3}
\end{sidewaysfigure}

\begin{sidewaysfigure}[htbp!]
\centering
\begin{subfigure}{0.32\textwidth}
  \includegraphics[width=\linewidth, height=0.22\textheight, keepaspectratio]{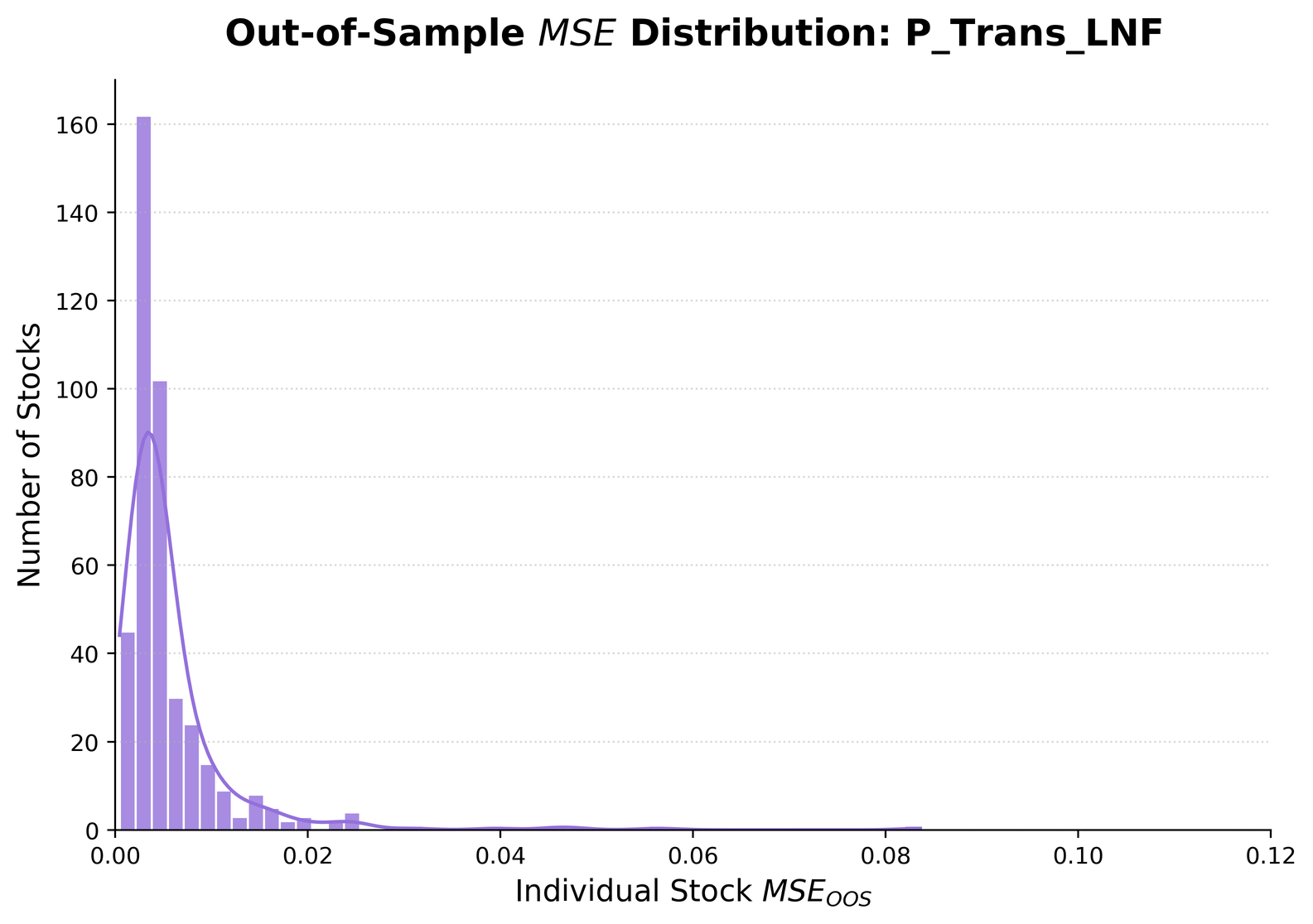}
  \caption{P\_Trans\_LNF(1911)}
\end{subfigure}
\hfill
\begin{subfigure}{0.32\textwidth}
  \includegraphics[width=\linewidth, height=0.22\textheight, keepaspectratio]{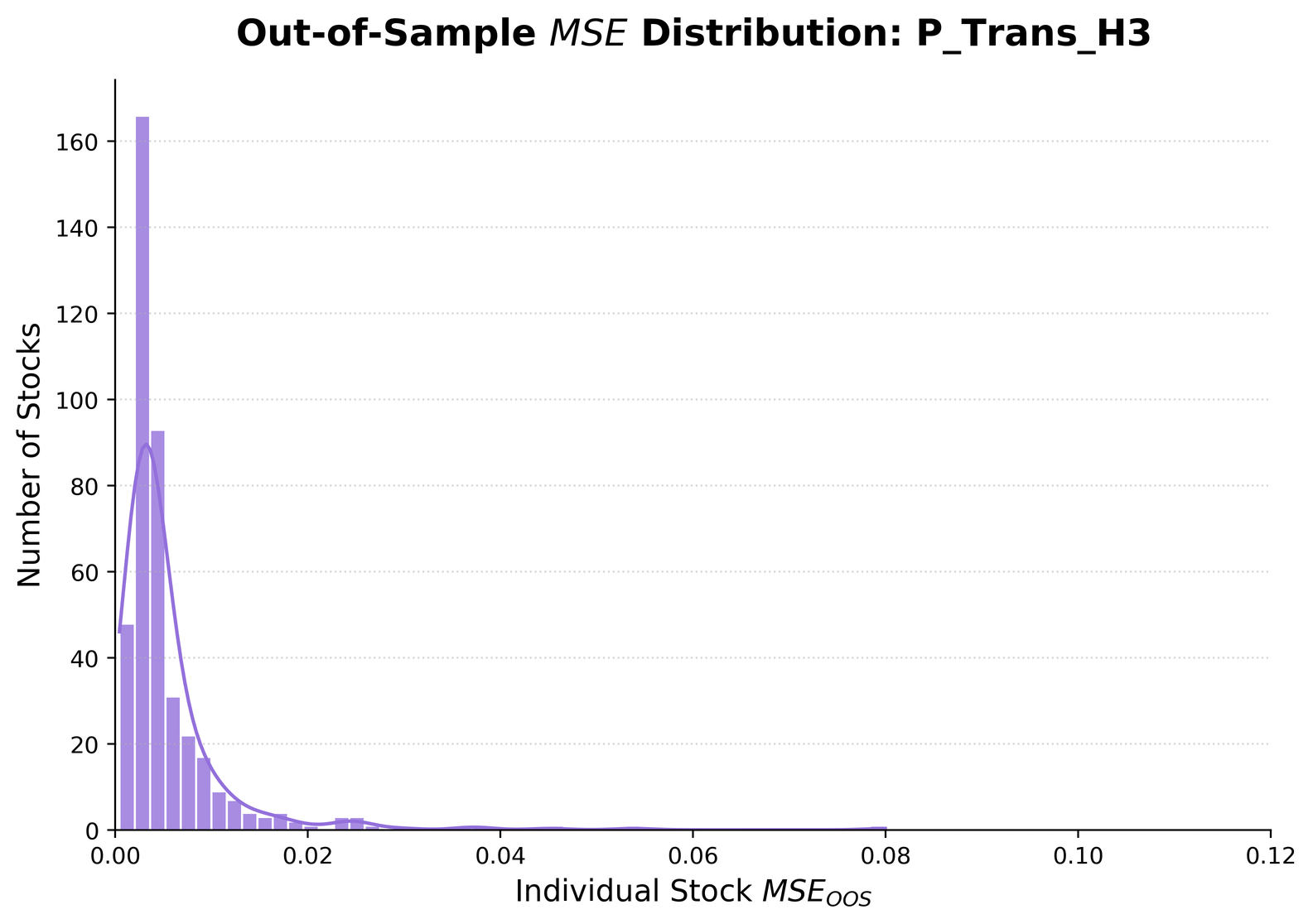}
  \caption{P\_Trans\_H3(1911)}
\end{subfigure}
\hfill
\begin{subfigure}{0.32\textwidth}
  \includegraphics[width=\linewidth, height=0.22\textheight, keepaspectratio]{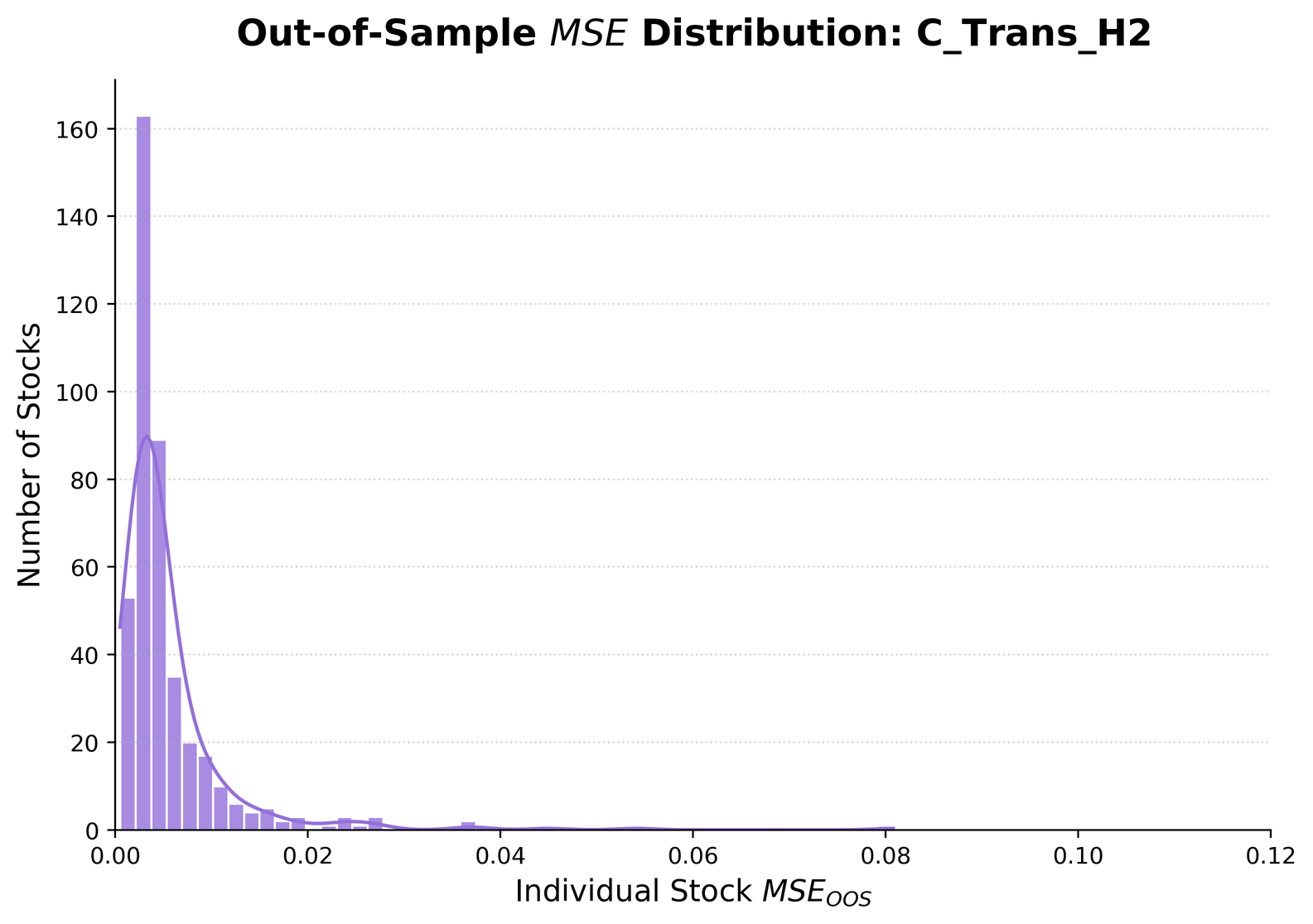}
  \caption{C\_Trans\_H2(1911)}
\end{subfigure}

\vspace{0em} 

\begin{subfigure}{0.32\textwidth}
  \includegraphics[width=\linewidth, height=0.22\textheight, keepaspectratio]{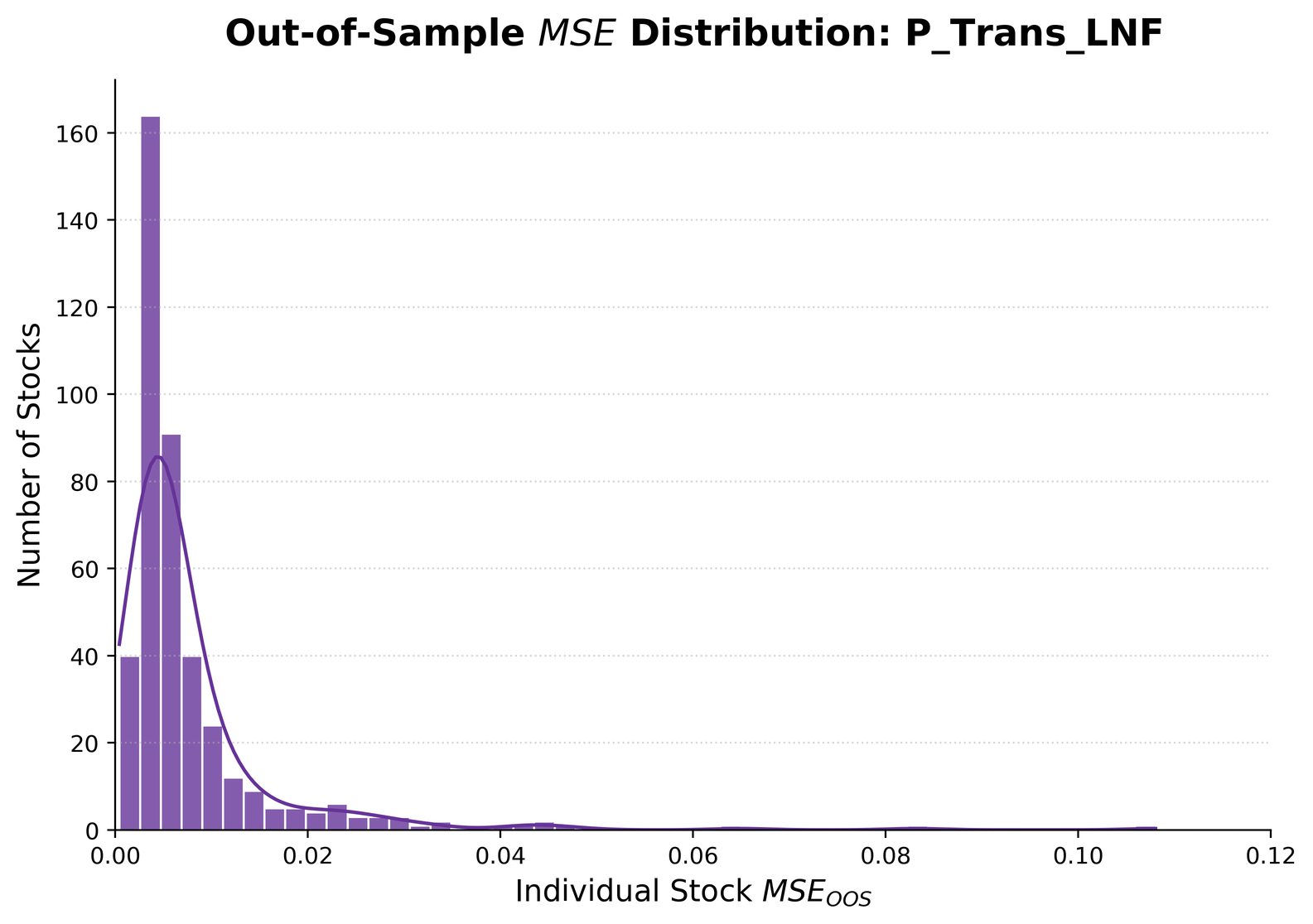}
  \caption{P\_Trans\_LNF(2112)}
\end{subfigure}
\hfill
\begin{subfigure}{0.32\textwidth}
  \includegraphics[width=\linewidth, height=0.22\textheight, keepaspectratio]{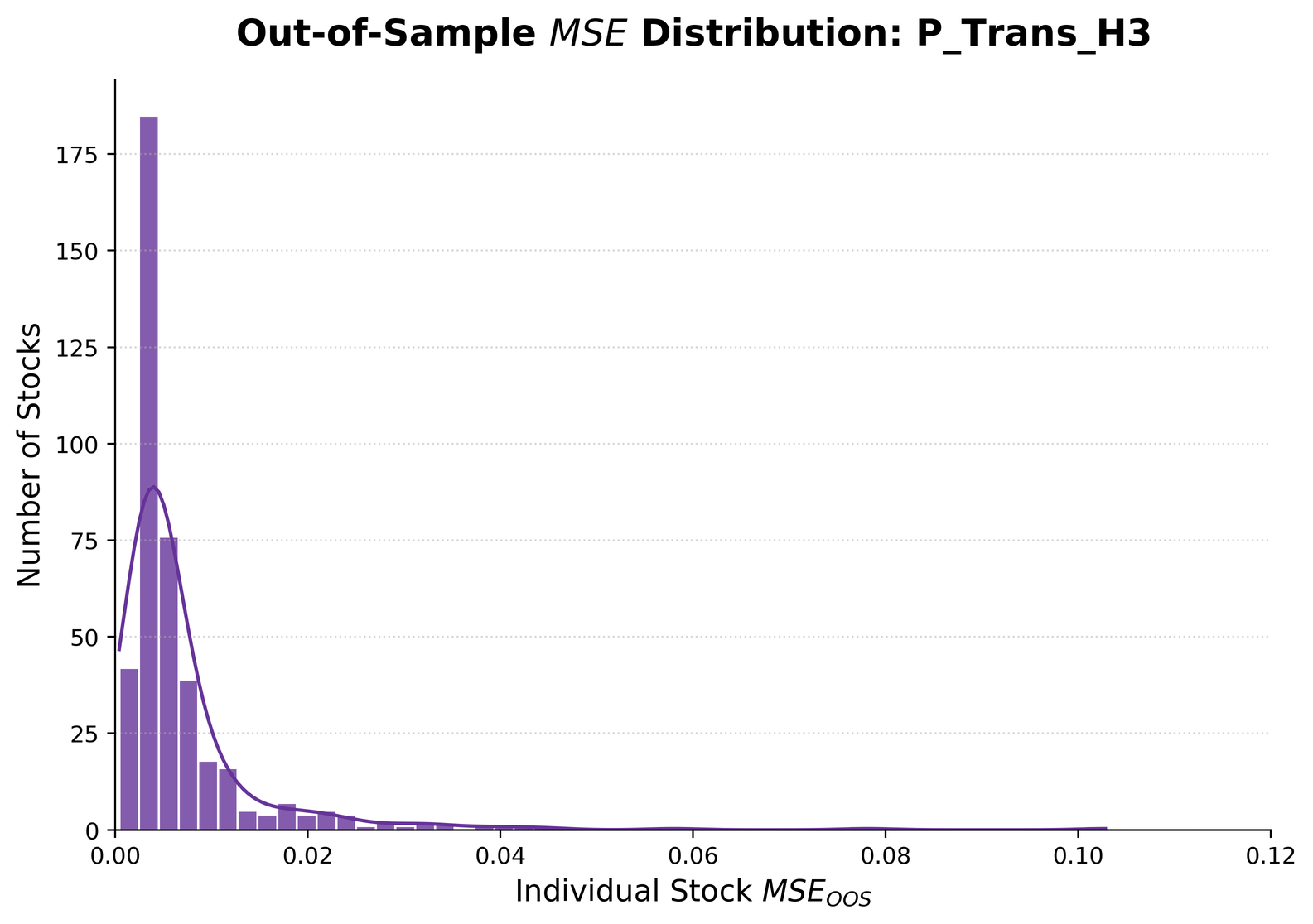}
  \caption{P\_Trans\_H3(2112)}
\end{subfigure}
\hfill
\begin{subfigure}{0.32\textwidth}
  \includegraphics[width=\linewidth, height=0.22\textheight, keepaspectratio]{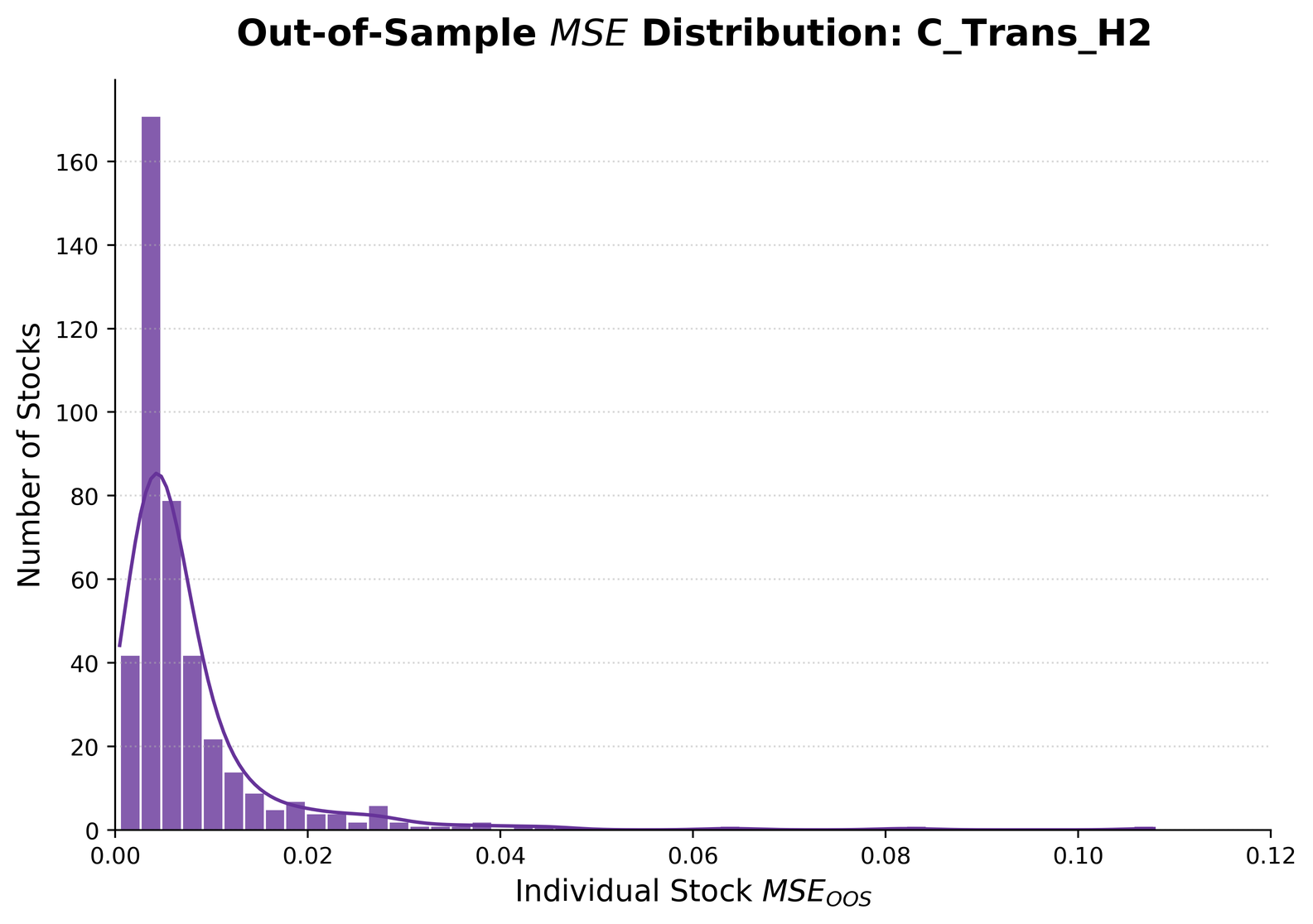}
  \caption{C\_Trans\_H2(2112)}
\end{subfigure}

\vspace{0em} 

\begin{subfigure}{0.32\textwidth}
  \includegraphics[width=\linewidth, height=0.22\textheight, keepaspectratio]{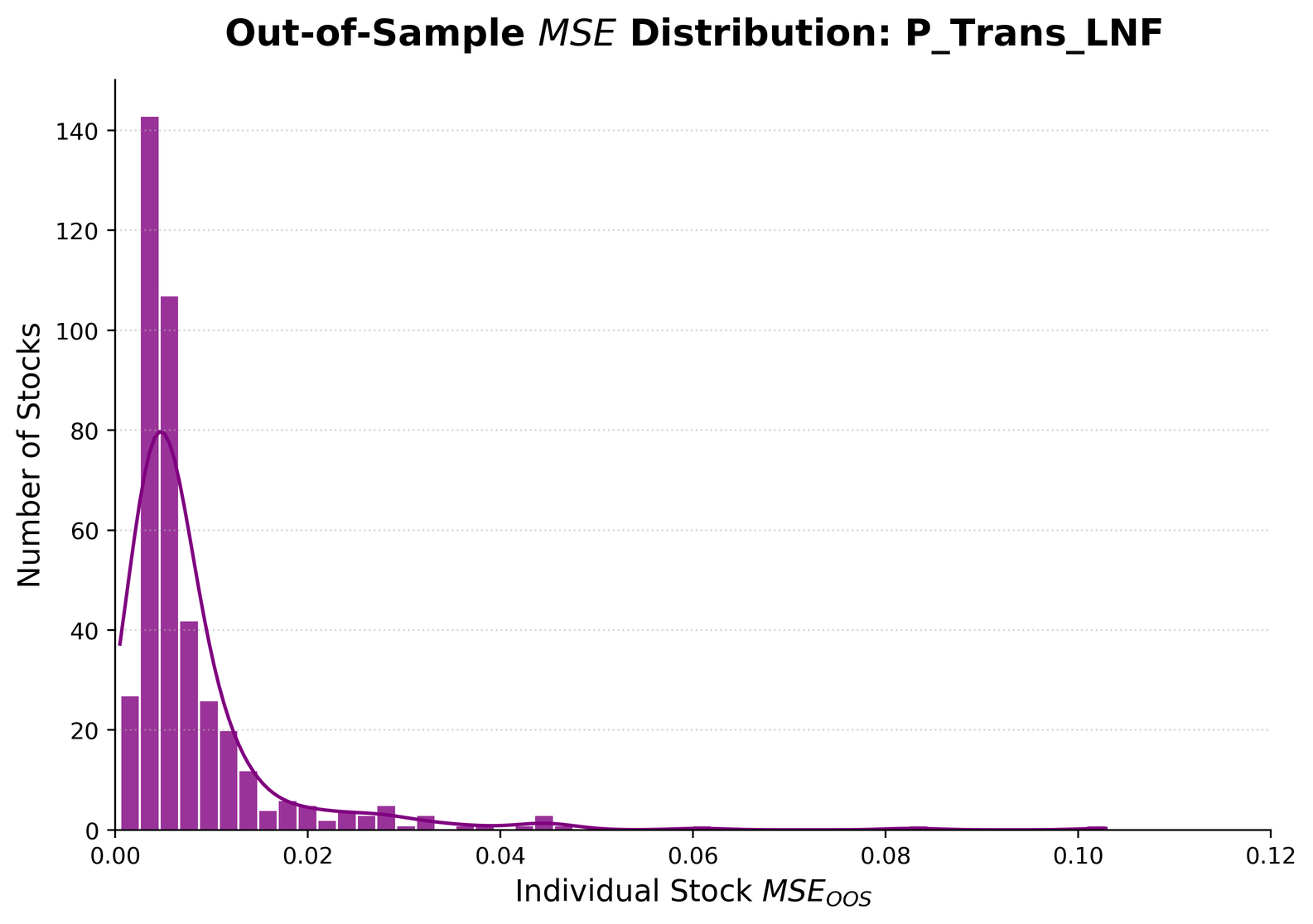}
  \caption{P\_Trans\_LNF(2212)}
\end{subfigure}
\hfill
\begin{subfigure}{0.32\textwidth}
  \includegraphics[width=\linewidth, height=0.22\textheight, keepaspectratio]{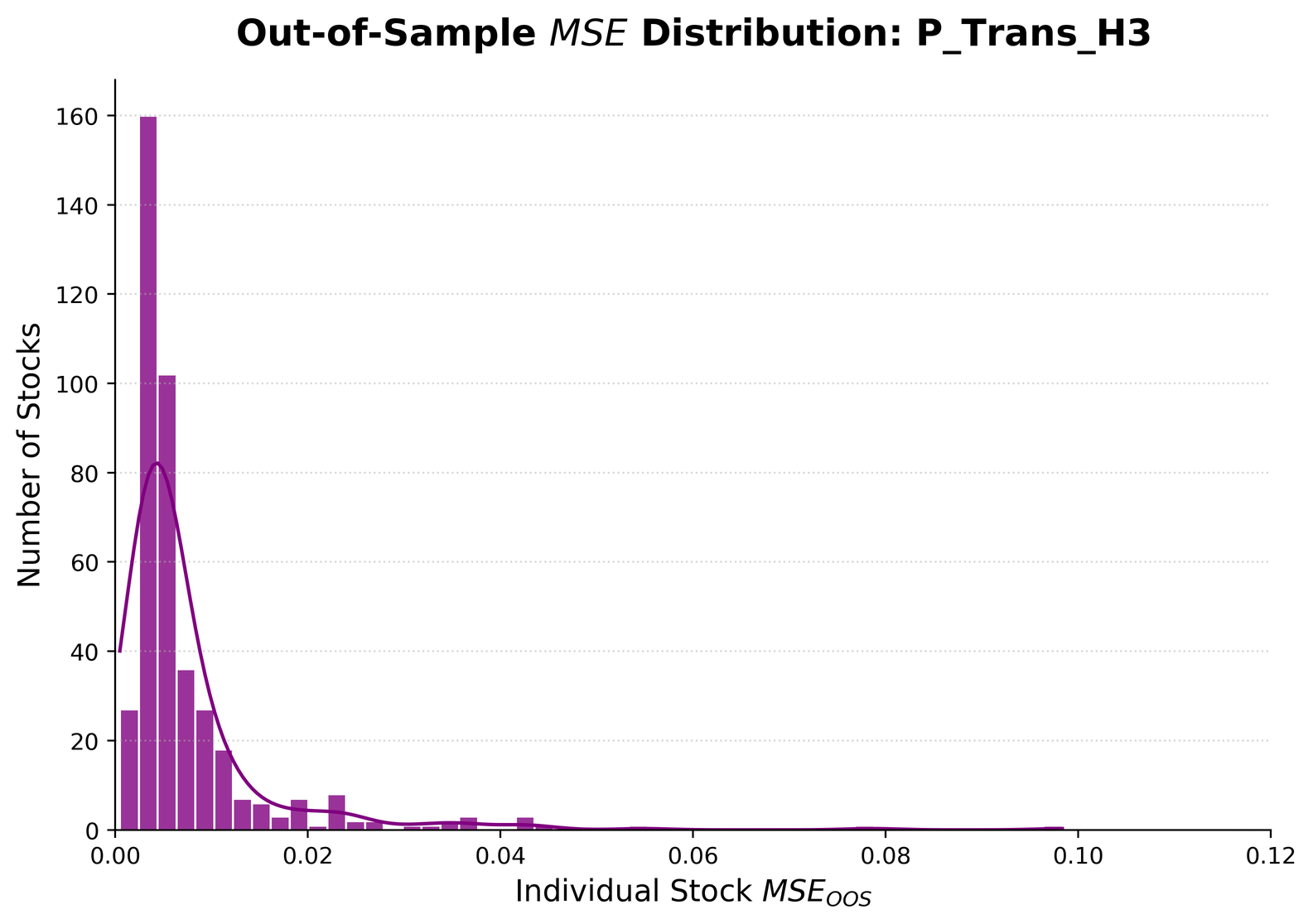}
  \caption{P\_Trans\_H3(2212)}
\end{subfigure}
\hfill
\begin{subfigure}{0.32\textwidth}
  \includegraphics[width=\linewidth, height=0.22\textheight, keepaspectratio]{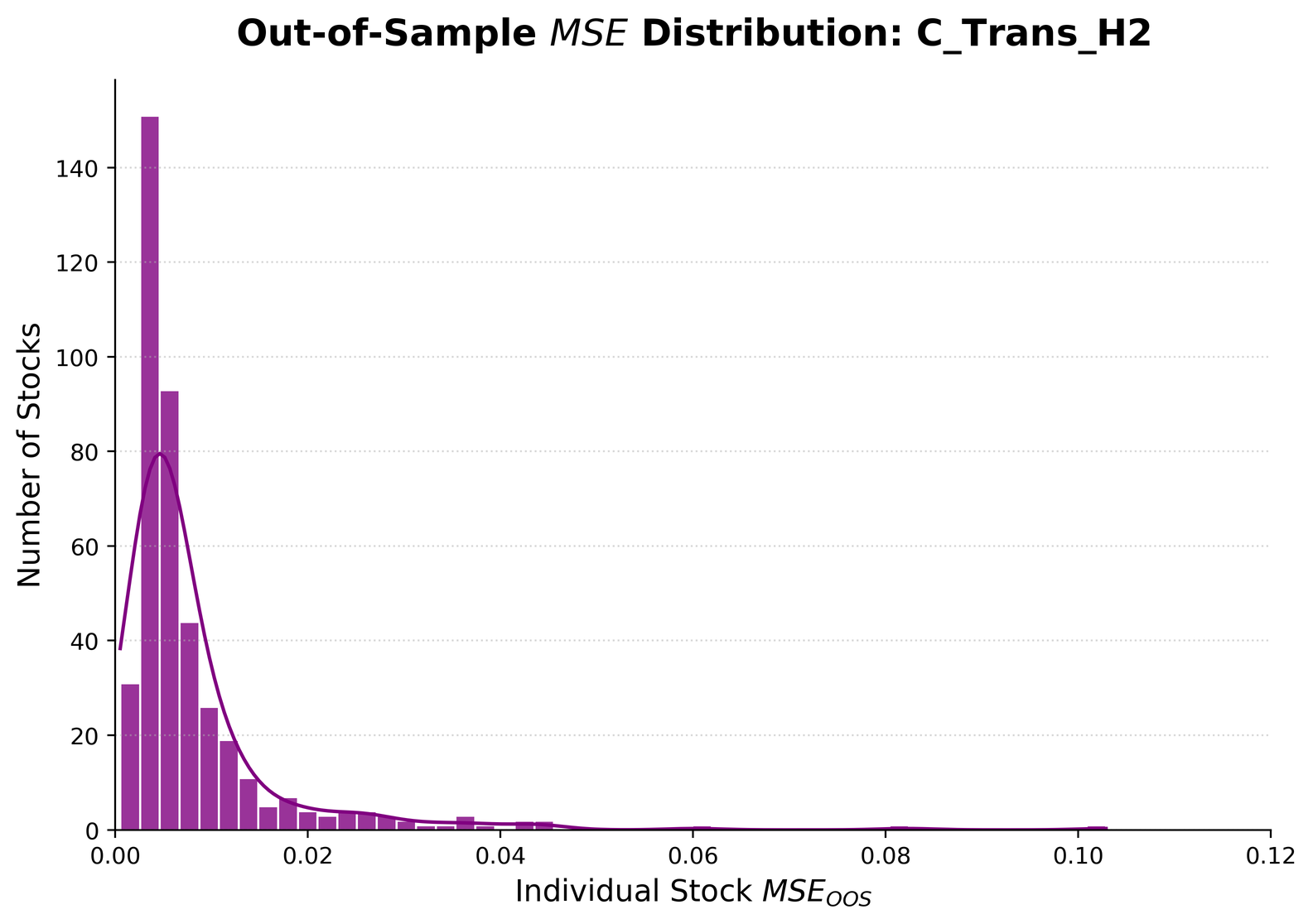}
  \caption{C\_Trans\_H2(2212)}
\end{subfigure}
 \caption[OOS $MSE$ distribution diagrams of best-performed Transformer model in each group for each period.]{OOS $MSE$ distribution diagrams of best-performed Transformer model in each group for each period. X-axis shows the OOS $MSE$ value ranges in each subgraph, while the Y-axis shows the stock numbers.}
  \label{fig:mse_distribution513_Trans_ch3}
\end{sidewaysfigure}

Figure~\ref{fig:r2_distribution512_Trans_ch3} exhibits the OOS $R^2$ distribution of the best model in each group. The distribution diagrams distinctly show that the pre-trained Transformer model with LNF significantly underperforms the best proposed Transformer model and the best standard Transformer model in all periods. The distribution charts support the conclusions drawn from other model performance indicators that the best proposed Transformer model insignificantly outperforms the best standard Transformer model in Period ‘1911’ but is noticeably superior to the benchmark models in Period ‘2112’ and Period ‘2212’ respectively. OOS MSE confirms the findings of OOS $R^2$ distribution chart which exhibits in Figure~\ref{fig:mse_distribution513_Trans_ch3}.\\

\subsection{Model performance of SERT}\label{subsec:Model performance of SERT_ch3}
Similarly, Model SERTs are examined in groups of proposed models of LNF (SERT\_LNF), with different heads (SERT\_H1 to SERT\_H7) and benchmark standard encoder-only Transformers (En\_Trans\_H1 to En\_Trans\_H4) in three periods. Table~\ref{tab:Model_perform521_SERT_ch3} and Table ~\ref{tab:dm_SERT522_ch3} show the OOS model performance and DM test results, respectively. By jointly considering the OOS $R^2$, MSE and DM statistics, the best model for each period is variable. In the pre-COVID period, SERT\_H2 (with 2 heads) has the highest average OOS $R^2$, which is 8.53\%, while in ‘2112’, SERT\_H7 (with 7 heads) has the highest value at 11.94\%. SERT\_H4 (with 4 heads) wins the model fitting competition in period ‘2212’ by a narrow margin. However, according to the DM test statistics, after controlling for the autocorrelation effect via the HAC estimator, the difference between the models in the pre-COVID period is insignificant, except for Model SERT\_LNF. However, during the extreme market fluctuation periods, models in different groups perform significantly differently from each other, although models still perform insignificantly within groups. This is similar to the findings in the DM results of the standard Transformer models. The same issue of the proposed pre-trained Transformer with the LNF (P\_Trans\_LNF) model exists in SERT with the LNF (SERT\_LNF) model and has a worse impact on OOS $R^2$. It fails to surpass the Transformer encoder-only models (En\_Trans\_H1 to En\_Trans\_H4).\\

In addition, by further investigating the OOS $R^2$ and DM statistics within each group, there is no obvious evidence supporting an optimal head number for SERT models in each period, as proposed by pre-trained Transformer models. This implies that the SERT models' training stability is inferior to the pre-trained Transformer models, although they reduce the parameter size for controlling the overfitting issues, whereas there is no significant difference between models with different numbers of attention heads apart from anomalies in period ‘2212’.\\

The mean $\alpha$ of all models in all periods is significant at a 99\% significance level, which indicates Model SERTs and their encoder-only Transformer benchmarks have positive excess returns beyond factors. The conclusion drawn on mean $\alpha$ here still stands on those drawn on Transformer models in Section~\ref{subsec:Model performance of Transformer_ch3}.\\

\begin{table}[htbp]
  \centering
  \begin{tabular}{lccccc}
    \toprule
    \textbf{Model} & \textbf{Avg\_R2} & \textbf{Avg\_MSE} & \textbf{avg\_$\alpha$} & \textbf{Ann.$\alpha$} & \textbf{$\alpha$\_t\_statistic} \\
    \midrule
    \multicolumn{6}{c}{\textbf{Pre-COVID-19 Period (1911)}} \\
    \midrule
    SERT\_LNF    & 0.0361 & 0.0057 & 0.0138 & 0.1658 & 31.9028*** \\
    SERT\_H1     & 0.0727 & 0.0055 & 0.0114 & 0.1370 & 29.6519*** \\
    SERT\_H2     & 0.0853 & 0.0054 & 0.0106 & 0.1269 & 27.9494*** \\
    SERT\_H3     & 0.0730 & 0.0055 & 0.0108 & 0.1297 & 28.4304*** \\
    SERT\_H4     & 0.0787 & 0.0054 & 0.0085 & 0.1024 & 22.7531*** \\
    SERT\_H6     & 0.0679 & 0.0055 & 0.0104 & 0.1246 & 27.5085*** \\
    SERT\_H7     & 0.0805 & 0.0054 & 0.0076 & 0.0909 & 19.7284*** \\
    En\_Trans\_H1 & 0.0643 & 0.0055 & 0.0106 & 0.1269 & 27.7645*** \\
    En\_Trans\_H2 & 0.0757 & 0.0055 & 0.0077 & 0.0923 & 19.8085*** \\
    En\_Trans\_H4 & 0.0781 & 0.0055 & 0.0052 & 0.0629 & 12.9971*** \\
    \midrule
    \multicolumn{6}{c}{\textbf{COVID-19-Inclusive Period (2112)}} \\
    \midrule
    SERT\_LNF    & 0.0168 & 0.0077 & 0.0147 & 0.1765 & 37.0430*** \\
    SERT\_H1     & 0.1089 & 0.0070 & 0.0114 & 0.1373 & 34.8619*** \\
    SERT\_H2     & 0.1066 & 0.0070 & 0.0115 & 0.1378 & 35.2856*** \\
    SERT\_H3     & 0.1070 & 0.0070 & 0.0109 & 0.1303 & 33.6458*** \\
    SERT\_H4     & 0.1095 & 0.0070 & 0.0091 & 0.1087 & 28.4759*** \\
    SERT\_H6     & 0.1133 & 0.0070 & 0.0091 & 0.1096 & 28.3948*** \\
    SERT\_H7     & 0.1194 & 0.0069 & 0.0077 & 0.0922 & 23.1957*** \\
    En\_Trans\_H1 & 0.0304 & 0.0077 & 0.0118 & 0.1421 & 35.6362*** \\
    En\_Trans\_H2 & 0.0368 & 0.0076 & 0.0097 & 0.1162 & 30.3021*** \\
    En\_Trans\_H4 & 0.0378 & 0.0077 & 0.0082 & 0.0987 & 25.5289*** \\
    \midrule
    \multicolumn{6}{c}{\textbf{Period Including COVID-19 and One-Year After (2212)}} \\
    \midrule
    SERT\_LNF    & 0.0221 & 0.0081 & 0.0135 & 0.1618 & 38.4693*** \\
    SERT\_H1     & 0.0956 & 0.0075 & 0.0100 & 0.1195 & 35.9572*** \\
    SERT\_H2     & 0.1135 & 0.0074 & 0.0100 & 0.1197 & 36.2795*** \\
    SERT\_H3     & 0.0939 & 0.0075 & 0.0094 & 0.1129 & 34.3404*** \\
    SERT\_H4     & 0.1147 & 0.0074 & 0.0080 & 0.0959 & 29.1307*** \\
    SERT\_H6     & 0.0991 & 0.0075 & 0.0079 & 0.0944 & 28.3412*** \\
    SERT\_H7     & 0.1043 & 0.0075 & 0.0066 & 0.0789 & 22.6214*** \\
    En\_Trans\_H1 & 0.0291 & 0.0081 & 0.0103 & 0.1238 & 36.5651*** \\
    En\_Trans\_H2 & 0.0345 & 0.0081 & 0.0084 & 0.1009 & 30.6498*** \\
    En\_Trans\_H4 & 0.0356 & 0.0081 & 0.0071 & 0.0850 & 25.2777*** \\
    \bottomrule
  \end{tabular}
\caption[Model performance indicators of SERTs and their benchmarks.]{Model performance indicators of SERTs and their benchmarks. Model SERT\_LNF is the proposed model with layer normalization first. SERT\_H1 to SERT\_H7 are proposed SERT models, and En\_Trans\_H1 to En\_Trans\_H4 are the benchmark Transformer models. In the ‘$\alpha$\_t\_statistic’ column, values with three stars mean it is significant at a 99\% level, two stars mean a 95\% significance level, and one star means a 90\% significance level.}
\label{tab:Model_perform521_SERT_ch3}
\end{table}

\clearpage
\begin{sidewaystable}[htbp]
\centering
\scriptsize
\renewcommand{\arraystretch}{1.2}

\begin{tabularx}{\textwidth}{l *{9}{>{\centering\arraybackslash}X}}
\toprule
\multicolumn{10}{c}{\textbf{Pre--COVID-19 Period (1911)}} \\
\midrule
 & SERT\_H1 & SERT\_H2 & SERT\_H3 & SERT\_H4 & SERT\_H6 & SERT\_H7 & SERT\_LNF & En\_H1 & En\_H2 \\
\midrule
SERT\_H1  &  &  &  &  &  &  &  &  & \\
SERT\_H2  & -0.59 &  &  &  &  &  &  &  & \\
SERT\_H3  & 0.09 & 0.64 &  &  &  &  &  &  & \\
SERT\_H4  & -0.29 & 0.43 & -0.37 &  &  &  &  &  & \\
SERT\_H6  & 1.29 & 1.05 & 0.87 & 1 &  &  &  &  & \\
SERT\_H7  & -0.85 & 0.24 & -0.96 & -0.42 & -1.52 &  &  &  & \\
SERT\_LNF & 2.21** & 2.32** & 2.10** & 2.22** & 1.43 & 2.43** &  &  & \\
En\_H1    & -0.24 & -0.55 & -0.22 & -0.38 & 0.14 & -0.49 & 1.55 &  & \\
En\_H2    & 0.28 & -0.77 & -0.26 & -0.42 & 0.14 & -0.52 & 1.54 & 0.01 & \\
En\_H4    & 0 & -0.22 & -0.03 & -0.12 & 0.38 & -0.24 & 1.53 & -0.43 & -0.22 \\
\bottomrule
\end{tabularx}

\vspace{0.3em}

\begin{tabularx}{\textwidth}{l *{9}{>{\centering\arraybackslash}X}}
\toprule
\multicolumn{10}{c}{\textbf{COVID--19--Inclusive Period (2112)}} \\
\midrule
 & SERT\_H1 & SERT\_H2 & SERT\_H3 & SERT\_H4 & SERT\_H6 & SERT\_H7 & SERT\_LNF & En\_H1 & En\_H2 \\
\midrule
SERT\_H1  &  &  &  &  &  &  &  &  & \\
SERT\_H2  & -0.23 &  &  &  &  &  &  &  & \\
SERT\_H3  & 0.45 & 0.41 &  &  &  &  &  &  & \\
SERT\_H4  & 0.09 & 0.29 & -0.21 &  &  &  &  &  & \\
SERT\_H6  & 0.05 & 0.22 & -0.15 & -0.01 &  &  &  &  & \\
SERT\_H7  & -1.12 & -0.43 & -1.42 & -1.18 & -1.22 &  &  &  & \\
SERT\_LNF & 2.97*** & 3.30*** & 2.98*** & 2.95*** & 2.87*** & 3.43*** &  &  & \\
En\_H1    & -1.65* & -1.79* & -1.63 & -1.69* & -1.61 & -2.05** & 1.51 &  & \\
En\_H2    & 1.74* & -2.11** & -1.73* & -1.78* & -1.71* & -2.12** & 1.43 & 0.06 & \\
En\_H4    & 1.55 & -0.2 & 1.52 & -1.59 & -1.52 & -1.96** & 1.42 & -0.29 & -0.2 \\
\bottomrule
\end{tabularx}

\vspace{0.3em}

\begin{tabularx}{\textwidth}{l *{9}{>{\centering\arraybackslash}X}}
\toprule
\multicolumn{10}{c}{\textbf{Period Including COVID--19 and One-Year After (2212)}} \\
\midrule
 & SERT\_H1 & SERT\_H2 & SERT\_H3 & SERT\_H4 & SERT\_H6 & SERT\_H7 & SERT\_LNF & En\_H1 & En\_H2 \\
\midrule
SERT\_H1  &  &  &  &  &  &  &  &  & \\
SERT\_H2  & -1.47 &  &  &  &  &  &  &  & \\
SERT\_H3  & 0.39 & 1.62 &  &  &  &  &  &  & \\
SERT\_H4  & -1.61 & 0.29 & -1.80* &  &  &  &  &  & \\
SERT\_H6  & 0.03 & 1.32 & -0.15 & 1.34 &  &  &  &  & \\
SERT\_H7  & -1.15 & 0.83 & -1.42 & 0.82 & -1.23 &  &  &  & \\
SERT\_LNF & 2.62*** & 3.82*** & 2.62*** & 3.48*** & 2.54** & 3.05*** &  &  & \\
En\_H1    & -1.65* & -2.43** & -1.64 & -2.34** & -1.62 & -2.05** & 1.02 &  & \\
En\_H2    & 1.73* & -2.79*** & -1.73* & -2.43** & -1.71* & -2.12** & 0.95 & 0.05 & \\
En\_H4    & 1.53 & -0.21 & 1.51 & -2.24** & -1.51 & -1.95* & 1.04 & -0.33 & -0.21 \\
\bottomrule
\end{tabularx}
\caption[DM test results of SERTs and their benchmarks.]{DM test results of SERTs and their benchmarks. The negative values demonstrate that the model with column labels as names performs worse than the one with row labels as names. `En\_H1', `En\_H2', `En\_H4' are short for `En\_Trans\_H1', `En\_Trans\_H2', `En\_Trans\_H4', respectively. Significance levels: * $p<0.1$, ** $p<0.05$, *** $p<0.01$.}
\label{tab:dm_SERT522_ch3}
\end{sidewaystable}

\clearpage
\begin{sidewaysfigure}[htbp!]
\centering
\begin{subfigure}{0.32\textwidth}
  \includegraphics[width=\linewidth, height=0.22\textheight, keepaspectratio]{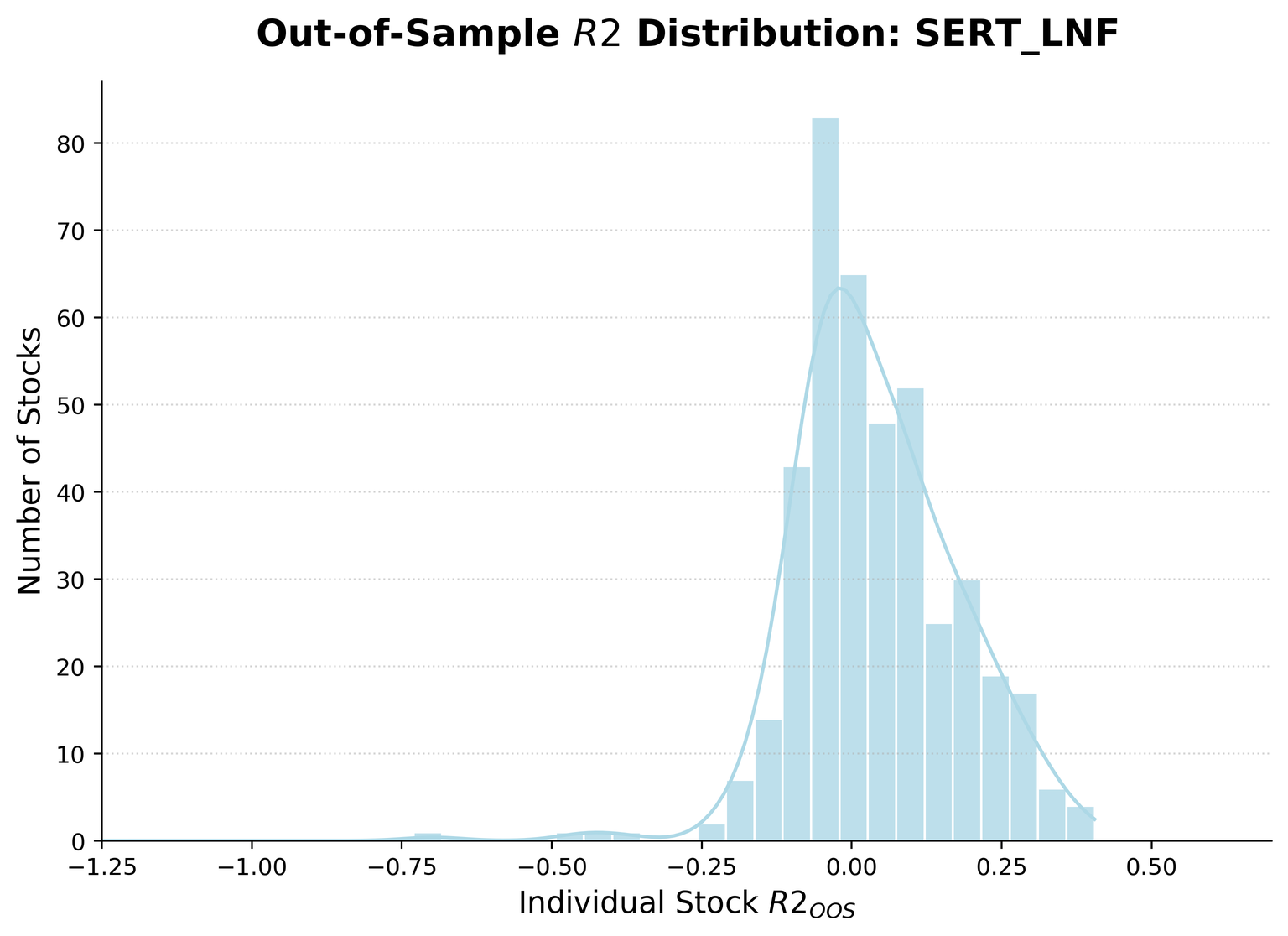}
  \caption{SERT\_LNF(1911)}
\end{subfigure}
\hfill
\begin{subfigure}{0.32\textwidth}
  \includegraphics[width=\linewidth, height=0.22\textheight, keepaspectratio]{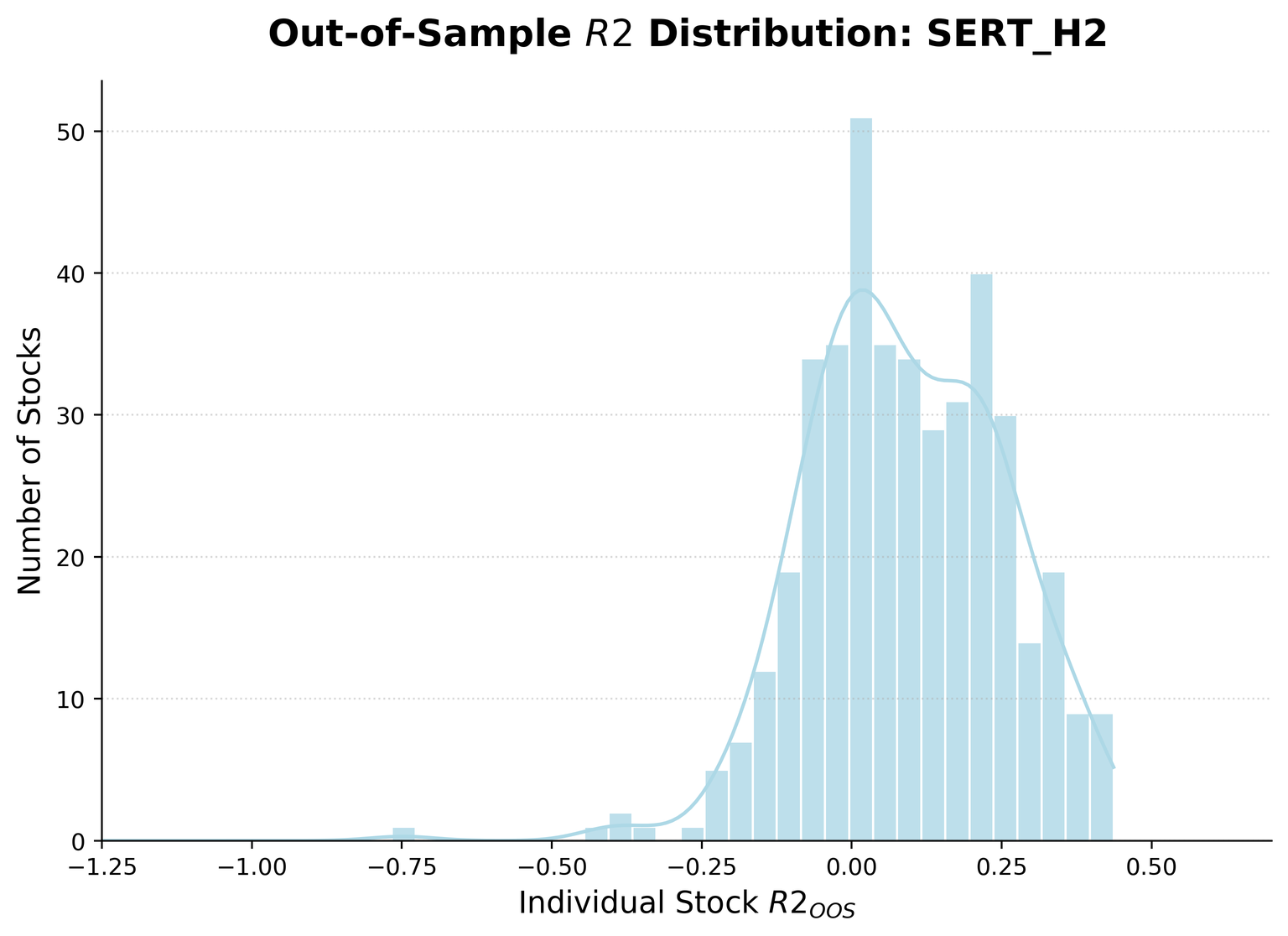}
  \caption{SERT\_H2(1911)}
\end{subfigure}
\hfill
\begin{subfigure}{0.32\textwidth}
  \includegraphics[width=\linewidth, height=0.22\textheight, keepaspectratio]{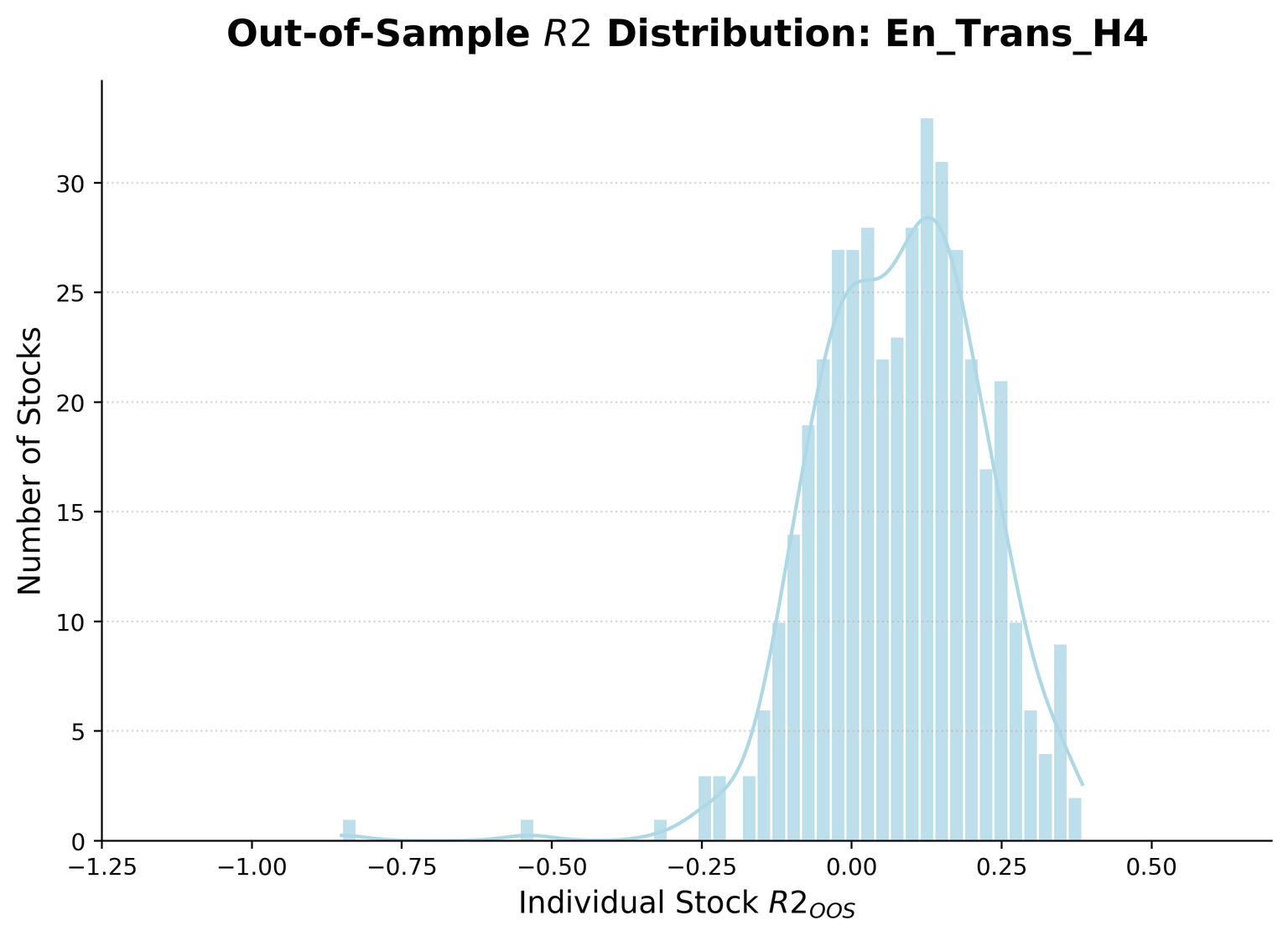}
  \caption{En\_Trans\_H4(1911)}
\end{subfigure}

\vspace{0.1em} 

\begin{subfigure}{0.32\textwidth}
  \includegraphics[width=\linewidth, height=0.22\textheight, keepaspectratio]{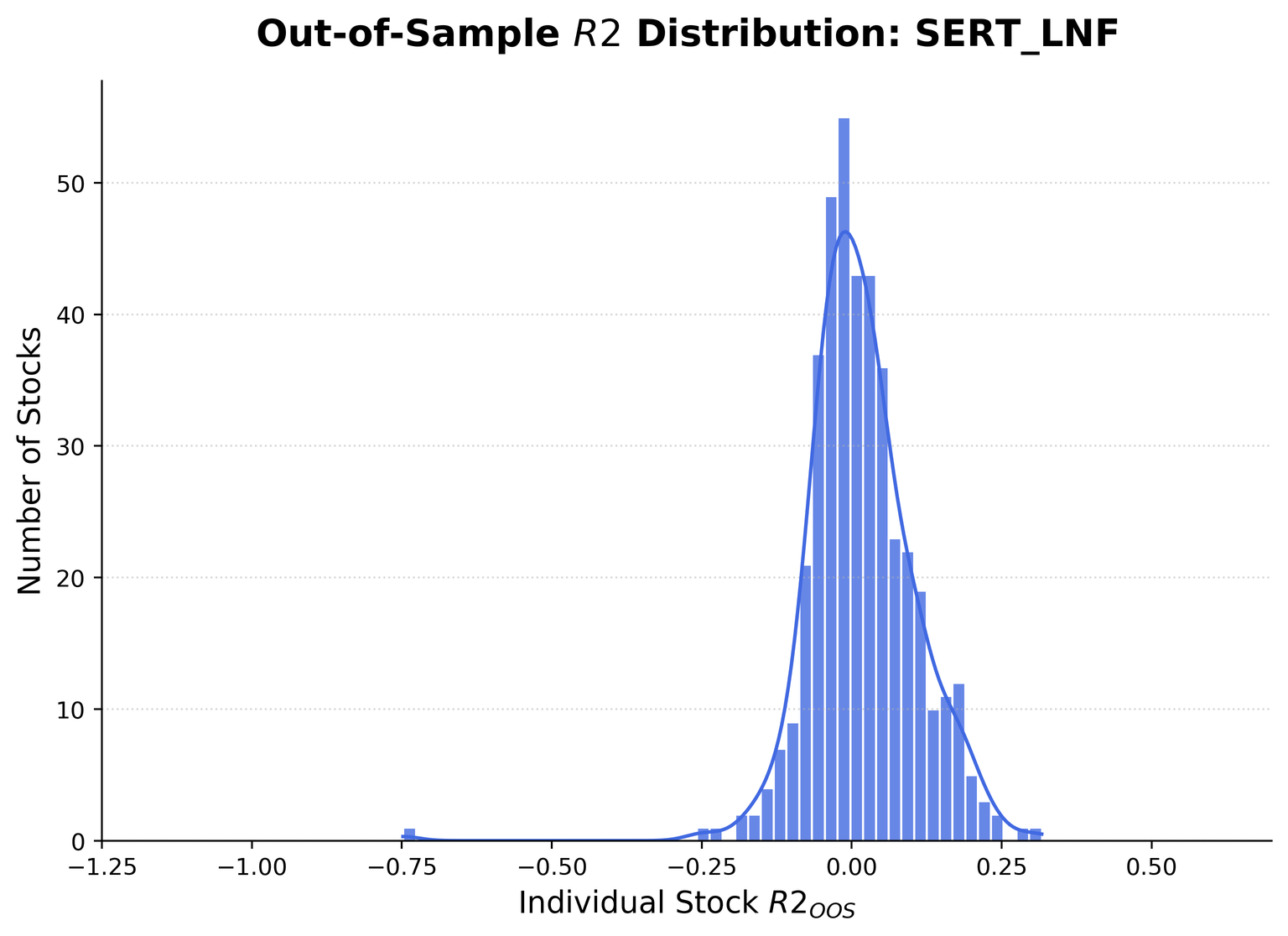}
  \caption{SERT\_LNF(2112)}
\end{subfigure}
\hfill
\begin{subfigure}{0.32\textwidth}
  \includegraphics[width=\linewidth, height=0.22\textheight, keepaspectratio]{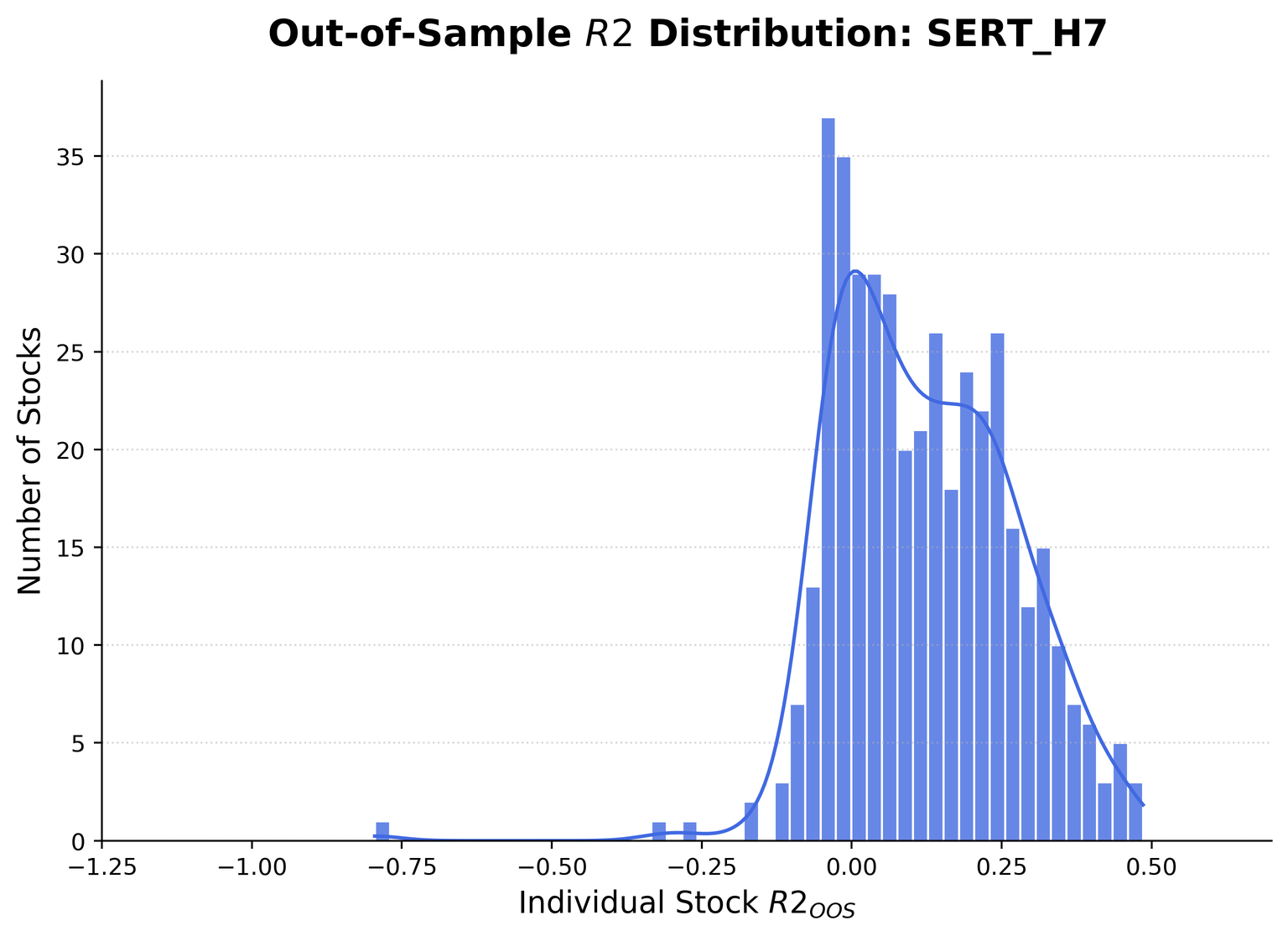}
  \caption{SERT\_H7(2112)}
\end{subfigure}
\hfill
\begin{subfigure}{0.32\textwidth}
  \includegraphics[width=\linewidth, height=0.22\textheight, keepaspectratio]{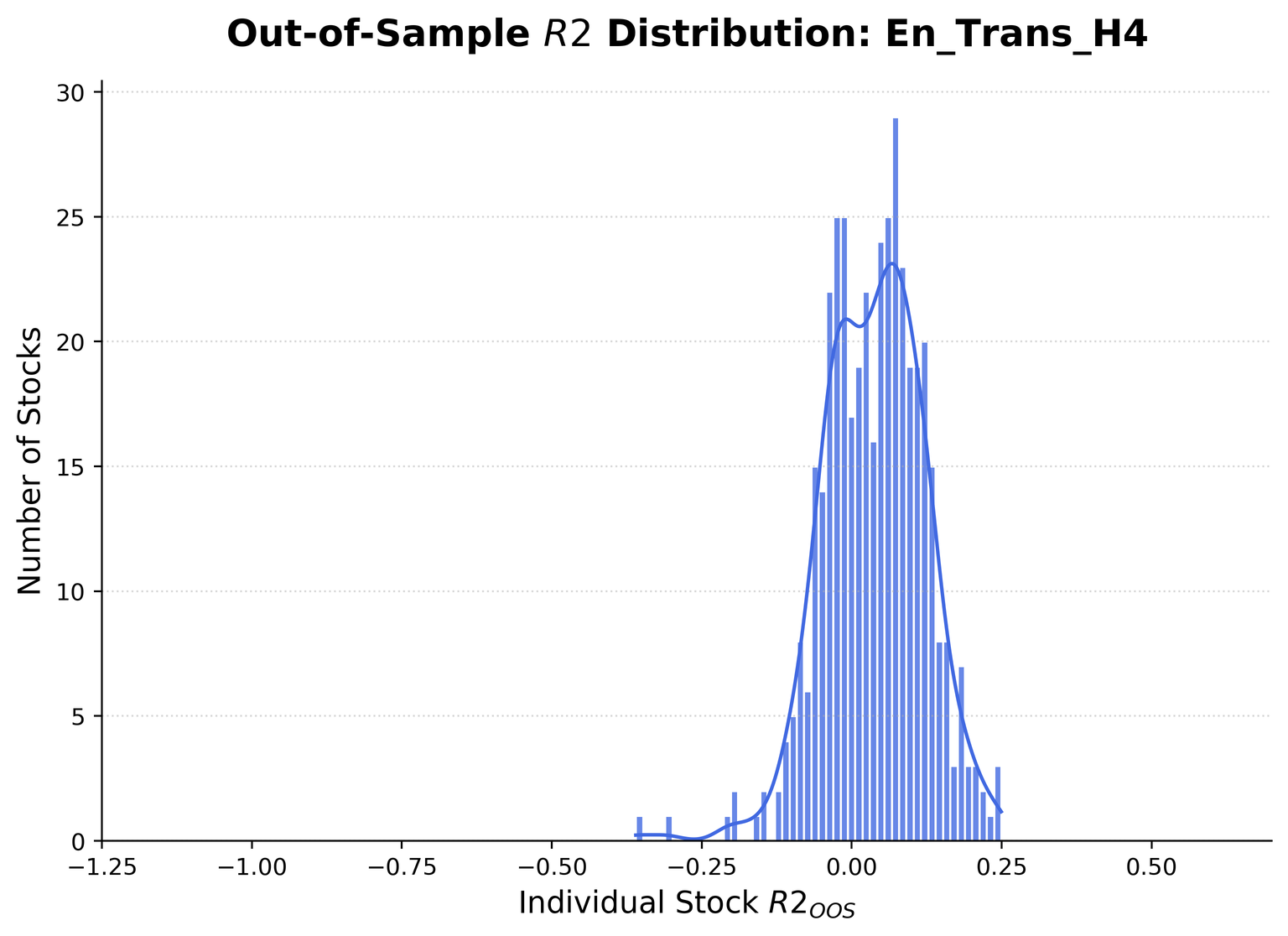}
  \caption{En\_Trans\_H4(2112)}
\end{subfigure}

\vspace{0.1em} 

\begin{subfigure}{0.32\textwidth}
  \includegraphics[width=\linewidth, height=0.22\textheight, keepaspectratio]{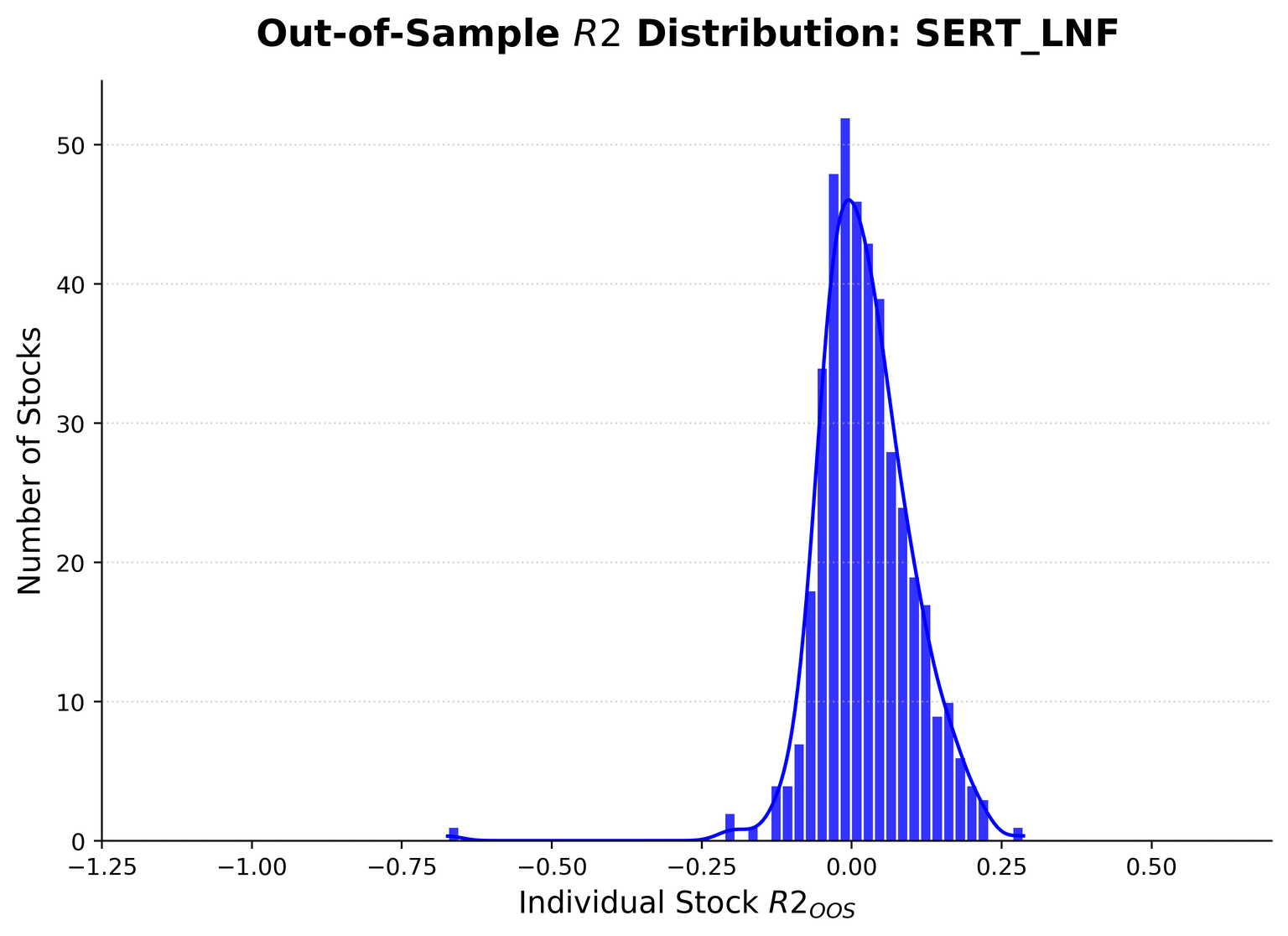}
  \caption{SERT\_LNF(2212)}
\end{subfigure}
\hfill
\begin{subfigure}{0.32\textwidth}
  \includegraphics[width=\linewidth, height=0.22\textheight, keepaspectratio]{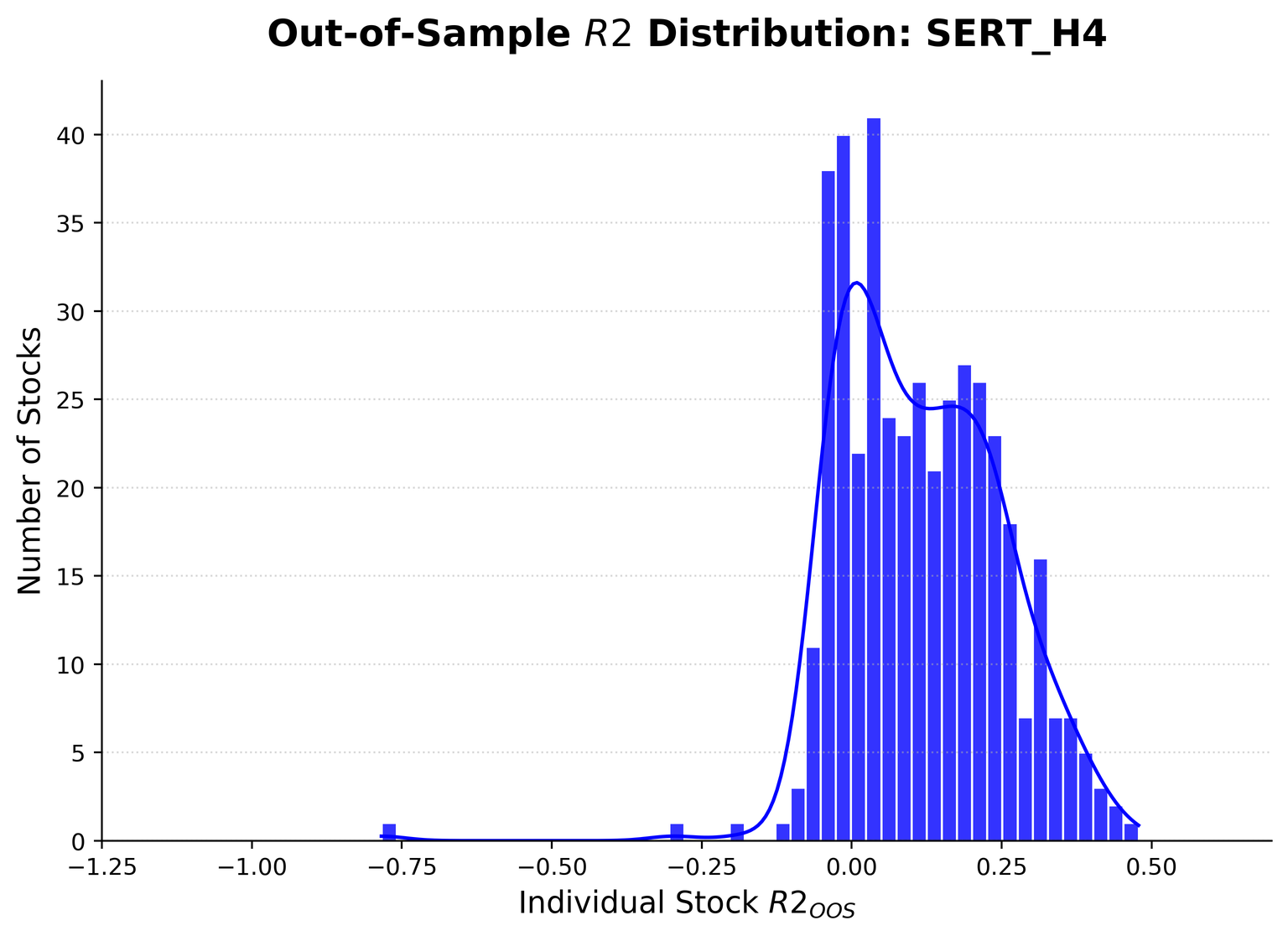}
  \caption{SERT\_H4(2212)}
\end{subfigure}
\hfill
\begin{subfigure}{0.32\textwidth}
  \includegraphics[width=\linewidth, height=0.22\textheight, keepaspectratio]{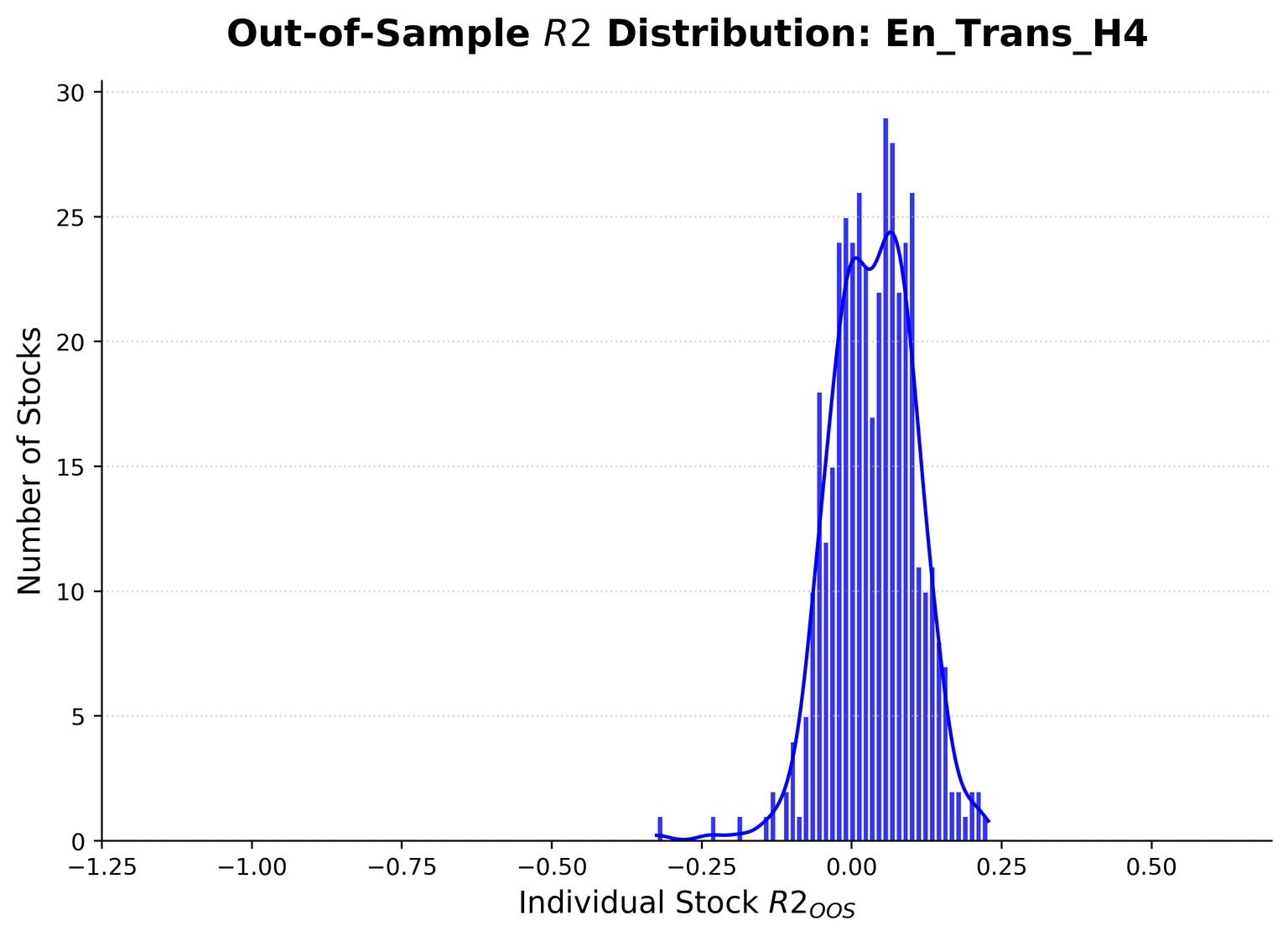}
  \caption{En\_Trans\_H4(2212)}
\end{subfigure}
\caption[OOS $R^2$ distribution diagrams of best-performed SERT models and their benchmarks.]{OOS $R^2$ distribution diagrams of best-performed SERT models and their benchmarks in each group for each period. The X-axis shows the OOS $R^2$ value ranges in each subgraph, while the Y-axis shows the stock numbers.}
\label{fig:r2_distribution521_SERT_ch3}
\end{sidewaysfigure}

\begin{sidewaysfigure}[htbp!]
\centering
\begin{subfigure}{0.32\textwidth}
  \includegraphics[width=\linewidth, height=0.22\textheight, keepaspectratio]{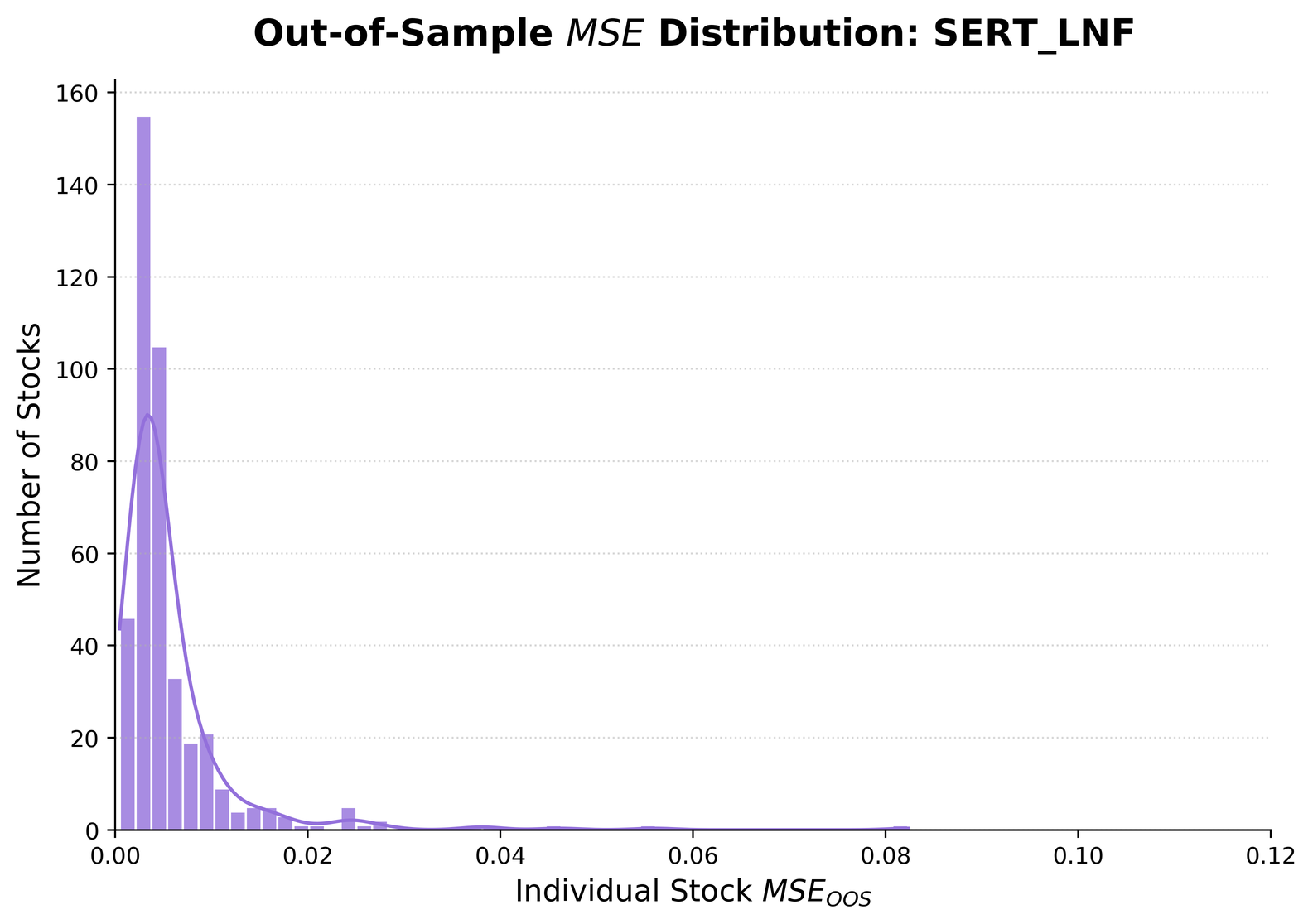}
  \caption{SERT\_LNF(1911)}
\end{subfigure}
\hfill
\begin{subfigure}{0.32\textwidth}
  \includegraphics[width=\linewidth, height=0.22\textheight, keepaspectratio]{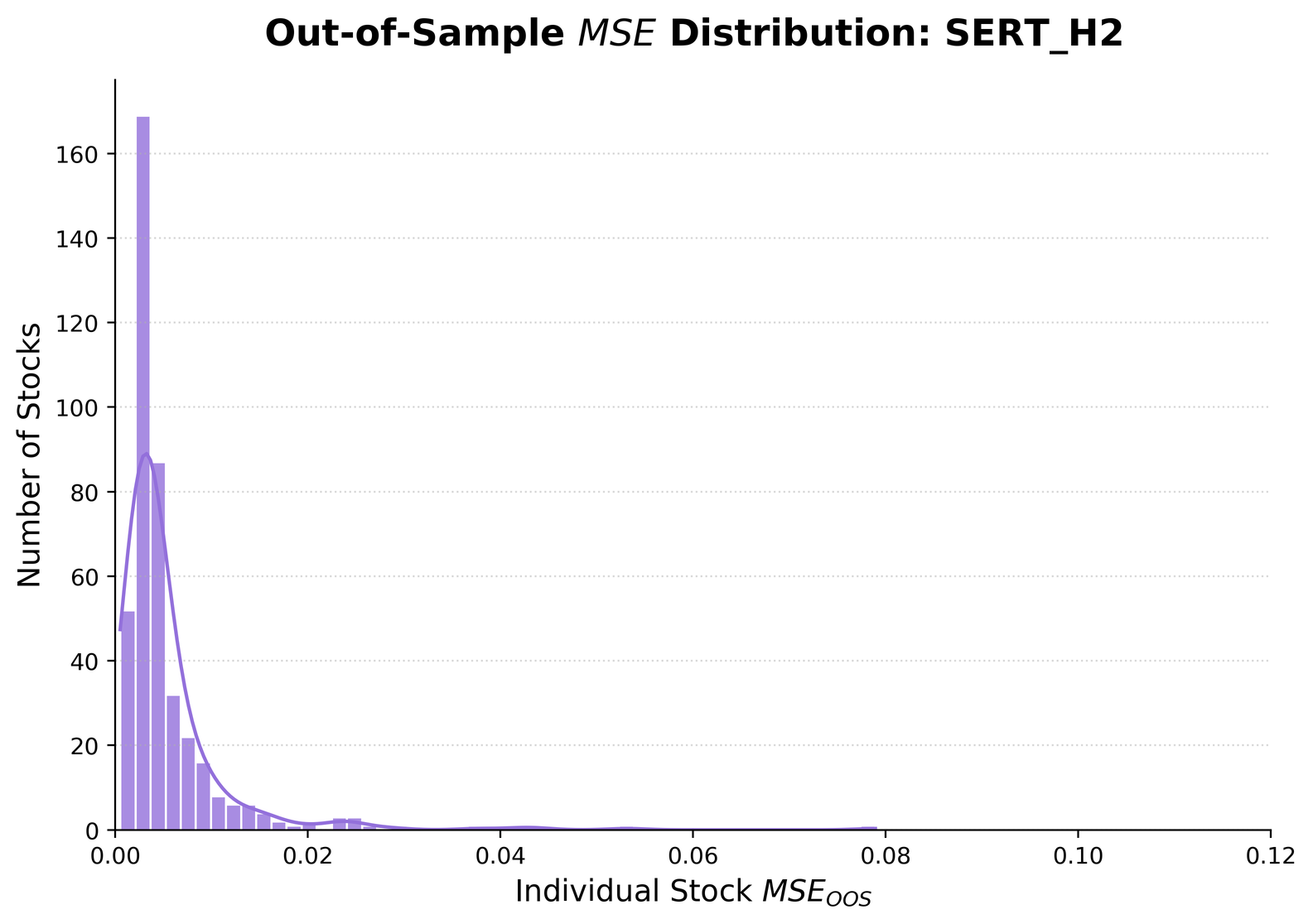}
  \caption{SERT\_H2(1911)}
\end{subfigure}
\hfill
\begin{subfigure}{0.32\textwidth}
  \includegraphics[width=\linewidth, height=0.22\textheight, keepaspectratio]{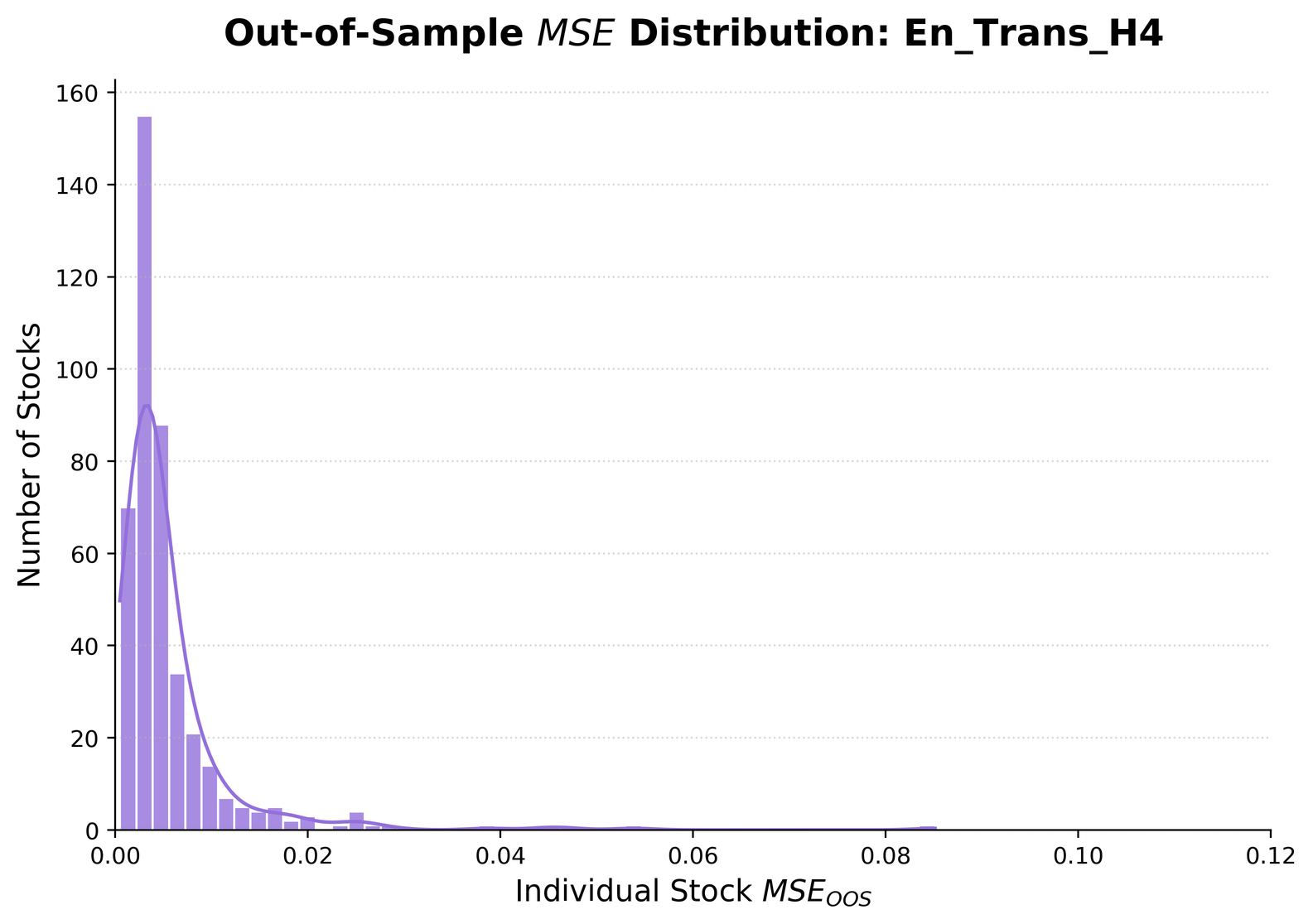}
  \caption{En\_Trans\_H4(1911)}
\end{subfigure}

\vspace{0.1em} 

\begin{subfigure}{0.32\textwidth}
  \includegraphics[width=\linewidth, height=0.22\textheight, keepaspectratio]{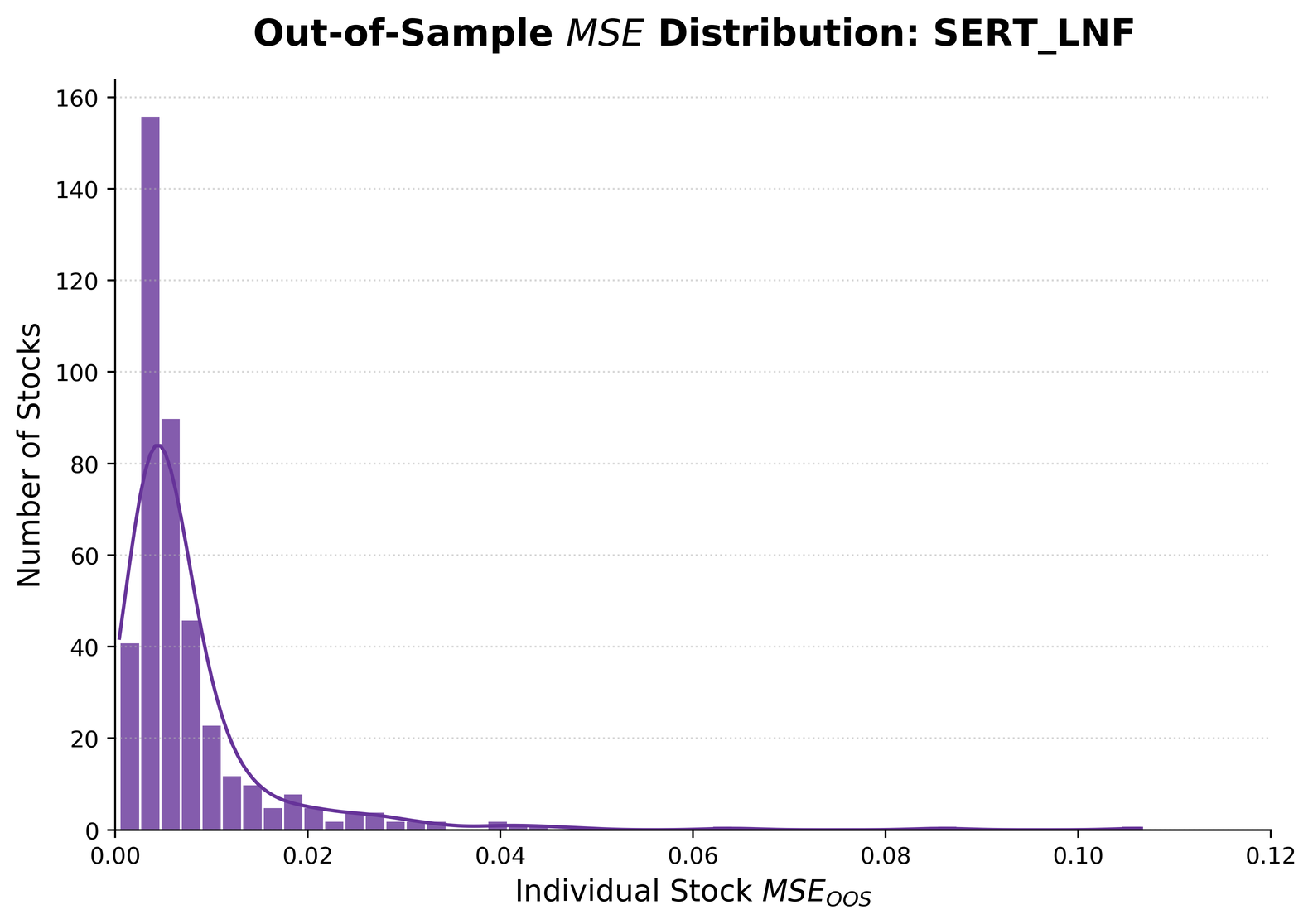}
  \caption{SERT\_LNF(2112)}
\end{subfigure}
\hfill
\begin{subfigure}{0.32\textwidth}
  \includegraphics[width=\linewidth, height=0.22\textheight, keepaspectratio]{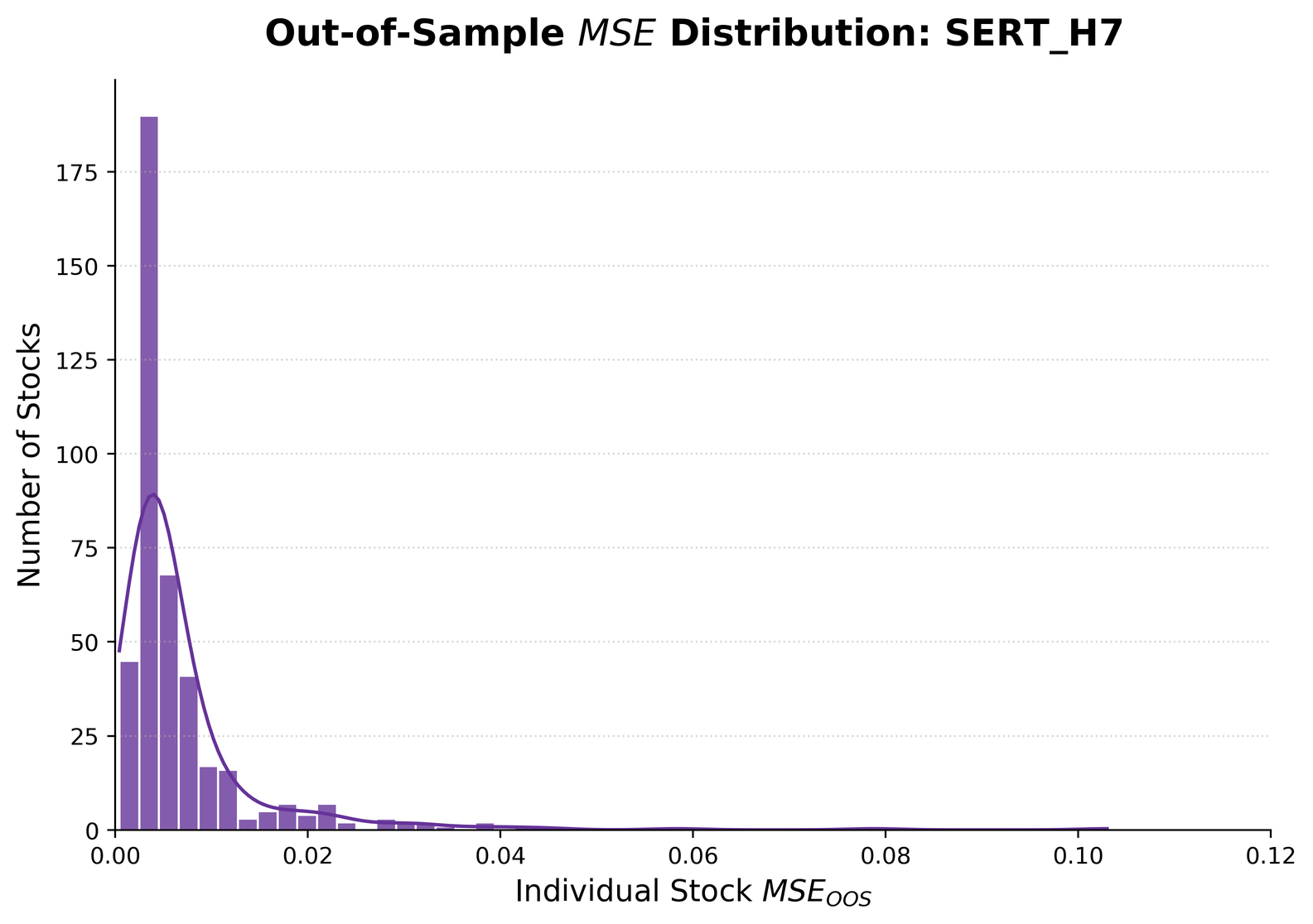}
  \caption{SERT\_H7(2112)}
\end{subfigure}
\hfill
\begin{subfigure}{0.32\textwidth}
  \includegraphics[width=\linewidth, height=0.22\textheight, keepaspectratio]{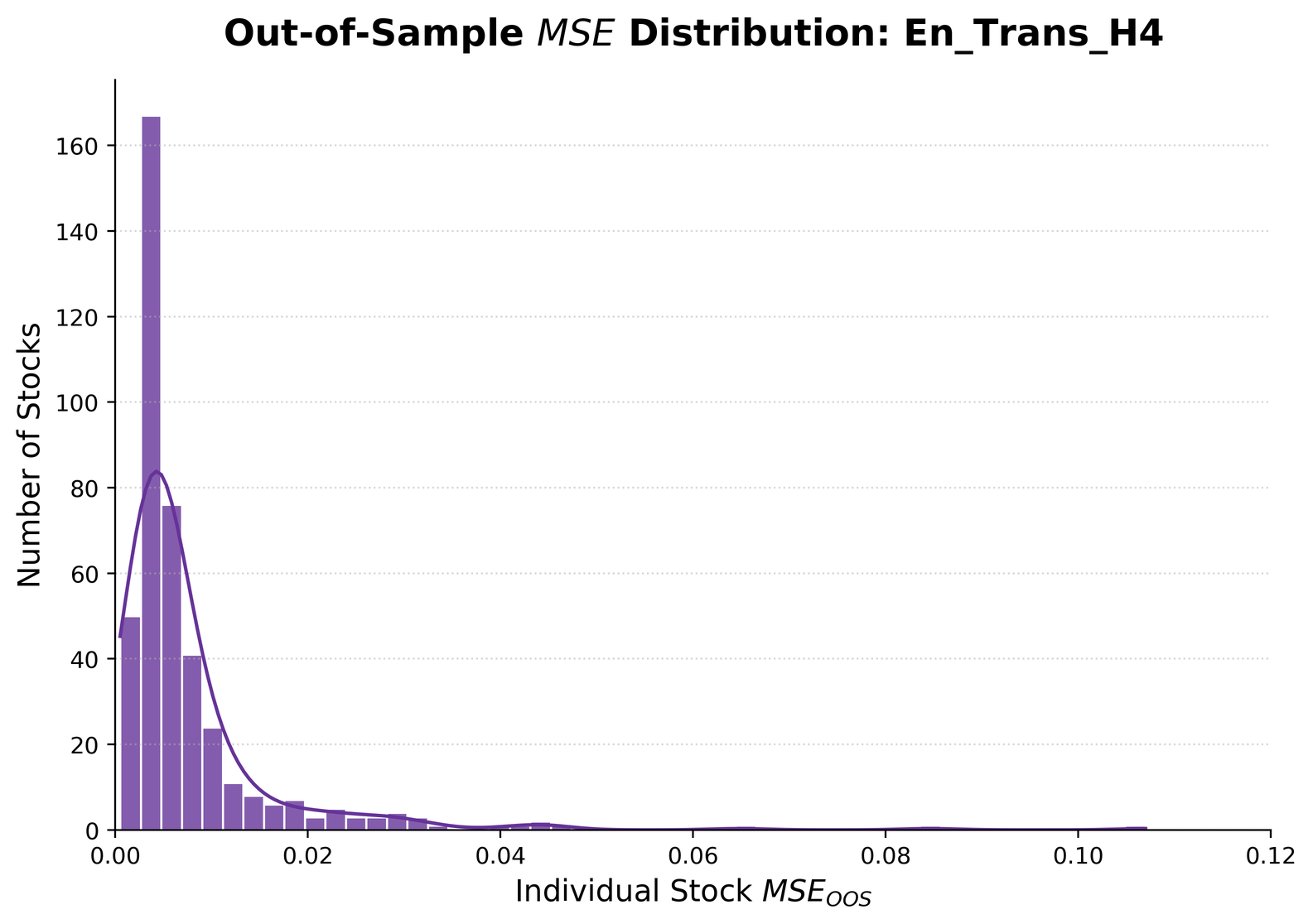}
  \caption{En\_Trans\_H4(2112)}
\end{subfigure}

\vspace{0.1em} 

\begin{subfigure}{0.32\textwidth}
  \includegraphics[width=\linewidth, height=0.22\textheight, keepaspectratio]{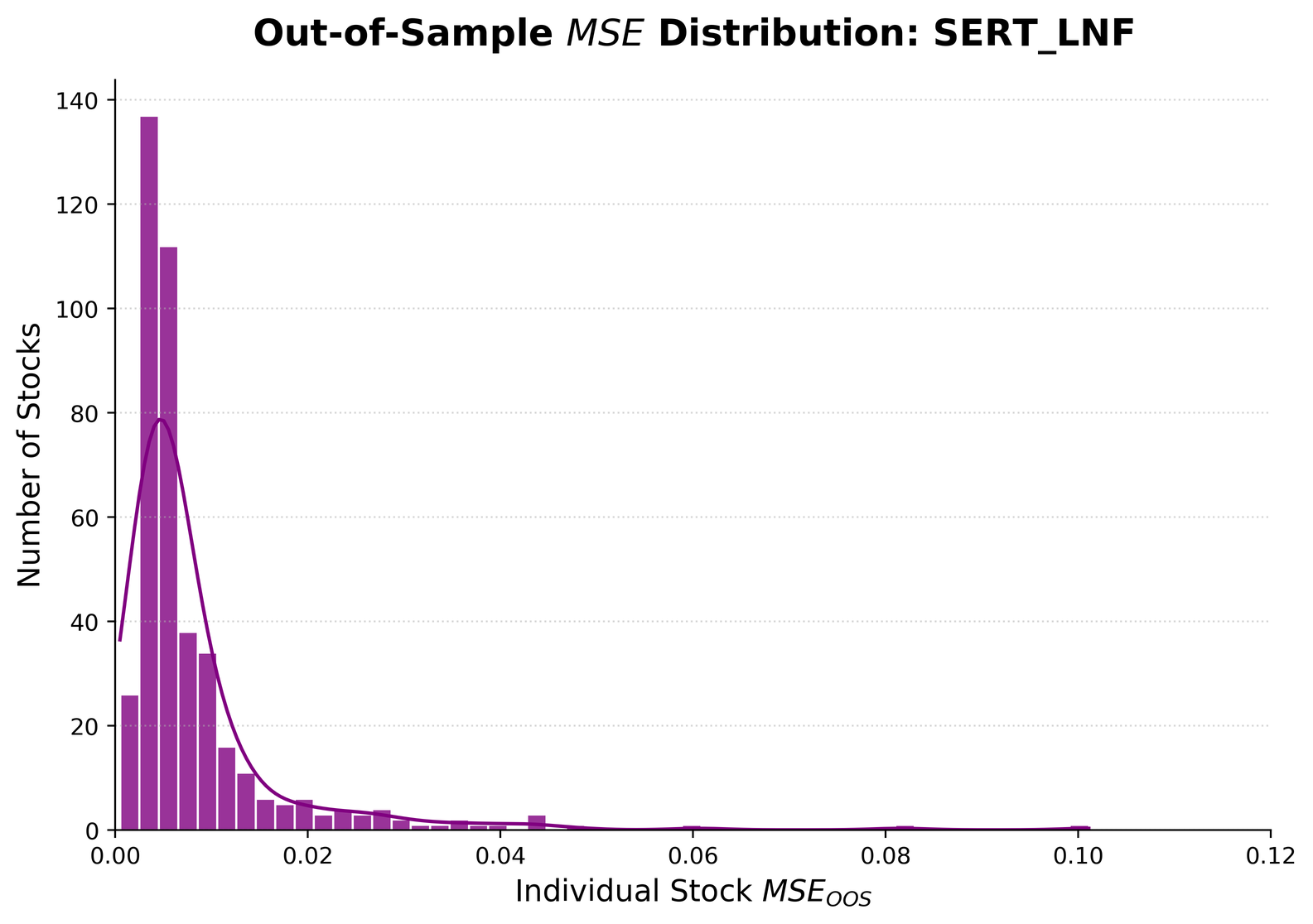}
  \caption{SERT\_LNF(2212)}
\end{subfigure}
\hfill
\begin{subfigure}{0.32\textwidth}
  \includegraphics[width=\linewidth, height=0.22\textheight, keepaspectratio]{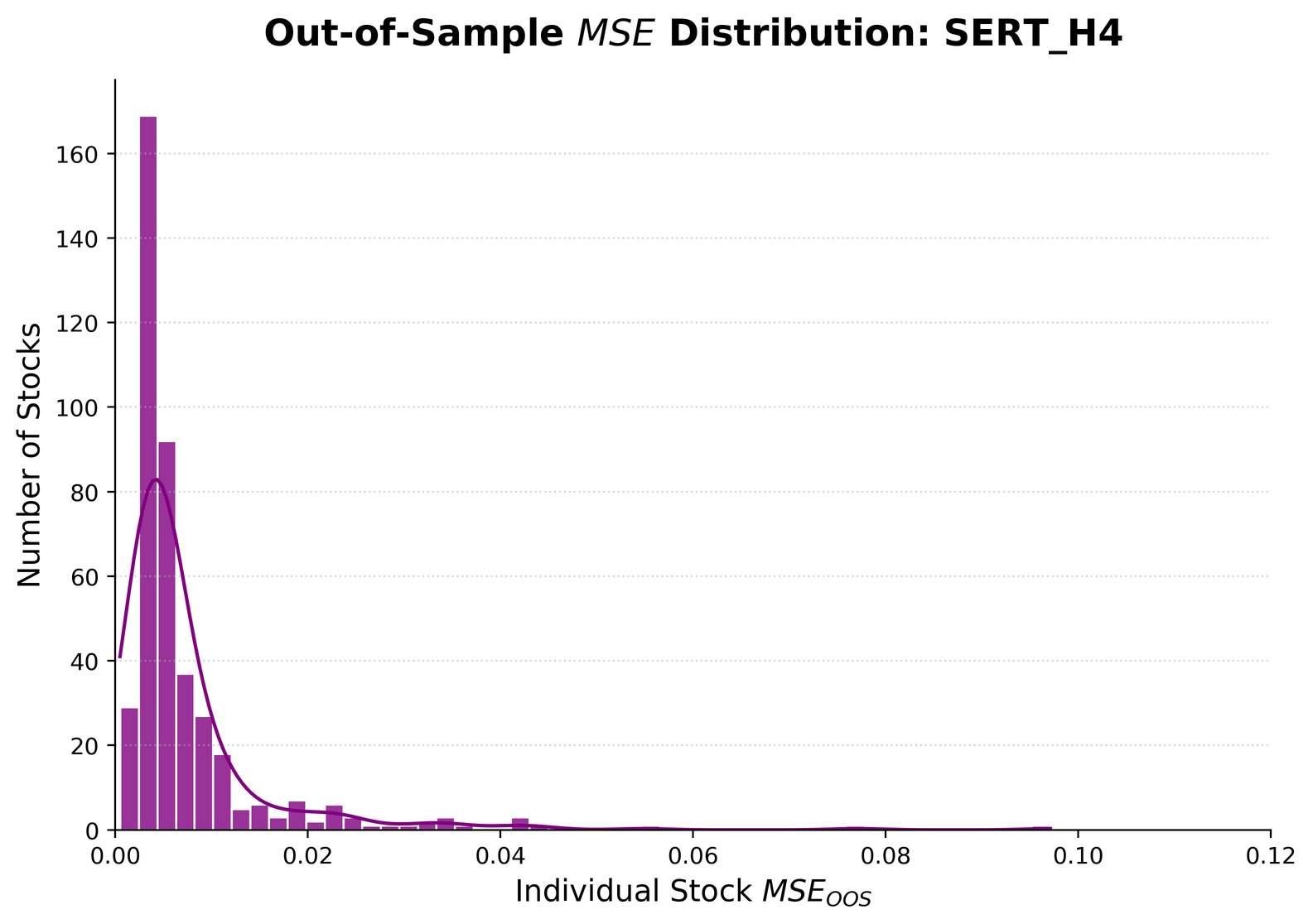}
  \caption{SERT\_H4(2212)}
\end{subfigure}
\hfill
\begin{subfigure}{0.32\textwidth}
  \includegraphics[width=\linewidth, height=0.22\textheight, keepaspectratio]{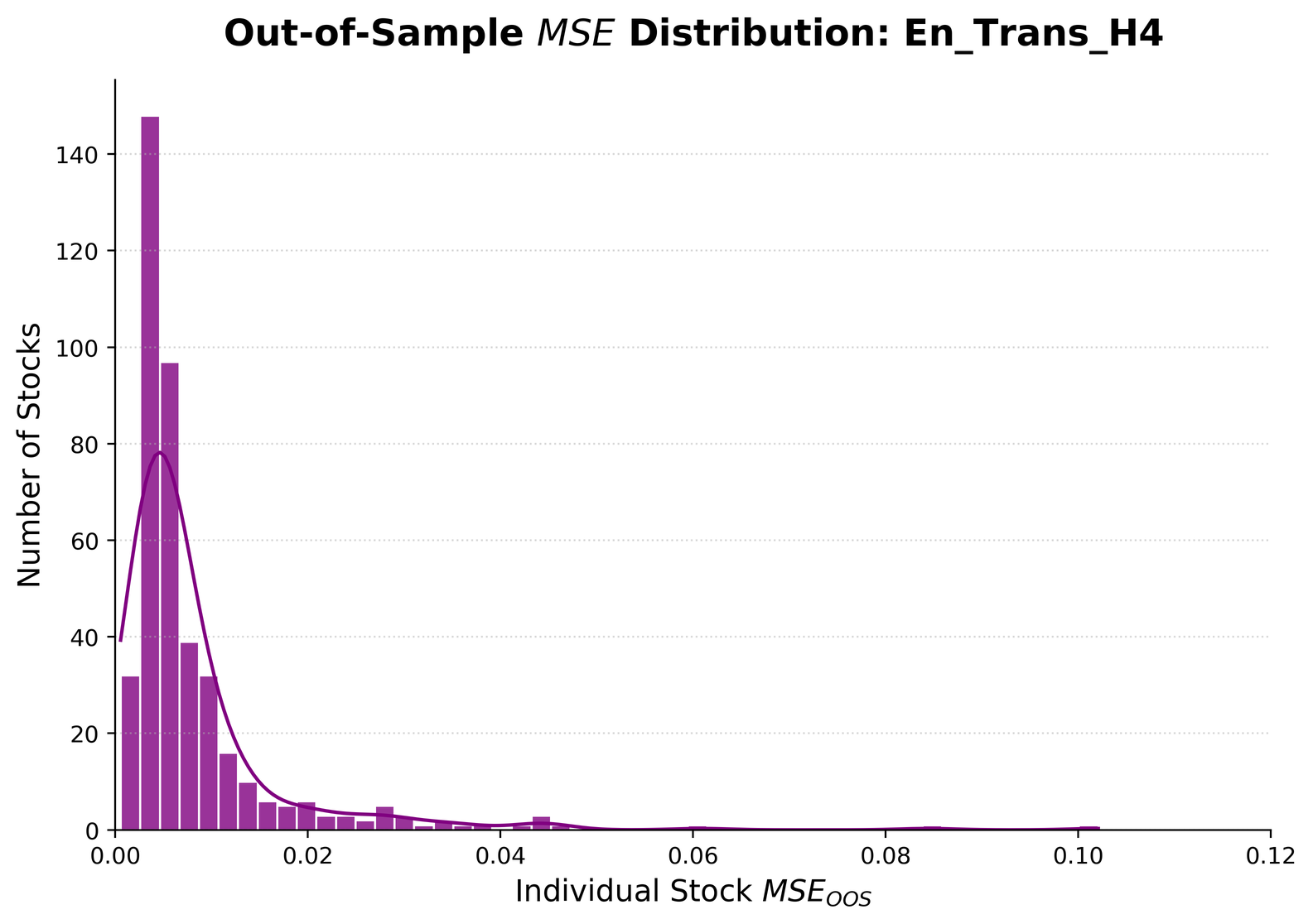}
  \caption{En\_Trans\_H4(2212)}
\end{subfigure}
 \caption[OOS $MSE$ distribution diagrams of best-performed SERT models and their benchmarks.]{[OOS $MSE$ distribution diagrams of best-performed SERT models and their benchmarks in each group for each period. The X-axis shows the OOS $MSE$ value ranges in each subgraph, while the Y-axis shows the stock numbers.}
  \label{fig:mse_distribution522_SERT_ch3}
\end{sidewaysfigure}

Figure~\ref{fig:r2_distribution521_SERT_ch3} shows the OOS $R^2$ distribution of the best proposed SERT model and its benchmarks: LNF SERT and the best standard encoder-only Transformer. The diagrams in the pre-COVID period show no significant difference between the best models in each group, but during the high market fluctuation periods, which are ‘2112’ and ‘2212’, the best proposed SERT model significantly surpasses the alternatives. This agrees with the findings from other model performance indicators. Although the diagrams also present the inconsistency of the best proposed SERT model in different periods, according to DM test results, the differences between the proposed SERT models are insignificant. Thus, even if the inconsistency exists between different periods, it cannot affect the robustness of the proposed SERT models’ outperformance towards benchmarks. The OOS MSE distribution diagrams support the findings from the OOS $R^2$ distribution diagrams, which are presented in Figure~\ref{fig:mse_distribution522_SERT_ch3}.\\

Compared with Transformer models, encoder-only Transformer models and SERT models have slightly higher model fitness at the expense of training stability. Similar to Transformer models, the proposed SERT models significantly exceed the encoder-only Transformer models during extreme market fluctuation periods.\\

\subsection{Backtesting performance}\label{subsec:Factor investing strategy-wise performance_ch3}
The Transformer models and encoder-only Transformer models are backtested with two approaches: sign signal (one of trend following trading strategies) and sign signal with softmax signal filter. They are analyzed in equal-weighted methods and value-weighted methods. Under the sign signal trading rules, the long-only strategy positions are opened at the point that both predicted return and actual return have a positive sign, which means the next value of the signal is the first value counted as the capital gain or loss. The position is closed when an identical negative sign is detected in predicted and actual returns. The softmax trading signal filter utilizes the softmax function on sign signals and filters out the worst 50\% of trading signals for 420 stocks each month. The best models are selected mainly based on the indicator of the Sortino ratio and annualized returns, the maximum drawdown is the reference for the extreme downside risk. Compared with the classic Sharpe ratio, the Sortino ratio is more reasonable since investors could be more sensitive to the downside risk from the investing psychology perspective. And the annualized returns are the absolute yearly capital gain of a strategy. Table~\ref{tab:BH531_ch3} exhibits the performance of equal-weighted (BHE) and value-weighted (BHV) buy-and-hold strategies as the most fundamental benchmarks commonly used for measuring investing performance. All models in this study are compared with these benchmarks.\\
\begin{table}[htbp!]
\centering
\small 
\begin{tabular}{llcccccc}
\toprule
\textbf{Strategy} & \textbf{Period} & \textbf{MDD} & \textbf{Ann. Return} & \textbf{Sharpe} & \textbf{Sortino} & \textbf{Ann. SR} & \textbf{Ann. SO} \\
\midrule
\multirow{3}{*}{BHE} 
& 1911 & -0.3701 & 0.1139 & 0.2588 & 0.3631 & 0.8965 & 1.2578 \\
& 2112 & -0.5310 & 0.1349 & 0.2472 & 0.2988 & 0.8563 & 1.0351 \\
& 2212 & -0.5308 & 0.1141 & 0.2148 & 0.2760 & 0.7441 & 0.9561 \\
\midrule
\multirow{3}{*}{BHV} 
& 1911 & -0.2963 & 0.1197 & 0.2956 & 0.4243 & 1.0240 & 1.4698 \\
& 2112 & -0.4513 & 0.1337 & 0.2673 & 0.3263 & 0.9260 & 1.1303 \\
& 2212 & -0.4513 & 0.1156 & 0.2311 & 0.3031 & 0.8006 & 1.0500 \\
\bottomrule
\end{tabular}
\caption[The backtesting performance of buy-and-hold strategies during the examined periods considering the static transaction cost.]{The backtesting performance of buy-and-hold strategies during the examined periods considering the static transaction cost. ‘MDD’ is the maximum drawdown, ‘sharpe’ means Sharpe ratio, ‘sortino’ means Sortino ratio. BHE and BHV are denoted as buy-and-hold strategies for the equal-weighted portfolio and the value-weighted portfolio, respectively. Ann.SR and Ann.SO are the annualized Sharpe and Sortino ratios.}
\label{tab:BH531_ch3}
\end{table}

\subsubsection{Transformer}\label{subsec:5.3.1	Transformer_ch3}
Table~\ref{tab:invest_perform5311_ch3} presents the investing performance of equal-weighted (EW) portfolios under the static transaction cost scenario, for both sign signal and sign signal with softmax filter strategies. For the strategy without the softmax filter, none of the models has an annualized return higher than the one in the buy-and-hold strategy, and there is no significant difference between the annualized returns of each model during the Pre–COVID-19 Period (1911). In addition, the Sortino ratio indicates that only Models ‘P\_Trans\_H2’, ‘P\_Trans\_H7’, and ‘C\_Trans\_H1’ surpass the buy-and-hold benchmark, and Model ‘P\_Trans\_H7’ (the proposed model with 7 heads) outperforms all other models during the pre-COVID testing period. However, the proposed pre-trained Transformer models significantly outperform the benchmark models as well as the buy-and-hold strategy (BHE) during the extreme market fluctuation (2112 and 2212) according to the Sortino ratio, which supports the findings in Section~\ref{subsec:Model performance of Transformer_ch3} that the proposed models perform better during the extreme market conditions. The ‘P\_Trans\_H6’ (with 6 heads) edges out Model ‘P\_Trans\_H7’ to become the best model in ‘2112’ and ‘2212’ in the sign signal for the equal-weighted portfolio group.\\

From the back testing results of the softmax filter for the equal-weighted portfolio considering the static transaction cost, the softmax filter generally increases the annualized returns for all models in all periods. With the signal filter, all models’ annualized returns perform slightly better than the buy-and-hold strategy in the testing period of ‘2112’. And most of the models' annualized returns in the rest of the two periods excel the buy-and-hold benchmarks. Whereas it does not significantly lower the downside risks, it eliminates the advantage of the proposed models and the difference between models. \\

\begin{table}[htbp!]
\centering
\small
\renewcommand{\arraystretch}{1.1}
\begin{tabularx}{\textwidth}{l *{6}{>{\centering\arraybackslash}X}}
\toprule
\multicolumn{7}{c}{\textbf{Sign Equal Weighted Portfolios In Static Transaction Cost}} \\
\midrule
\multicolumn{7}{c}{\textbf{Pre–COVID-19 Period (1911)}} \\
\midrule
\textbf{Model} & AR & Ann.SR & SR & Ann.SO & SO & MDD \\
\midrule
P\_Trans\_LNF & 0.0916 & 0.8421 & 0.2431 & 1.1220 & 0.3239 & -0.2949 \\
P\_Trans\_H1  & 0.0901 & 0.7985 & 0.2305 & 1.0247 & 0.2958 & -0.3211 \\
P\_Trans\_H2  & 0.0927 & 0.8726 & 0.2519 & 1.3306 & 0.3841 & -0.3503 \\
P\_Trans\_H3  & 0.0935 & 0.8293 & 0.2394 & 1.0538 & 0.3042 & -0.3093 \\
P\_Trans\_H4  & 0.0885 & 0.7777 & 0.2245 & 1.0084 & 0.2911 & -0.3415 \\
P\_Trans\_H6  & 0.0979 & 0.8851 & 0.2555 & 1.2100 & 0.3493 & -0.3511 \\
P\_Trans\_H7  & 0.1011 & 0.9131 & 0.2636 & 1.3493 & 0.3895 & -0.3523 \\
C\_Trans\_H1  & 0.1005 & 0.9166 & 0.2646 & 1.2758 & 0.3683 & -0.2349 \\
C\_Trans\_H2  & 0.1014 & 0.8691 & 0.2509 & 1.1778 & 0.3400 & -0.2916 \\
C\_Trans\_H4  & 0.0946 & 0.7850 & 0.2266 & 1.0843 & 0.3130 & -0.4201 \\
\midrule
\multicolumn{7}{c}{\textbf{COVID-19–Inclusive Period (2112)}} \\
\midrule
\textbf{Model} & AR & Ann.SR & SR & Ann.SO & SO & MDD \\
\midrule
P\_Trans\_LNF & 0.1124 & 0.8293 & 0.2394 & 0.9789 & 0.2826 & -0.5589 \\
P\_Trans\_H1  & 0.1241 & 0.9256 & 0.2672 & 1.3216 & 0.3815 & -0.4768 \\
P\_Trans\_H2  & 0.1255 & 0.9384 & 0.2709 & 1.3645 & 0.3939 & -0.4946 \\
P\_Trans\_H3  & 0.1259 & 0.9391 & 0.2711 & 1.3101 & 0.3782 & -0.4625 \\
P\_Trans\_H4  & 0.1160 & 0.9014 & 0.2602 & 1.2495 & 0.3607 & -0.4631 \\
P\_Trans\_H6  & 0.1284 & 0.9838 & 0.2840 & 1.4341 & 0.4140 & -0.4218 \\
P\_Trans\_H7  & 0.1250 & 0.9627 & 0.2779 & 1.3735 & 0.3965 & -0.4184 \\
C\_Trans\_H1  & 0.1271 & 0.8504 & 0.2455 & 1.0212 & 0.2948 & -0.5789 \\
C\_Trans\_H2  & 0.1231 & 0.8276 & 0.2389 & 0.9727 & 0.2808 & -0.5659 \\
C\_Trans\_H4  & 0.1208 & 0.7922 & 0.2287 & 0.9644 & 0.2784 & -0.5928 \\
\midrule
\multicolumn{7}{c}{\textbf{Period Including COVID-19 and One-Year After (2212)}} \\
\midrule
\textbf{Model} & AR & Ann.SR & SR & Ann.SO & SO & MDD \\
\midrule
P\_Trans\_LNF & 0.0886 & 0.6267 & 0.1809 & 0.7763 & 0.2241 & -0.5589 \\
P\_Trans\_H1  & 0.1056 & 0.7462 & 0.2154 & 1.0791 & 0.3115 & -0.4768 \\
P\_Trans\_H2  & 0.1075 & 0.7486 & 0.2161 & 1.0915 & 0.3151 & -0.4946 \\
P\_Trans\_H3  & 0.1055 & 0.7441 & 0.2148 & 1.0590 & 0.3057 & -0.4625 \\
P\_Trans\_H4  & 0.0958 & 0.6984 & 0.2016 & 0.9900 & 0.2858 & -0.4631 \\
P\_Trans\_H6  & 0.1070 & 0.7635 & 0.2204 & 1.1307 & 0.3264 & -0.4218 \\
P\_Trans\_H7  & 0.1074 & 0.7656 & 0.2210 & 1.1141 & 0.3216 & -0.4184 \\
C\_Trans\_H1  & 0.1066 & 0.6921 & 0.1998 & 0.8823 & 0.2547 & -0.5789 \\
C\_Trans\_H2  & 0.1047 & 0.6790 & 0.1960 & 0.8421 & 0.2431 & -0.5659 \\
C\_Trans\_H4  & 0.1017 & 0.6523 & 0.1883 & 0.8432 & 0.2434 & -0.5928 \\
\bottomrule
\end{tabularx}
\caption[The backtesting performance for the equal-weighted portfolios of Transformer models considering static transaction cost.]{The investing performance for the equal-weighted portfolios of Transformer models considering static transaction cost. P\_Trans\_LNF to C\_Trans\_H4 are the names of models. The ‘Sign\_equal\_weighted’ means Sign signal strategy for equal-weighted portfolio performance. MDD is the maximum drawdown, `SR' means Sharpe ratio, `Ann.SR' is the annualized Sharpe ratio, `SO' means Sortino ratio and `Ann.SO' is the annualized Sortino ratio.}
\label{tab:invest_perform5311_ch3}
\end{table}

For the value-weighted (VW) portfolio considering the static transaction cost, which is shown in Figure~\ref{tab:invest_vwperform5312_ch3} in the sign signal trading strategy, the annualized returns of all models are surprisingly lowered. The profitability under downside risks, which is shown in the Sortino ratio, of all models becomes more variable between models, but during the 2112 and 2212 period, the proposed models still significantly outperform the alternatives, and most of them are better than the ‘buy-and-hold’ strategy. The Model ‘P\_Trans\_H2’ shows the best performance during the large market fluctuation. The value-weighted method moderates the downside risks to a degree but removes the absolute capital gain considerably. This phenomenon does not happen in MLP pricing model studies. The possible reason for that is the Transformer models have more strength for detecting and modelling data with high fluctuations, which means higher fluctuations, and higher capital gains, but under the value-weighted condition, higher weights added on the large capital stocks with lower fluctuations derive less profit. The softmax filter still rules out the difference between models and increases the annualized returns as well as downside risks, but none of them excels in the buy-and-hold (BHV) benchmark in the high market fluctuation period. However, it significantly increases performance during the pre-COVID period, which implies that it can rule out unprofitable signals and reduce the downside risks when market fluctuation is not incredibly significant. It adjusts the weight distribution bias as well.\\

\begin{table}[htbp!]
\centering
\small
\renewcommand{\arraystretch}{1.1}
\begin{tabularx}{\textwidth}{l *{6}{>{\centering\arraybackslash}X}}
\toprule
\multicolumn{7}{c}{\textbf{Sign Value-Weighted Portfolios In Static Transaction Cost}} \\
\midrule
\multicolumn{7}{c}{\textbf{Pre–COVID-19 Period (1911)}} \\
\midrule
\textbf{Model} & AR & Ann.SR & SR & Ann.SO & SO & MDD \\
\midrule
P\_Trans\_LNF & 0.0532 & 0.7981 & 0.2304 & 0.9183 & 0.2651 & -0.3061 \\
P\_Trans\_H1  & 0.0455 & 0.6267 & 0.1809 & 0.6745 & 0.1949 & -0.3289 \\
P\_Trans\_H2  & 0.0535 & 0.8359 & 0.2413 & 1.0126 & 0.2923 & -0.3260 \\
P\_Trans\_H3  & 0.0494 & 0.6765 & 0.1953 & 0.7534 & 0.2175 & -0.3044 \\
P\_Trans\_H4  & 0.0441 & 0.6322 & 0.1825 & 0.6960 & 0.2009 & -0.3202 \\
P\_Trans\_H6  & 0.0538 & 0.8224 & 0.2374 & 0.9526 & 0.2750 & -0.2939 \\
P\_Trans\_H7  & 0.0540 & 0.8525 & 0.2461 & 1.0316 & 0.2978 & -0.3242 \\
C\_Trans\_H1  & 0.0548 & 0.8872 & 0.2561 & 1.0721 & 0.3095 & -0.2883 \\
C\_Trans\_H2  & 0.0527 & 0.7171 & 0.2070 & 0.8428 & 0.2433 & -0.3649 \\
C\_Trans\_H4  & 0.0566 & 0.8179 & 0.2361 & 1.0234 & 0.2954 & -0.3061 \\
\midrule
\multicolumn{7}{c}{\textbf{COVID-19–Inclusive Period (2112)}} \\
\midrule
\textbf{Model} & AR & Ann.SR & SR & Ann.SO & SO & MDD \\
\midrule
P\_Trans\_LNF & 0.0648 & 0.8768 & 0.2531 & 1.1012 & 0.3179 & -0.3415 \\
P\_Trans\_H1  & 0.0691 & 0.8757 & 0.2528 & 1.1251 & 0.3248 & -0.3473 \\
P\_Trans\_H2  & 0.0801 & 1.0785 & 0.3113 & 1.5935 & 0.4600 & -0.3260 \\
P\_Trans\_H3  & 0.0748 & 0.9444 & 0.2726 & 1.2223 & 0.3529 & -0.3044 \\
P\_Trans\_H4  & 0.0617 & 0.8369 & 0.2415 & 1.0475 & 0.3026 & -0.3202 \\
P\_Trans\_H6  & 0.0785 & 1.0538 & 0.3042 & 1.4805 & 0.4274 & -0.2939 \\
P\_Trans\_H7  & 0.0709 & 1.0205 & 0.2946 & 1.3926 & 0.4020 & -0.3242 \\
C\_Trans\_H1  & 0.0744 & 0.8636 & 0.2493 & 1.0945 & 0.3160 & -0.4398 \\
C\_Trans\_H2  & 0.0658 & 0.7825 & 0.2259 & 0.9422 & 0.2720 & -0.4353 \\
C\_Trans\_H4  & 0.0745 & 0.8693 & 0.2509 & 1.1120 & 0.3210 & -0.4205 \\
\midrule
\multicolumn{7}{c}{\textbf{Period Including COVID-19 and One-Year After (2212)}} \\
\midrule
\textbf{Model} & AR & Ann.SR & SR & Ann.SO & SO & MDD \\
\midrule
P\_Trans\_LNF & 0.0495 & 0.6357 & 0.1835 & 0.8268 & 0.2387 & -0.3415 \\
P\_Trans\_H1  & 0.0584 & 0.6700 & 0.1934 & 0.8715 & 0.2516 & -0.3473 \\
P\_Trans\_H2  & 0.0672 & 0.8446 & 0.2436 & 1.2308 & 0.3553 & -0.3260 \\
P\_Trans\_H3  & 0.0629 & 0.7558 & 0.2180 & 1.0022 & 0.2893 & -0.3044 \\
P\_Trans\_H4  & 0.0509 & 0.6284 & 0.1814 & 0.8089 & 0.2335 & -0.3202 \\
P\_Trans\_H6  & 0.0683 & 0.8168 & 0.2358 & 1.1411 & 0.3294 & -0.2939 \\
P\_Trans\_H7  & 0.0611 & 0.8106 & 0.2340 & 1.1179 & 0.3227 & -0.3242 \\
C\_Trans\_H1  & 0.0634 & 0.7095 & 0.2048 & 0.9412 & 0.2717 & -0.4398 \\
C\_Trans\_H2  & 0.0556 & 0.6422 & 0.1855 & 0.8186 & 0.2363 & -0.4353 \\
C\_Trans\_H4  & 0.0653 & 0.7191 & 0.2076 & 0.9692 & 0.2797 & -0.4205 \\
\bottomrule
\end{tabularx}
\caption[The investing performance for the value-weighted portfolios of Transformer models considering the static transaction cost.]{The investing performance for the value-weighted portfolios of Transformer models considering the static transaction cost. P\_Trans\_LNF to C\_Trans\_H4 are model names. MDD is maximum drawdown, SR (Sharpe) means Sharpe ratio, Ann.SR is the annualized Sharpe ratio, SO (Sortino) means Sortino ratio and Ann.SO is the annualized Sortino ratio. Negative values are denoted by a minus sign.}
\label{tab:invest_vwperform5312_ch3}
\end{table}

The robustness examination is conducted to investigate how the turnover rate would affect the profitability of these models. It applies the turnover rate considering the 20bps transaction deduction rate for large-cap stocks. Table ~\ref{tab:invest_perform5311_ch4_ew_trans_turnover} and Table ~\ref{tab:invest_vwperform5312_ch3_turnover} show the dynamic transaction cost robustness backtesting results for equal-weighted and value-weighted portfolios, respectively. Comparing the backtesting results of equal-weighted portfolios in both transaction cost scenarios, the conservative penalty (the static transaction cost of 50bps) significantly lowers the absolute profitability (AR) of the models in all periods, but the high turnover rates deteriorate the models' profitability when considering the downside risks (SO). However, the high turnover moderates the maximum drawdown (MDD). This indicates that models with high predictive power are highly sensitive to capturing the noise. In the dynamic transaction cost scenario, none of the models' absolute capital returns is higher than the buy-and-hold (BHE) benchmarks. Similar conclusion draws in the comparison study of value-weighted portfolios, and the static transaction cost (50bps) over-penalizes the models' profitability as well, according to annulized returns (AR) increasing and decreasing maximum drawdown (MDD) in dynamic transaction case. \\

\begin{table}[htbp]
\centering
\small
\renewcommand{\arraystretch}{1.1}

\begin{tabularx}{\textwidth}{l *{7}{X}}
\toprule
\multicolumn{8}{c}{\textbf{Pre--COVID-19 Period (1911)}} \\
\midrule
\textbf{Model} & \textbf{AR} & \textbf{Ann.SR} & \textbf{SR} & \textbf{Ann.SO} & \textbf{SO} & \textbf{MDD} & \textbf{Turnover} \\
\midrule
P\_Trans\_LNF & 0.0733 & 0.6683 & 0.1929 & 0.8658 & 0.2499 & -0.1511 & 0.7229 \\
P\_Trans\_H1  & 0.0680 & 0.6002 & 0.1733 & 0.7497 & 0.2164 & -0.1772 & 0.8748 \\
P\_Trans\_H2  & 0.0709 & 0.6641 & 0.1917 & 0.9958 & 0.2875 & -0.1481 & 0.8667 \\
P\_Trans\_H3  & 0.0718 & 0.6344 & 0.1831 & 0.7810 & 0.2255 & -0.1730 & 0.8619 \\
P\_Trans\_H4  & 0.0653 & 0.5726 & 0.1653 & 0.7190 & 0.2076 & -0.1773 & 0.9185 \\
P\_Trans\_H6  & 0.0787 & 0.7087 & 0.2046 & 0.9390 & 0.2711 & -0.1394 & 0.7589 \\
P\_Trans\_H7  & 0.0784 & 0.7053 & 0.2036 & 1.0189 & 0.2941 & -0.1368 & 0.8968 \\
C\_Trans\_H1         & 0.0837 & 0.7586 & 0.2190 & 1.0321 & 0.2979 & -0.1297 & 0.6654 \\
C\_Trans\_H2         & 0.0828 & 0.7064 & 0.2039 & 0.9383 & 0.2709 & -0.1472 & 0.7340 \\
C\_Trans\_H4         & 0.0826 & 0.6820 & 0.1969 & 0.9079 & 0.2621 & -0.1463 & 0.4798 \\
BHE           & 0.1141 & 0.8892 & 0.2567 & 1.2303 & 0.3552 & -0.1483 & 0.0120 \\
\midrule

\multicolumn{8}{c}{\textbf{COVID--19--Inclusive Period (2112)}} \\
\midrule
P\_Trans\_LNF & 0.0944 & 0.6968 & 0.2011 & 0.8052 & 0.2324 & -0.2956 & 0.6980 \\
P\_Trans\_H1  & 0.1037 & 0.7726 & 0.2230 & 1.0770 & 0.3109 & -0.2778 & 0.7861 \\
P\_Trans\_H2  & 0.1053 & 0.7862 & 0.2270 & 1.1211 & 0.3236 & -0.2823 & 0.7829 \\
P\_Trans\_H3  & 0.1059 & 0.7891 & 0.2278 & 1.0688 & 0.3085 & -0.2731 & 0.7729 \\
P\_Trans\_H4  & 0.0948 & 0.7358 & 0.2124 & 0.9890 & 0.2855 & -0.2674 & 0.8229 \\
P\_Trans\_H6  & 0.1107 & 0.8454 & 0.2441 & 1.1946 & 0.3449 & -0.2428 & 0.6875 \\
P\_Trans\_H7  & 0.1041 & 0.8007 & 0.2311 & 1.1174 & 0.3226 & -0.2548 & 0.8051 \\
C\_Trans\_H1         & 0.1123 & 0.7503 & 0.2166 & 0.8926 & 0.2577 & -0.3105 & 0.5733 \\
C\_Trans\_H2         & 0.1071 & 0.7198 & 0.2078 & 0.8385 & 0.2420 & -0.2883 & 0.6220 \\
C\_Trans\_H4         & 0.1102 & 0.7204 & 0.2080 & 0.8632 & 0.2492 & -0.2904 & 0.4162 \\
BHE           & 0.1353 & 0.8512 & 0.2457 & 1.0181 & 0.2939 & -0.2890 & 0.0093 \\
\midrule

\multicolumn{8}{c}{\textbf{Period Including COVID--19 and One-Year After (2212)}} \\
\midrule
P\_Trans\_LNF & 0.0705 & 0.5036 & 0.1454 & 0.6125 & 0.1768 & -0.2956 & 0.7144 \\
P\_Trans\_H1  & 0.0866 & 0.6146 & 0.1774 & 0.8743 & 0.2524 & -0.2778 & 0.7414 \\
P\_Trans\_H2  & 0.0884 & 0.6183 & 0.1785 & 0.8892 & 0.2567 & -0.2823 & 0.7498 \\
P\_Trans\_H3  & 0.0870 & 0.6160 & 0.1778 & 0.8544 & 0.2466 & -0.2731 & 0.7270 \\
P\_Trans\_H4  & 0.0762 & 0.5585 & 0.1612 & 0.7721 & 0.2229 & -0.2674 & 0.7707 \\
P\_Trans\_H6  & 0.0906 & 0.6479 & 0.1870 & 0.9366 & 0.2704 & -0.2428 & 0.6434 \\
P\_Trans\_H7  & 0.0879 & 0.6288 & 0.1815 & 0.9015 & 0.2602 & -0.2548 & 0.7634 \\
C\_Trans\_H1         & 0.0927 & 0.6036 & 0.1742 & 0.7622 & 0.2200 & -0.3105 & 0.5460 \\
C\_Trans\_H2         & 0.0897 & 0.5838 & 0.1685 & 0.7177 & 0.2072 & -0.2883 & 0.5938 \\
C\_Trans\_H4         & 0.0917 & 0.5880 & 0.1697 & 0.7462 & 0.2154 & -0.2904 & 0.3995 \\
BHE           & 0.1143 & 0.7017 & 0.2026 & 0.8883 & 0.2564 & -0.2890 & 0.0083 \\
\bottomrule
\end{tabularx}
\caption[The backtesting results for the equal-weighted portfolios of Transformer models considering the dynamic transaction cost.]{The backtesting results for the equal-weighted portfolios of Transformer models considering the dynamic transaction cost. 'BHE' represents the buy-and-hold benchmark strategy under equal-weighted method.}
\label{tab:invest_perform5311_ch4_ew_trans_turnover}
\end{table}

\clearpage
\begin{table}[htbp]
\centering
\small
\renewcommand{\arraystretch}{1.2}

\begin{tabularx}{\textwidth}{l *{7}{X}}
\toprule
\multicolumn{8}{c}{\textbf{Pre--COVID-19 Period (1911)}} \\
\midrule
\textbf{Model} & \textbf{AR} & \textbf{Ann.SR} & \textbf{SR} & \textbf{Ann.SO} & \textbf{SO} & \textbf{MDD} & \textbf{Turnover} \\
\midrule
P\_Trans\_LNF & 0.1125 & 1.0716 & 0.3094 & 1.3302 & 0.3840 & -0.1234 & 0.6952 \\
P\_Trans\_H1  & 0.0935 & 0.8719 & 0.2517 & 1.0950 & 0.3161 & -0.1281 & 0.8439 \\
P\_Trans\_H2  & 0.0939 & 0.8869 & 0.2560 & 1.1551 & 0.3334 & -0.1463 & 0.8166 \\
P\_Trans\_H3  & 0.0861 & 0.8096 & 0.2337 & 0.9607 & 0.2773 & -0.1353 & 0.8094 \\
P\_Trans\_H4  & 0.0904 & 0.8425 & 0.2432 & 1.0353 & 0.2989 & -0.1315 & 0.8541 \\
P\_Trans\_H6  & 0.0998 & 0.9457 & 0.2730 & 1.1738 & 0.3388 & -0.1436 & 0.7247 \\
P\_Trans\_H7  & 0.0983 & 0.9460 & 0.2731 & 1.3033 & 0.3762 & -0.1342 & 0.8495 \\
C\_Trans\_H1  & 0.1117 & 1.0511 & 0.3034 & 1.3977 & 0.4035 & -0.0961 & 0.6430 \\
C\_Trans\_H2  & 0.1017 & 0.9289 & 0.2681 & 1.1851 & 0.3421 & -0.1294 & 0.6911 \\
C\_Trans\_H4  & 0.1256 & 1.0954 & 0.3162 & 1.4692 & 0.4241 & -0.1138 & 0.4638 \\
BHV           & 0.1388 & 1.1795 & 0.3405 & 1.6054 & 0.4634 & -0.1297 & 0.0120 \\
\midrule

\multicolumn{8}{c}{\textbf{COVID--19--Inclusive Period (2112)}} \\
\midrule
P\_Trans\_LNF & 0.1315 & 1.0401 & 0.3002 & 1.2939 & 0.3735 & -0.1959 & 0.6775 \\
P\_Trans\_H1  & 0.1268 & 0.9590 & 0.2768 & 1.3431 & 0.3877 & -0.1784 & 0.7638 \\
P\_Trans\_H2  & 0.1213 & 0.9453 & 0.2729 & 1.2489 & 0.3605 & -0.1931 & 0.7400 \\
P\_Trans\_H3  & 0.1214 & 0.9703 & 0.2801 & 1.3288 & 0.3836 & -0.1471 & 0.7300 \\
P\_Trans\_H4  & 0.1176 & 0.9712 & 0.2804 & 1.3255 & 0.3826 & -0.1479 & 0.7819 \\
P\_Trans\_H6  & 0.1267 & 1.0279 & 0.2967 & 1.4166 & 0.4090 & -0.1583 & 0.6647 \\
P\_Trans\_H7  & 0.1235 & 1.0156 & 0.2932 & 1.4776 & 0.4266 & -0.1596 & 0.7696 \\
C\_Trans\_H1  & 0.1348 & 0.9832 & 0.2838 & 1.2351 & 0.3566 & -0.2237 & 0.5539 \\
C\_Trans\_H2  & 0.1276 & 0.9425 & 0.2721 & 1.1958 & 0.3452 & -0.2142 & 0.5896 \\
C\_Trans\_H4  & 0.1431 & 1.0245 & 0.2957 & 1.2709 & 0.3669 & -0.2279 & 0.4060 \\
BHV           & 0.1540 & 1.0688 & 0.3085 & 1.3057 & 0.3769 & -0.2352 & 0.0093 \\
\midrule

\multicolumn{8}{c}{\textbf{Period Including COVID-19 and One-Year After (2212)}} \\
\midrule
P\_Trans\_LNF & 0.1055 & 0.7924 & 0.2287 & 1.0260 & 0.2962 & -0.1959 & 0.6967 \\
P\_Trans\_H1  & 0.1060 & 0.7585 & 0.2190 & 1.0760 & 0.3106 & -0.1784 & 0.7109 \\
P\_Trans\_H2  & 0.0998 & 0.7169 & 0.2070 & 0.9831 & 0.2838 & -0.1931 & 0.7053 \\
P\_Trans\_H3  & 0.0990 & 0.7470 & 0.2157 & 1.0397 & 0.3001 & -0.1756 & 0.6827 \\
P\_Trans\_H4  & 0.0968 & 0.7447 & 0.2150 & 1.0379 & 0.2996 & -0.1661 & 0.7352 \\
P\_Trans\_H6  & 0.1086 & 0.8012 & 0.2313 & 1.1222 & 0.3240 & -0.1756 & 0.6295 \\
P\_Trans\_H7  & 0.1050 & 0.7752 & 0.2238 & 1.1152 & 0.3219 & -0.1656 & 0.7357 \\
C\_Trans\_H1  & 0.1126 & 0.7825 & 0.2259 & 1.0495 & 0.3030 & -0.2237 & 0.5278 \\
C\_Trans\_H2  & 0.1060 & 0.7367 & 0.2127 & 0.9748 & 0.2814 & -0.2142 & 0.5660 \\
C\_Trans\_H4  & 0.1232 & 0.8376 & 0.2418 & 1.0988 & 0.3172 & -0.2279 & 0.3875 \\
BHV           & 0.1308 & 0.8654 & 0.2498 & 1.1270 & 0.3253 & -0.2352 & 0.0083 \\
\bottomrule
\end{tabularx}
\caption[The backtesting results for the value-weighted portfolios of Transformer models considering the dynamic transaction cost.]{The backtesting results for the value-weighted portfolios of Transformer models considering the dynamic transaction cost. 'BHV' represents the buy-and-hold benchmark of the value-weighted method.}
\label{tab:invest_vwperform5312_ch3_turnover}
\end{table}

\subsubsection{Encoder-only Transformer and SERT models}\label{subsec:532Encoder-only Transformer and SERT models}
Some findings from Transformer models agree with the findings of the proposed SERT models and encoder-only Transformer benchmarks in the static transaction cost scenario. The proposed SERT models show the absolute advantage according to the Sharpe ratio and Sortino ratio during the high fluctuation periods in the equal-weighted portfolios with a non-filter trading signal system. Model SERT\_H4 is the best-performing model during the 2112 and 2212 periods, which outperforms the BHE benchmark in annualized return, Sharpe ratio and Sortino ratio during the pandemic. Furthermore, during the pandemic period of 2112, all proposed SERT models outperformed the encoder-only Transformer models and benchmark models concerning the Sharpe ratio and Sortino ratio. However, all models in the 2212 period have no better performance than BHE in the equal-weighted portfolio without a signal filter. The softmax filter still removes the difference between models, but it makes all models slightly exceed the BHE benchmark performance in the first two periods in most cases. In period 2212, it increases the annualized returns for most models, but not the profit under the risks. Table~\ref{tab:invest_SERT_ewperform5321_ch3} shows the investment performance for the equal-weighted portfolios of the encoder-only Transformer models considering the static transaction cost (50bps). \\
\begin{table}[htbp!]
\centering
\small
\renewcommand{\arraystretch}{1.1}
\begin{tabularx}{\textwidth}{l *{6}{>{\centering\arraybackslash}X}}
\toprule
\multicolumn{7}{c}{\textbf{Sign Equal Weighted Portfolios In Static Transaction Cost}} \\
\midrule
\multicolumn{7}{c}{\textbf{Pre–COVID-19 Period (1911)}} \\
\midrule
\textbf{Model} & AR & Ann.SR & SR & Ann.SO & SO & MDD \\
\midrule
SERT\_LNF     & 0.0968 & 0.9083 & 0.2622 & 1.2724 & 0.3673 & -0.3566 \\
SERT\_H1      & 0.0939 & 0.8646 & 0.2496 & 1.2284 & 0.3546 & -0.3066 \\
SERT\_H2      & 0.0839 & 0.7597 & 0.2193 & 0.9339 & 0.2696 & -0.3586 \\
SERT\_H3      & 0.0894 & 0.8137 & 0.2349 & 1.1255 & 0.3249 & -0.3604 \\
SERT\_H4      & 0.1066 & 0.9481 & 0.2737 & 1.1563 & 0.3338 & -0.2956 \\
SERT\_H6      & 0.0908 & 0.7843 & 0.2264 & 1.0403 & 0.3003 & -0.3651 \\
SERT\_H7      & 0.0890 & 0.7857 & 0.2268 & 1.0042 & 0.2899 & -0.3321 \\
En\_Trans\_H1 & 0.0940 & 0.8775 & 0.2533 & 1.2928 & 0.3732 & -0.2941 \\
En\_Trans\_H2 & 0.0964 & 0.8324 & 0.2403 & 1.2284 & 0.3546 & -0.3904 \\
En\_Trans\_H4 & 0.0960 & 0.8380 & 0.2419 & 1.0753 & 0.3104 & -0.3059 \\
\midrule
\multicolumn{7}{c}{\textbf{COVID-19–Inclusive Period (2112)}} \\
\midrule
\textbf{Model} & AR & Ann.SR & SR & Ann.SO & SO & MDD \\
\midrule
SERT\_LNF     & 0.1263 & 0.9083 & 0.2622 & 1.2124 & 0.3500 & -0.5154 \\
SERT\_H1      & 0.1252 & 0.9478 & 0.2736 & 1.3773 & 0.3976 & -0.4498 \\
SERT\_H2      & 0.1162 & 0.8889 & 0.2566 & 1.2138 & 0.3504 & -0.5137 \\
SERT\_H3      & 0.1201 & 0.9284 & 0.2680 & 1.3801 & 0.3984 & -0.4354 \\
SERT\_H4      & 0.1374 & 1.0351 & 0.2988 & 1.4186 & 0.4095 & -0.4119 \\
SERT\_H6      & 0.1184 & 0.8560 & 0.2471 & 1.1331 & 0.3271 & -0.5083 \\
SERT\_H7      & 0.1214 & 0.9055 & 0.2614 & 1.2523 & 0.3615 & -0.4993 \\
En\_Trans\_H1 & 0.1222 & 0.8411 & 0.2428 & 1.0264 & 0.2963 & -0.6018 \\
En\_Trans\_H2 & 0.1252 & 0.8189 & 0.2364 & 1.0278 & 0.2967 & -0.6110 \\
En\_Trans\_H4 & 0.1197 & 0.8151 & 0.2353 & 0.9800 & 0.2829 & -0.5944 \\
\midrule
\multicolumn{7}{c}{\textbf{Period Including COVID-19 and One-Year After (2212)}} \\
\midrule
\textbf{Model} & AR & Ann.SR & SR & Ann.SO & SO & MDD \\
\midrule
SERT\_LNF     & 0.1016 & 0.7070 & 0.2041 & 0.9793 & 0.2827 & -0.5154 \\
SERT\_H1      & 0.1053 & 0.7489 & 0.2162 & 1.0863 & 0.3136 & -0.4498 \\
SERT\_H2      & 0.0897 & 0.6495 & 0.1875 & 0.8770 & 0.2532 & -0.5137 \\
SERT\_H3      & 0.1006 & 0.7229 & 0.2087 & 1.0697 & 0.3088 & -0.4354 \\
SERT\_H4      & 0.1089 & 0.7791 & 0.2249 & 1.0562 & 0.3049 & -0.4119 \\
SERT\_H6      & 0.0968 & 0.6651 & 0.1920 & 0.9104 & 0.2628 & -0.5083 \\
SERT\_H7      & 0.0978 & 0.6876 & 0.1985 & 0.9699 & 0.2800 & -0.4993 \\
En\_Trans\_H1 & 0.1024 & 0.6772 & 0.1955 & 0.8667 & 0.2502 & -0.6018 \\
En\_Trans\_H2 & 0.1044 & 0.6641 & 0.1917 & 0.8698 & 0.2511 & -0.6110 \\
En\_Trans\_H4 & 0.1009 & 0.6637 & 0.1916 & 0.8515 & 0.2458 & -0.5944 \\
\bottomrule
\end{tabularx}
\caption[The investing performance for the equal-weighted portfolios of the encoder-only Transformer models considering the static transaction cost.]{The investing performance for the equal-weighted portfolios of the encoder-only Transformer models considering the static transaction cost. SERT\_LNF to SERT\_H7 and En\_Trans\_H1 to En\_Trans\_H4 are the names of the models. SERT\_LNF is the proposed SERT with LNF and SERT\_H1 to SERT\_H7 are proposed SERTs with different attention heads. En\_Trans\_H1 to En\_Trans\_H4 are standard encoder-only Transformer models with different heads as benchmarks. MDD is the maximum drawdown, `SR' means Sharpe ratio, `Ann.SR' is the annualized Sharpe ratio, `SO' means Sortino ratio and `Ann.SO' is the annualized Sortino ratio. Negative values are denoted by a minus sign. }
\label{tab:invest_SERT_ewperform5321_ch3}
\end{table}

For the value-weighted portfolios, the annualized returns of all models are reduced in all periods. Most SERT models show the advantage during the pandemic period according to the Sortino ratio. Model SERT\_H1 outperforms all alternatives during the periods ‘2112’ and ‘2212’, achieving the highest returns under downside risk. The softmax signal filter moderates the difference between models as well. Similar to the function of the softmax filter in pre-trained Transformer models, the filter adjusts the weight distribution bias. With a softmax filter, SERT\_H1 becomes the model with the highest Sortino ratio, which is better than BHV. It also makes SERT\_LNF the best-performing model in Period 2112 and Period 2212 if only considering the Sortino ratio. SERT\_LNF is also better than the BHV benchmark. Table~\ref{tab:invest_SERT_vwperform5322_ch3} shows the investment performance for the value-weighed portfolios of encoder-only Transformer models. \\

\begin{table}[htbp!]
\centering
\small
\renewcommand{\arraystretch}{1.1}
\begin{tabularx}{\textwidth}{l *{6}{>{\centering\arraybackslash}X}}
\toprule
\multicolumn{7}{c}{\textbf{Sign Value-Weighted Portfolios In Static Transaction Cost}} \\
\midrule
\multicolumn{7}{c}{\textbf{Pre–COVID-19 Period (1911)}} \\
\midrule
\textbf{Model} & AR & Ann.SR & SR & Ann.SO & SO & MDD \\
\midrule
SERT\_LNF     & 0.0546 & 0.9114 & 0.2631 & 1.0108 & 0.2918 & -0.2475 \\
SERT\_H1      & 0.0540 & 0.8123 & 0.2345 & 0.9544 & 0.2755 & -0.2998 \\
SERT\_H2      & 0.0394 & 0.5355 & 0.1546 & 0.5473 & 0.1580 & -0.3715 \\
SERT\_H3      & 0.0445 & 0.6564 & 0.1895 & 0.7510 & 0.2168 & -0.4871 \\
SERT\_H4      & 0.0532 & 0.7216 & 0.2083 & 0.7427 & 0.2144 & -0.3211 \\
SERT\_H6      & 0.0499 & 0.6810 & 0.1966 & 0.7659 & 0.2211 & -0.3132 \\
SERT\_H7      & 0.0498 & 0.6551 & 0.1891 & 0.7084 & 0.2045 & -0.3096 \\
En\_Trans\_H1 & 0.0575 & 0.9194 & 0.2654 & 1.0538 & 0.3042 & -0.3684 \\
En\_Trans\_H2 & 0.0605 & 0.8986 & 0.2594 & 1.1217 & 0.3238 & -0.2395 \\
En\_Trans\_H4 & 0.0576 & 0.7645 & 0.2207 & 0.8168 & 0.2358 & -0.3000 \\
\midrule
\multicolumn{7}{c}{\textbf{COVID-19–Inclusive Period (2112)}} \\
\midrule
\textbf{Model} & AR & Ann.SR & SR & Ann.SO & SO & MDD \\
\midrule
SERT\_LNF     & 0.0737 & 0.9540 & 0.2754 & 1.2502 & 0.3609 & -0.3382 \\
SERT\_H1      & 0.0760 & 1.0070 & 0.2907 & 1.4348 & 0.4142 & -0.2998 \\
SERT\_H2      & 0.0670 & 0.8456 & 0.2441 & 1.0240 & 0.2956 & -0.3715 \\
SERT\_H3      & 0.0714 & 0.9339 & 0.2696 & 1.2762 & 0.3684 & -0.4871 \\
SERT\_H4      & 0.0780 & 0.9724 & 0.2807 & 1.1854 & 0.3422 & -0.3211 \\
SERT\_H6      & 0.0751 & 0.8979 & 0.2592 & 1.2336 & 0.3561 & -0.3170 \\
SERT\_H7      & 0.0743 & 0.8896 & 0.2568 & 1.1528 & 0.3328 & -0.3096 \\
En\_Trans\_H1 & 0.0691 & 0.8664 & 0.2501 & 0.9526 & 0.2750 & -0.4391 \\
En\_Trans\_H2 & 0.0678 & 0.8719 & 0.2517 & 1.0902 & 0.3147 & -0.3838 \\
En\_Trans\_H4 & 0.0666 & 0.7514 & 0.2169 & 0.8622 & 0.2489 & -0.5050 \\
\midrule
\multicolumn{7}{c}{\textbf{Period Including COVID-19 and One-Year After (2212)}} \\
\midrule
\textbf{Model} & AR & Ann.SR & SR & Ann.SO & SO & MDD \\
\midrule
SERT\_LNF     & 0.0528 & 0.6277 & 0.1812 & 0.7773 & 0.2244 & -0.3382 \\
SERT\_H1      & 0.0670 & 0.8196 & 0.2366 & 1.1802 & 0.3407 & -0.2998 \\
SERT\_H2      & 0.0419 & 0.4749 & 0.1371 & 0.5466 & 0.1578 & -0.4023 \\
SERT\_H3      & 0.0585 & 0.6876 & 0.1985 & 0.9111 & 0.2630 & -0.4871 \\
SERT\_H4      & 0.0519 & 0.5879 & 0.1697 & 0.6873 & 0.1984 & -0.3584 \\
SERT\_H6      & 0.0609 & 0.6772 & 0.1955 & 0.9457 & 0.2730 & -0.3170 \\
SERT\_H7      & 0.0636 & 0.7219 & 0.2084 & 0.9893 & 0.2856 & -0.3096 \\
En\_Trans\_H1 & 0.0566 & 0.6589 & 0.1902 & 0.7756 & 0.2239 & -0.4391 \\
En\_Trans\_H2 & 0.0551 & 0.6606 & 0.1907 & 0.8525 & 0.2461 & -0.3838 \\
En\_Trans\_H4 & 0.0537 & 0.5816 & 0.1679 & 0.7077 & 0.2043 & -0.5050 \\
\bottomrule
\end{tabularx}
\caption[The investing performance for the value-weighted portfolios of encoder-only Transformer models considering the static transaction cost.]{The investing performance for the value-weighted portfolios of encoder-only Transformer models considering the static transaction cost. MDD is maximum drawdown, `SR (Sharpe)' means Sharpe ratio, `Ann.SR' is the annualized Sharpe ratio, `SO (Sortino)' means Sortino ratio and `Ann.SO' is the annualized Sortino ratio. Negative values are denoted by a minus sign. The values in bold font indicate values that outperform the buy-and-hold benchmark.}
\label{tab:invest_SERT_vwperform5322_ch3}
\end{table}

A robust examination of the dynamic transaction cost effect is also conducted, which is presented in Table ~\ref{tab:invest_SERT_ewperform5321_ch3_turnover} (equal-weighted) and Table ~\ref{tab:invest_SERT_vwperform5322_ch3_turnover} (value-weighted), respectively. In equal-weighted portfolios, the static transaction cost does not show significant over-penalization based on the annualized returns, but the dynamic transaction cost, which counts on the turnover rate with 20bps, highly reduces the maximum drawdown, Sharpe ratio and Sortino ratio. This indicates that high turnover rates erode the profitability of these models. The encoder-only Transformer models in the equal-weighted method, high turnover rate impact higher than the heavy static transaction cost penalty. Model SERT\_H4 outperforms alternatives in profitability in both transaction cost methods. In value-weighted portfolios, the static transaction cost shows significant over-penalization by penalizing the annualized returns and decreasing the maximum drawdown. However, models, which are less sensitive to forecasting signals (lower turnover rates) derives higher Sharpe ratio and Sortino ratio, especially during the high fluctuation phrase. \\ 

\begin{table}[htbp]
\centering
\small 
\renewcommand{\arraystretch}{1.2}

\begin{tabularx}{\textwidth}{l *{7}{X}}
\toprule
\multicolumn{8}{c}{\textbf{Pre--COVID-19 Period (1911)}} \\
\midrule
\textbf{Model} & \textbf{AR} & \textbf{Ann.SR} & \textbf{SR} & \textbf{Ann.SO} & \textbf{SO} & \textbf{MDD} & \textbf{Turnover} \\
\midrule
SERT\_LNF    & 0.0782 & 0.7278 & 0.2101 & 1.0021 & 0.2893 & -0.1399 & 0.7301 \\
SERT\_H1     & 0.0717 & 0.6578 & 0.1899 & 0.9092 & 0.2625 & -0.1625 & 0.8792 \\
SERT\_H2     & 0.0615 & 0.5541 & 0.1600 & 0.6645 & 0.1918 & -0.1939 & 0.8899 \\
SERT\_H3     & 0.0667 & 0.6049 & 0.1746 & 0.8163 & 0.2357 & -0.1580 & 0.9013 \\
SERT\_H4     & 0.0856 & 0.7588 & 0.2190 & 0.8953 & 0.2585 & -0.1664 & 0.8212 \\
SERT\_H6     & 0.0690 & 0.5940 & 0.1715 & 0.7682 & 0.2218 & -0.1812 & 0.8632 \\
SERT\_H7     & 0.0679 & 0.5969 & 0.1723 & 0.7463 & 0.2154 & -0.1825 & 0.8372 \\
En\_Trans\_H1 & 0.0786 & 0.7247 & 0.2092 & 1.0519 & 0.3037 & -0.1483 & 0.6129 \\
En\_Trans\_H2 & 0.0785 & 0.6765 & 0.1953 & 0.9550 & 0.2757 & -0.1654 & 0.7037 \\
En\_Trans\_H4 & 0.0789 & 0.6810 & 0.1966 & 0.8479 & 0.2448 & -0.1732 & 0.6734 \\
BHE          & 0.1141 & 0.8892 & 0.2567 & 1.2303 & 0.3552 & -0.1483 & 0.0120 \\
\midrule

\multicolumn{8}{c}{\textbf{COVID--19--Inclusive Period (2112)}} \\
\midrule
SERT\_LNF    & 0.1088 & 0.7804 & 0.2253 & 1.0253 & 0.2960 & -0.2921 & 0.6748 \\
SERT\_H1     & 0.1049 & 0.7932 & 0.2290 & 1.1241 & 0.3245 & -0.2616 & 0.7834 \\
SERT\_H2     & 0.0957 & 0.7306 & 0.2109 & 0.9769 & 0.2820 & -0.2914 & 0.7977 \\
SERT\_H3     & 0.0997 & 0.7693 & 0.2221 & 1.1181 & 0.3228 & -0.2497 & 0.7925 \\
SERT\_H4     & 0.1182 & 0.8876 & 0.2562 & 1.1803 & 0.3407 & -0.2543 & 0.7371 \\
SERT\_H6     & 0.0982 & 0.7099 & 0.2049 & 0.9173 & 0.2648 & -0.2842 & 0.7835 \\
SERT\_H7     & 0.1010 & 0.7522 & 0.2171 & 1.0183 & 0.2940 & -0.2841 & 0.7910 \\
En\_Trans\_H1 & 0.1089 & 0.7456 & 0.2152 & 0.9017 & 0.2603 & -0.3263 & 0.5202 \\
En\_Trans\_H2 & 0.1095 & 0.7162 & 0.2068 & 0.8819 & 0.2546 & -0.3458 & 0.6082 \\
En\_Trans\_H4 & 0.1045 & 0.7089 & 0.2047 & 0.8377 & 0.2418 & -0.3358 & 0.5893 \\
BHE          & 0.1353 & 0.8512 & 0.2457 & 1.0181 & 0.2939 & -0.2890 & 0.0093 \\
\midrule

\multicolumn{8}{c}{\textbf{Period Including COVID--19 and One-Year After (2212)}} \\
\midrule
SERT\_LNF    & 0.0840 & 0.5875 & 0.1696 & 0.8044 & 0.2322 & -0.2921 & 0.6905 \\
SERT\_H1     & 0.0865 & 0.6176 & 0.1783 & 0.8795 & 0.2539 & -0.2616 & 0.7367 \\
SERT\_H2     & 0.0687 & 0.5026 & 0.1451 & 0.6659 & 0.1922 & -0.2914 & 0.8296 \\
SERT\_H3     & 0.0818 & 0.5905 & 0.1705 & 0.8161 & 0.2487 & -0.2497 & 0.7395 \\
SERT\_H4     & 0.0892 & 0.6410 & 0.1850 & 0.8450 & 0.2439 & -0.2543 & 0.7708 \\
SERT\_H6     & 0.0778 & 0.5393 & 0.1557 & 0.7258 & 0.2095 & -0.2842 & 0.7432 \\
SERT\_H7     & 0.0788 & 0.5574 & 0.1609 & 0.7762 & 0.2241 & -0.2841 & 0.7460 \\
En\_Trans\_H1 & 0.0898 & 0.5941 & 0.1715 & 0.7540 & 0.2177 & -0.3263 & 0.4956 \\
En\_Trans\_H2 & 0.0895 & 0.5725 & 0.1653 & 0.7365 & 0.2126 & -0.3458 & 0.5835 \\
En\_Trans\_H4 & 0.0865 & 0.5700 & 0.1645 & 0.7214 & 0.2082 & -0.3358 & 0.5657 \\
BHE          & 0.1143 & 0.7017 & 0.2026 & 0.8883 & 0.2564 & -0.2890 & 0.0083 \\
\bottomrule
\end{tabularx}
\caption[The backtesting results for the equal-weighted portfolios of the proposed SERT and the encoder-only Transformer models considering the dynamic transaction cost.]{The backtesting results for the equal-weighted portfolios of the proposed SERT and the encoder-only Transformer considering the dynamic transaction cost. 'MDD' is the maximum drawdown, 'SR' means Sharpe ratio, 'SO' means Sortino ratio. 'BHE' represents the buy-and-hold benchmark strategy under the equal-weighted method. 'Ann' is the notation for 'annulized'.}
\label{tab:invest_SERT_ewperform5321_ch3_turnover}
\end{table}

\clearpage
\begin{table}[htbp]
\centering
\small
\renewcommand{\arraystretch}{1.2}

\begin{tabularx}{\textwidth}{l *{7}{X}}
\toprule
\multicolumn{8}{c}{\textbf{Pre--COVID-19 Period (1911)}} \\
\midrule
\textbf{Model} & \textbf{AR} & \textbf{Ann.SR} & \textbf{SR} & \textbf{Ann.SO} & \textbf{SO} & \textbf{MDD} & \textbf{Turnover} \\
\midrule
SERT\_LNF    & 0.1145 & 1.1008 & 0.3178 & 1.5609 & 0.4506 & -0.1297 & 0.6941 \\
SERT\_H1     & 0.0938 & 0.8959 & 0.2586 & 1.1700 & 0.3377 & -0.1481 & 0.8402 \\
SERT\_H2     & 0.1023 & 0.9654 & 0.2787 & 1.1489 & 0.3317 & -0.1182 & 0.8240 \\
SERT\_H3     & 0.0824 & 0.7830 & 0.2260 & 0.9822 & 0.2835 & -0.1460 & 0.8503 \\
SERT\_H4     & 0.1159 & 1.0813 & 0.3122 & 1.2697 & 0.3665 & -0.1305 & 0.7783 \\
SERT\_H6     & 0.0932 & 0.8528 & 0.2462 & 1.0662 & 0.3078 & -0.1523 & 0.8197 \\
SERT\_H7     & 0.0871 & 0.7986 & 0.2305 & 1.0021 & 0.2893 & -0.1466 & 0.7932 \\
En\_Trans\_H1 & 0.1085 & 1.0516 & 0.3036 & 1.5570 & 0.4495 & -0.0983 & 0.5974 \\
En\_Trans\_H2 & 0.1188 & 1.0585 & 0.3056 & 1.5451 & 0.4460 & -0.1169 & 0.6541 \\
En\_Trans\_H4 & 0.1190 & 1.0620 & 0.3066 & 1.2624 & 0.3644 & -0.1277 & 0.6486 \\
BHV          & 0.1388 & 1.1795 & 0.3405 & 1.6054 & 0.4634 & -0.1297 & 0.0120 \\
\midrule

\multicolumn{8}{c}{\textbf{COVID--19--Inclusive Period (2112)}} \\
\midrule
SERT\_LNF    & 0.1406 & 1.0701 & 0.3089 & 1.5182 & 0.4383 & -0.1914 & 0.6393 \\
SERT\_H1     & 0.1149 & 0.9101 & 0.2627 & 1.2083 & 0.3488 & -0.1920 & 0.7529 \\
SERT\_H2     & 0.1344 & 1.0828 & 0.3126 & 1.4585 & 0.4210 & -0.1413 & 0.7468 \\
SERT\_H3     & 0.1147 & 0.9394 & 0.2712 & 1.3114 & 0.3786 & -0.1485 & 0.7590 \\
SERT\_H4     & 0.1430 & 1.1621 & 0.3355 & 1.5470 & 0.4466 & -0.1373 & 0.7028 \\
SERT\_H6     & 0.1094 & 0.7819 & 0.2257 & 0.8566 & 0.2473 & -0.2653 & 0.7481 \\
SERT\_H7     & 0.1149 & 0.8939 & 0.2580 & 1.1837 & 0.3417 & -0.2030 & 0.7530 \\
En\_Trans\_H1 & 0.1335 & 0.9933 & 0.2868 & 1.2334 & 0.3560 & -0.2300 & 0.5204 \\
En\_Trans\_H2 & 0.1397 & 1.0041 & 0.2899 & 1.3237 & 0.3821 & -0.2206 & 0.5801 \\
En\_Trans\_H4 & 0.1368 & 0.9880 & 0.2852 & 1.1886 & 0.3431 & -0.2252 & 0.5694 \\
BHV          & 0.1540 & 1.0688 & 0.3085 & 1.3057 & 0.3769 & -0.2352 & 0.0093 \\
\midrule

\multicolumn{8}{c}{\textbf{Period Including COVID--19 and One-Year After (2212)}} \\
\midrule
SERT\_LNF    & 0.1102 & 0.8095 & 0.2337 & 1.1912 & 0.3439 & -0.1963 & 0.6597 \\
SERT\_H1     & 0.0969 & 0.7180 & 0.2073 & 0.9830 & 0.2838 & -0.1920 & 0.7052 \\
SERT\_H2     & 0.0981 & 0.7419 & 0.2142 & 0.9537 & 0.2753 & -0.2628 & 0.7871 \\
SERT\_H3     & 0.0931 & 0.7081 & 0.2044 & 0.9985 & 0.2882 & -0.1801 & 0.7018 \\
SERT\_H4     & 0.1073 & 0.8154 & 0.2354 & 1.0600 & 0.3060 & -0.2485 & 0.7390 \\
SERT\_H6     & 0.0848 & 0.5880 & 0.1698 & 0.7022 & 0.2027 & -0.2653 & 0.7063 \\
SERT\_H7     & 0.0927 & 0.6808 & 0.1965 & 0.9623 & 0.2778 & -0.2030 & 0.7089 \\
En\_Trans\_H1 & 0.1094 & 0.7681 & 0.2217 & 1.0088 & 0.2912 & -0.2300 & 0.4994 \\
En\_Trans\_H2 & 0.1118 & 0.7591 & 0.2191 & 1.0262 & 0.2963 & -0.2206 & 0.5654 \\
En\_Trans\_H4 & 0.1122 & 0.7826 & 0.2259 & 1.0042 & 0.2899 & -0.2252 & 0.5462 \\
BHV          & 0.1308 & 0.8654 & 0.2498 & 1.1270 & 0.3253 & -0.2352 & 0.0083 \\
\bottomrule
\end{tabularx}
\caption[The backtesting results for the value-weighted portfolios of the proposed SERT and the encoder-only Transformer models considering the dynamic transaction cost.]{The backtesting results for the value-weighted portfolios of the proposed SERT and the encoder-only Transformer models considering the dynamic transaction cost. MDD is the maximum drawdown, 'SR' means Sharpe ratio, 'SO' means Sortino ratio. 'BHV' represents the buy-and-hold benchmark strategy under the value-weighted method. 'Ann' is the notation for 'annulized'.}
\label{tab:invest_SERT_vwperform5322_ch3_turnover}
\end{table}

\subsubsection{Comparative analysis}\label{subsec:533Comparative analysis}
For comparing the Transformer models and encoder-only Transformer models, we extract the best model in each group and evaluate them in the equal-weighted and value-weighted portfolios in the static transaction cost scenario. Table~\ref{tab:best_ew5321_ch3} shows the backtesting performance of the equal-weighted portfolio between the best Transformer models and the best encoder-only Transformer models in each group, while Table~\ref{tab:best_vw5322_ch3} shows the backtesting performance in the value-weighted portfolio. Interestingly, from the periodical horizontal comparison, the equal-weighted portfolio shows insignificant differences between the best pre-trained Transformer model (P\_Trans\_H6) and the best SERT model (SERT\_H4). By only considering the absolute capital gain (Annualized return), SERT performs better than the pre-trained Transformer model, but when jointly considering the downside risks (Sortino ratio), the pre-trained Transformer model performs better, especially during the extreme market periods (2112 and 2212). Moreover, LNF SERT (SERT\_LNF) performs better than the LNF pre-trained Transformer (P\_Trans\_LNF) in all periods, and there is no significant difference between the standard encoder-only Transformer (En\_Trans\_H1) and the standard Transformer model (C\_Trans\_H1). \\

\begin{figure}[htbp!]
\centering
\includegraphics[width=0.95\columnwidth, height=0.8\textheight, keepaspectratio]{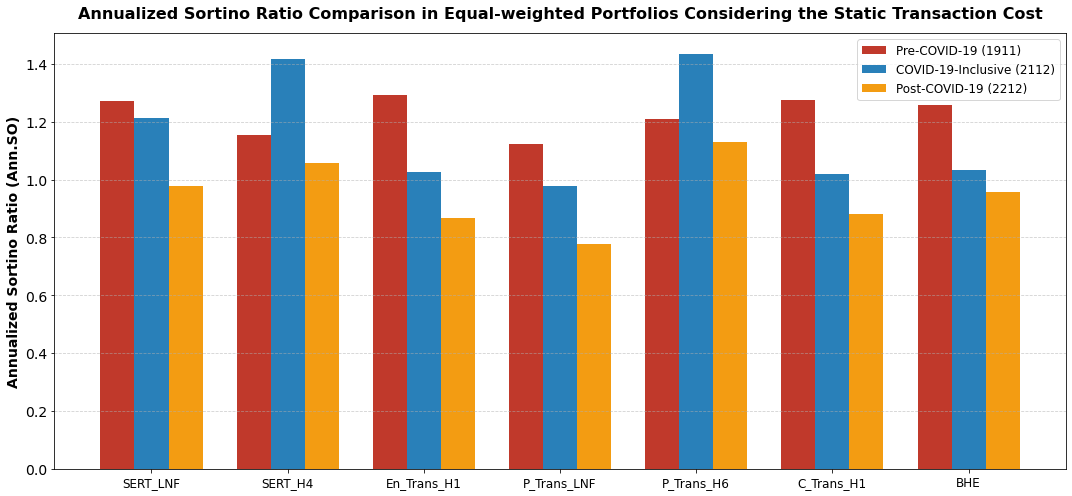}
\caption[Comparison of Sortino ratios for equal-weighted portfolios considering the static transaction cost.]{Comparison of Sortino ratios for equal-weighted portfolios considering the static transaction cost. The Y-axis presents the value of Sortino Ratio, and the X-axis presents the models. The red, blue and orange bars indicate the Sortino Ratios of each model during the period of ‘1911’, ‘2112’ and ‘2212’ respectively. ‘BHE’ is the notation of the buy-and-hold strategy for the equal-weighted portfolio.}
\label{fig:sortino_ew5321_bar_ch3}
\end{figure}

\begin{table}[htbp!]
\centering
\small
\begin{tabular}{lcccccc}
\hline
\multicolumn{7}{c}{\textbf{Pre–COVID-19 Period (1911)}} \\
\hline
\textbf{Model} & \textbf{MDD} & \textbf{AR} & \textbf{sharpe (SR)} & \textbf{Ann.SR} & \textbf{sortino (SO)} & \textbf{Ann.SO} \\
\hline
SERT\_LNF     & -0.3566 & 0.0968 & 0.2622 & 0.9082 & 0.3673 & 1.2725 \\
SERT\_H4      & -0.2956 & 0.1066 & 0.2737 & 0.9481 & 0.3338 & 1.1561 \\
En\_Trans\_H1 & -0.2941 & 0.0940 & 0.2533 & 0.8773 & 0.3732 & 1.2928 \\
P\_Trans\_LNF & -0.2949 & 0.0916 & 0.2431 & 0.8421 & 0.3239 & 1.1222 \\
P\_Trans\_H6  & -0.3511 & 0.0979 & 0.2555 & 0.8850 & 0.3493 & 1.2100 \\
C\_Trans\_H1  & -0.2349 & 0.1005 & 0.2646 & 0.9166 & 0.3683 & 1.2759 \\
BHE & -0.3701 & 0.1139 & 0.2588 & 0.8965 & 0.3631 & 1.2578\\
\hline
\multicolumn{7}{c}{\textbf{COVID-19–Inclusive Period (2112)}} \\
\hline
\textbf{Model} & \textbf{MDD} & \textbf{AR} & \textbf{sharpe (SR)} & \textbf{Ann.SR} & \textbf{sortino (SO)} & \textbf{Ann.SO} \\
\hline
SERT\_LNF     & -0.5154 & 0.1263 & 0.2622 & 0.9082 & 0.3500 & 1.2124 \\
SERT\_H4      & -0.4119 & 0.1374 & 0.2988 & 1.0350 & 0.4095 & 1.4185 \\
En\_Trans\_H1 & -0.6018 & 0.1222 & 0.2428 & 0.8412 & 0.2963 & 1.0263 \\
P\_Trans\_LNF & -0.5589 & 0.1124 & 0.2394 & 0.8294 & 0.2826 & 0.9788 \\
P\_Trans\_H6  & -0.4218 & 0.1284 & 0.2840 & 0.9837 & 0.4140 & 1.4341 \\
C\_Trans\_H1  & -0.5789 & 0.1271 & 0.2455 & 0.8505 & 0.2948 & 1.0211 \\
BHE & -0.5310 & 0.1349 & 0.2472 & 0.8563 & 0.2988 & 1.0351\\
\hline
\multicolumn{7}{c}{\textbf{Period Including COVID-19 and One-Year After (2212)}} \\
\hline
\textbf{Model} & \textbf{MDD} & \textbf{AR} & \textbf{sharpe (SR)} & \textbf{Ann.SR} & \textbf{sortino (SO)} & \textbf{Ann.SO} \\
\hline
SERT\_LNF     & -0.5154 & 0.1016 & 0.2041 & 0.7072 & 0.2827 & 0.9794 \\
SERT\_H4      & -0.4119 & 0.1089 & 0.2249 & 0.7791 & 0.3049 & 1.0563 \\
En\_Trans\_H1 & -0.6018 & 0.1024 & 0.1955 & 0.6771 & 0.2502 & 0.8669 \\
P\_Trans\_LNF & -0.5589 & 0.0886 & 0.1809 & 0.6268 & 0.2241 & 0.7765 \\
P\_Trans\_H6  & -0.4218 & 0.1070 & 0.2204 & 0.7636 & 0.3264 & 1.1308 \\
C\_Trans\_H1  & -0.5789 & 0.1066 & 0.1998 & 0.6920 & 0.2547 & 0.8822 \\
BHE & -0.5308 & 0.1141 & 0.2148 &  0.7441 & 0.2760 & 0.9561 \\
\hline
\end{tabular}
\caption[Back-testing performance of the equal-weighted portfolio between the best models in each group considering the static transaction cost.]{Back-testing performance of the equal-weighted portfolio between the best models in each group considering the static transaction cost. The notation of 'Ann' means annualized. SR and SO are denoted as Sharpe Ratio and Sortino Ratio, respectively.}
\label{tab:best_ew5321_ch3}
\end{table}
More specifically, the comparison of Sortino ratio for each model in different periods gives a clear picture of the trading strategy-wise model performance, which is shown in Figure~\ref{fig:sortino_ew5321_bar_ch3}. From Figure~\ref{fig:sortino_ew5321_bar_ch3}, the best proposed models and their benchmarks show no advantage or insignificant advantage over the buy-and-hold strategy for the equal-weighted portfolio (BHE) during the period of ‘1911’ (blue bars). However, the best proposed models (SERT\_H4 and P\_Trans\_H6) illustrate the great advantage compared with the BHE during the period of ‘2112’. Their Sortino ratios are approximately 47\% higher than BHE, followed by the LNF SERT (SERT\_LNF), 17\% higher than BHE. And during the period of ‘2212’, the best proposed models still demonstrate superiority to the BHE benchmark. This supports the conclusion that the best proposed models achieve significantly higher profitability by considering downside risks during extreme market fluctuations. \\

\begin{sidewaysfigure}[htbp!]
\centering
\includegraphics[width=0.95\columnwidth, height=0.8\textheight, keepaspectratio]{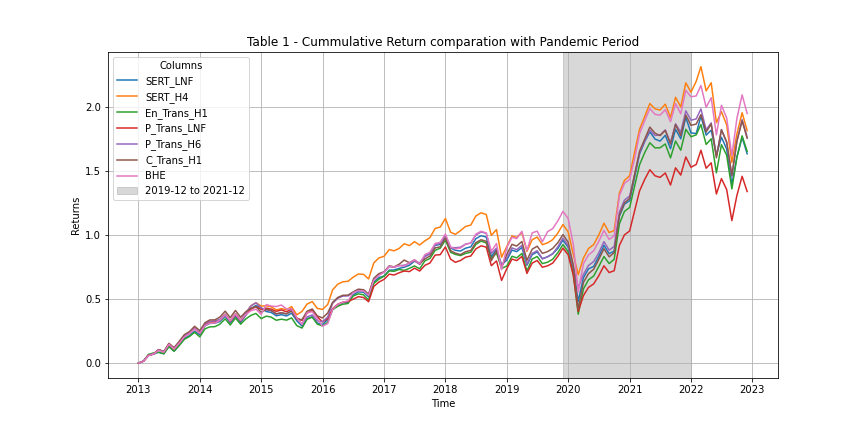}
\caption[Sign equal-weighted accumulative return plots considering the static transaction cost.]{Sign equal-weighted accumulative return plots considering the static transaction cost. The area covered by a grey shadow indicates the pandemic period. ‘BHE’ is the buy-and-hold strategy for the equal-weighted portfolio.}
\label{fig:cum_ew5322_ch3}
\end{sidewaysfigure}

Figure~\ref{fig:cum_ew5322_ch3} is the cumulative return plot of selected models under the equal-weighted method. It illustrates that Model SERT\_H4 outperforms the BHE strategy and achieves the highest cumulative returns in most periods. The best proposed pre-trained Transformer model, P\_Trans\_H6, is right below the BHE strategy and results into the second-best model for better cumulative returns. On the other hand, the LNF pre-trained Transformer model (P\_Trans\_LNF) performs noticeably worse than alternatives during the extreme market fluctuation period, which is followed by the single-head standard encoder-only Transformer model (En\_Trans\_H1). The LNF SERT model (SERT\_LNF) performs slightly higher than the Model En\_Trans\_H1 but is still worse than the average. This concludes that the LNF models are not ideal in the factor investing context of the equal-weighted method in the static transaction cost scenario. Moreover, amid the incredible difference between SERT\_H4 and En\_Trans\_H1, it presents the solid evidence of how the pre-trained method in the proposed SERT can improve investment performance. The similar phenomenon happens between the best standard Transformer model and the pre-trained Transformer models as well, which confirms the advantage of pre-training structure in Transformers for economic models. \\

Similar findings are shown in the value-weighted portfolios in the static transaction cost method, although the value-weighted method highly reduces the absolute capital gain of all models. Nevertheless, the difference between the two models is significant for both the absolute capital gain and the gain under the downside risk during the pandemic period ‘2112’. Concretely, the performance of the pre-trained Transformer model (P\_Trans\_H2) is better than that of the SERT model (SERT\_H1). Model P\_Trans\_H2 has a Sortino ratio of 0.46, which is 0.0458 higher than that of SERT\_H1. Furthermore, the LNF SERT outperforms the LNF pre-trained Transformer model, while the standard encoder-only Transformer underperforms the standard Transformer model. \\
\begin{table}[htbp!]
\centering
\small
\begin{tabular}{lcccccc}
\hline
\multicolumn{7}{c}{\textbf{Pre–COVID-19 Period (1911)}} \\
\hline
\textbf{Model} & \textbf{MDD} & \textbf{AR} & \textbf{sharpe (SR)} & \textbf{Ann.SR} & \textbf{sortino (SO)} & \textbf{Ann.SO} \\
\hline
SERT\_LNF     & -0.2475 & 0.0546 & 0.2631 & 0.9115 & 0.2918 & 1.0108 \\
SERT\_H1      & -0.2998 & 0.0540 & 0.2345 & 0.8124 & 0.2755 & 0.9542 \\
En\_Trans\_H2 & -0.2395 & 0.0605 & 0.2594 & 0.8985 & 0.3238 & 1.1216 \\
P\_Trans\_LNF & -0.3061 & 0.0532 & 0.2304 & 0.7980 & 0.2651 & 0.9185 \\
P\_Trans\_H2  & -0.3260 & 0.0535 & 0.2413 & 0.8359 & 0.2923 & 1.0126 \\
C\_Trans\_H4  & -0.3061 & 0.0566 & 0.2361 & 0.8180 & 0.2954 & 1.0235 \\
BHV & -0.2963 & 0.1197 & 0.2956 & 1.0240 & 0.4243 & 1.4698\\
\hline
\multicolumn{7}{c}{\textbf{COVID-19–Inclusive Period (2112)}} \\
\hline
\textbf{Model} & \textbf{MDD} & \textbf{AR} & \textbf{sharpe (SR)} & \textbf{Ann.SR} & \textbf{sortino (SO)} & \textbf{Ann.SO} \\
\hline
SERT\_LNF     & -0.3382 & 0.0737 & 0.2754 & 0.9540 & 0.3609 & 1.2503 \\
SERT\_H1      & -0.2998 & 0.0760 & 0.2907 & 1.0069 & 0.4142 & 1.4347 \\
En\_Trans\_H2 & -0.3838 & 0.0678 & 0.2517 & 0.8721 & 0.3147 & 1.0902 \\
P\_Trans\_LNF & -0.3415 & 0.0648 & 0.2531 & 0.8767 & 0.3179 & 1.1013 \\
P\_Trans\_H2  & -0.3260 & 0.0801 & 0.3113 & 1.0785 & 0.4600 & 1.5934 \\
C\_Trans\_H4  & -0.4205 & 0.0745 & 0.2509 & 0.8693 & 0.3210 & 1.1121 \\
BHV & -0.4513 & 0.1337 & 0.2673 & 0.9260 & 0.3263 & 1.1303\\
\hline
\multicolumn{7}{c}{\textbf{Period Including COVID-19 and One-Year After (2212)}} \\
\hline
\textbf{Model} & \textbf{MDD} & \textbf{AR} & \textbf{sharpe (SR)} & \textbf{Ann.SR} & \textbf{sortino (SO)} & \textbf{Ann.SO} \\
\hline
SERT\_LNF     & -0.3382 & 0.0528 & 0.1812 & 0.6277 & 0.2244 & 0.7773 \\
SERT\_H1      & -0.2998 & 0.0670 & 0.2366 & 0.8197 & 0.3407 & 1.1801 \\
En\_Trans\_H2 & -0.3838 & 0.0551 & 0.1907 & 0.6607 & 0.2461 & 0.8525 \\
P\_Trans\_LNF & -0.3415 & 0.0495 & 0.1835 & 0.6357 & 0.2387 & 0.8269 \\
P\_Trans\_H2  & -0.3260 & 0.0672 & 0.2436 & 0.8440 & 0.3553 & 1.2307 \\
C\_Trans\_H4  & -0.4205 & 0.0653 & 0.2076 & 0.7191 & 0.2797 & 0.9689 \\
BHV & -0.4513 & 0.1156 & 0.2311 & 0.8006 & 0.3031 & 1.0500\\
\hline
\end{tabular}
\caption[Back-testing performance of the value-weighted portfolio between the best models in each group considering the static transaction cost.]{Back-testing performance of the value-weighted portfolio between the best models in each group considering the static transaction cost. The notation of `Ann' means annualized. SR and SO are denoted as Sharpe Ratio and Sortino Ratio, respectively. AR means annualized return.}
\label{tab:best_vw5322_ch3}
\end{table}

\begin{figure}[htbp!]
\centering
\includegraphics[width=0.95\columnwidth, height=0.8\textheight, keepaspectratio]{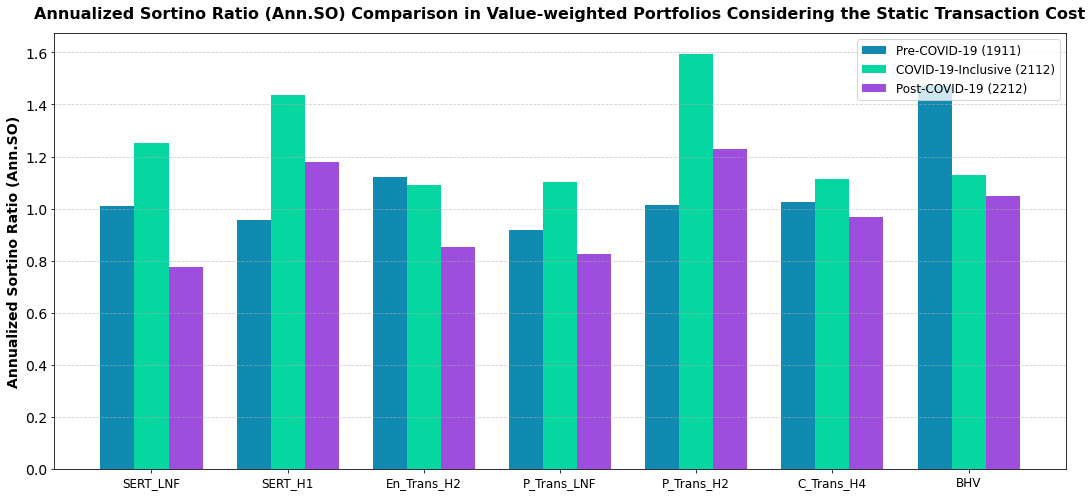}
\caption[Value-weighted accumulative return plots considering the static transaction cost.]{Value-weighted accumulative return plots considering the static transaction cost. ‘BHV’ is the buy-and-hold strategy for the value-weighted portfolio.}
\label{fig:sortino_vw5323_bar_ch3}
\end{figure}

Figure~\ref{fig:sortino_vw5323_bar_ch3} shows the comparison of Sortino ratios for value-weighted (VW) portfolios during the three periods. The Sortino ratios of value-weighted portfolios agree with the findings in equal-weighted portfolios that the best proposed models (SERT\_H1 and P\_Trans\_H2) dramatically outperform the buy-and-hold (BHV) benchmark during the period of ‘2112’ (28\% and 44\% higher than BHV, respectively), followed by LNF SERT (SERT\_LNF) and ‘2212’ (13\% and 20\% higher than BHV, respectively). Whereas the proposed models and their benchmarks fail to demonstrate superiority over the BHV strategy during the period of ‘1911’. It is also interesting to find that the LNF configuration performs better on the SERT model compared with the pre-trained Transformer model in both equal-weighted and value-weighted methods for capturing the extreme market fluctuations.\\

\begin{sidewaysfigure}[htbp!]
\centering
\includegraphics[width=0.95\columnwidth, height=0.8\textheight, keepaspectratio]{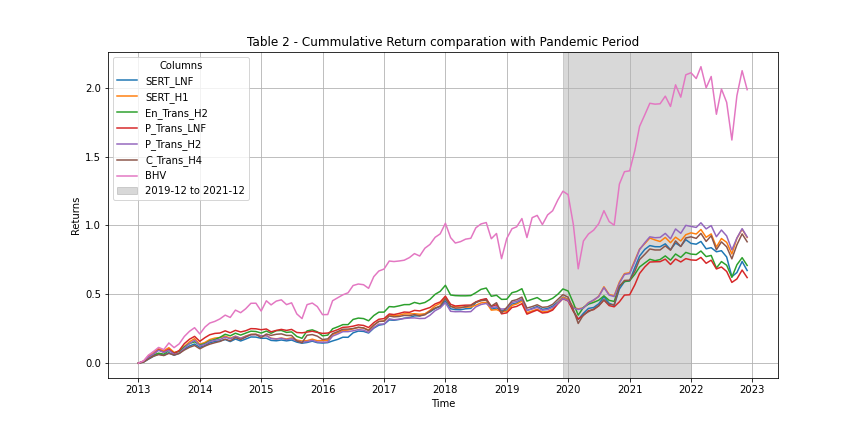}
\caption[Sign value-weighted accumulative return plots considering the static transaction cost.]{Sign value-weighted accumulative return plots considering the static transaction cost. The area covered by a grey shadow indicates the pandemic period. ‘BHE’ is the buy-and-hold strategy for the equal-weighted portfolio.}
\label{fig:cum_vw5322_ch3}
\end{sidewaysfigure}

Figure~\ref{fig:cum_vw5322_ch3} shows the cumulative return plot of the best models in value-weighted portfolios in the static transaction cost method. It shows all models have cumulative returns considerably lower (almost half lower) than the BHV strategy. This phenomenon may be caused by the weights distribution of the value-weighted method or over-penalization of high static transaction costs (50bps). The weight distribution is based on the market capitalization size of each stock. It means larger weights contribute to the large capital stocks which have lower volatilities. However, from the robust examination of the dynamic transaction cost method, there is a high probability that the difference is caused by over-penalization of static transaction costs. Additionally, under the value-weighted conditions, the standard encoder-only Transformer model (En\_Trans\_H2) has the highest returns, but during the high fluctuation, the proposed models, P\_Trans\_H2 and SERT\_H1, excel all alternative models. P\_Trans\_LNF, En\_Trans\_H2, and SERT\_LNF are still the last three models in this case. \\

\begin{sidewaysfigure}[htbp!]
\centering
\includegraphics[width=0.95\columnwidth, height=0.8\textheight, keepaspectratio]{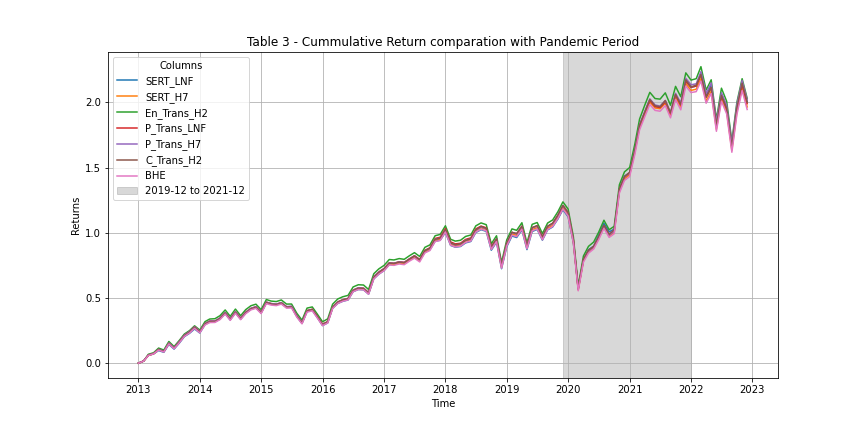}
\caption[Sign equal-weighted accumulative return plots with softmax signal filter considering the static transaction cost.]{Sign equal-weighted accumulative return plots with softmax signal filter considering the static transaction cost. The area covered by a grey shadow indicates the pandemic period. ‘BHE’ is the buy-and-hold strategy for the equal-weighted portfolio.}
\label{fig:cum_ew5322_ch3_softmax}
\end{sidewaysfigure}

\begin{sidewaysfigure}[htbp!]
\centering
\includegraphics[width=0.95\columnwidth, height=0.8\textheight, keepaspectratio]{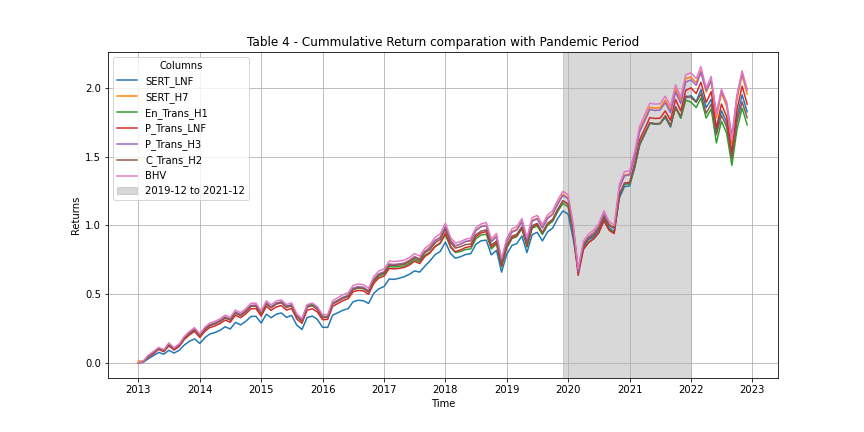}
\caption[Sign value-weighted accumulative return plots with softmax signal filter considering the static transaction cost.]{Sign value-weighted accumulative return plots with softmax signal filter considering the static transaction cost. The area covered by a grey shadow indicates the pandemic period. ‘BHE’ is the buy-and-hold strategy for the equal-weighted portfolio.}
\label{fig:cum_vw5322_ch3_softmax}
\end{sidewaysfigure}

Figure~\ref{fig:cum_ew5322_ch3_softmax} and Figure~\ref{fig:cum_vw5322_ch3_softmax} show the accumulative return plots for the models with softmax filters of the equal-weighted and value-weighted methods, respectively considering the static transaction cost. In the equal-weighted portfolio, after filtering out bottom 50\% signals, the standard encoder-only Transformer model with 2 attention heads (En\_Trans\_H2) outperforms all other models and gains the highest returns, while almost all models perform the BHE insignificantly apart from Model En\_Trans\_H1. Whereas En\_Trans\_H1 works better than BHE before and at the beginning of the pandemic, but not in the middle of the pandemic and one year afterwards. On the other hand, from Figure~\ref{fig:cum_vw5322_ch3_softmax}, it suggests that the softmax filter works more efficiently in the value-weighted circumstance. The softmax filter further excludes the less profitable trades after the value weighting process, significantly improving the accumulated returns of all models. Also, it does not make all models surpass the BHV strategy. However, the worst model varies in different periods, concretely, SERT\_LNF before the pandemic, P\_Trans\_LNF during the pandemic and En\_Trans\_H1 after the pandemic. Figure~\ref{fig:cum_ew5322_ch3_softmax} and Figure~\ref{fig:cum_vw5322_ch3_softmax} demonstrate that the softmax filter tends to eliminate the difference between models and make them perform approaching the BHV. This supports the conclusion drawn from Section~\ref{subsec:5.3.1 Transformer_ch3} and Section~\ref{subsec:532Encoder-only Transformer and SERT models} that the softmax signal filter adjusts weight distribution for value-weighted portfolios.\\

\clearpage

\begin{figure}[htbp!]
\centering
\includegraphics[width=0.95\columnwidth, height=0.8\textheight, keepaspectratio]{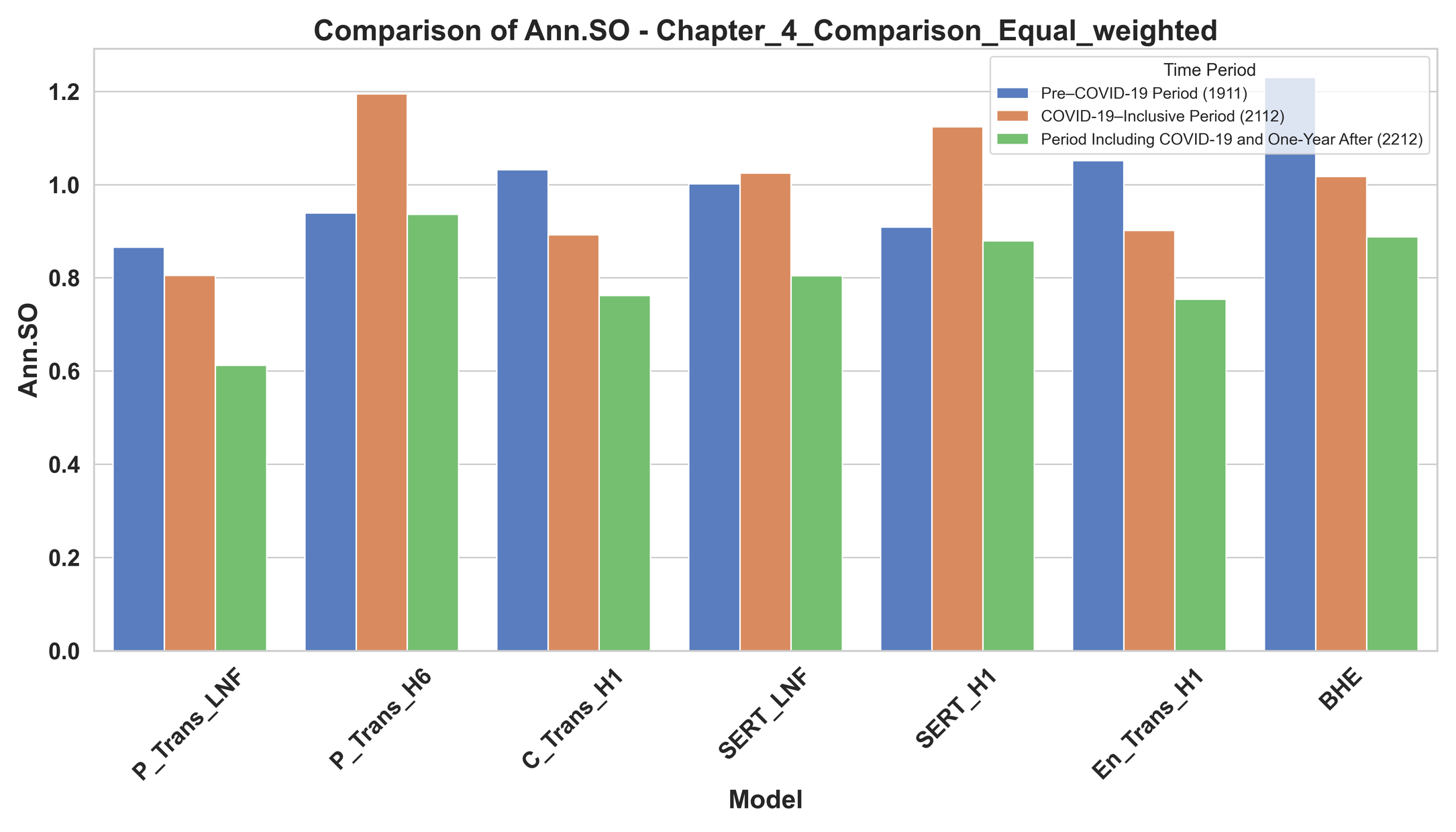}
\caption[Comparison of sortinos for equal-weighted portfolios considering the dynamic transaction cost.]{Comparison of sortinos for equal-weighted portfolios considering the dynamic transaction cost. The Y-axis presents the value of Sortino Ratio, and the X-axis presents the models. The blue, orange and grey bars indicate the Sortino Ratios of each model during the period of ‘1911’, ‘2112’ and ‘2212’ respectively. ‘BHE’ is the notation of the buy-and-hold strategy for an equal-weighted portfolio.}
\label{fig:sortino_ew5321_bar_ch3_turnover}
\end{figure}

\begin{sidewaysfigure}[htbp!]
\centering
\includegraphics[width=0.95\columnwidth, height=0.8\textheight, keepaspectratio]{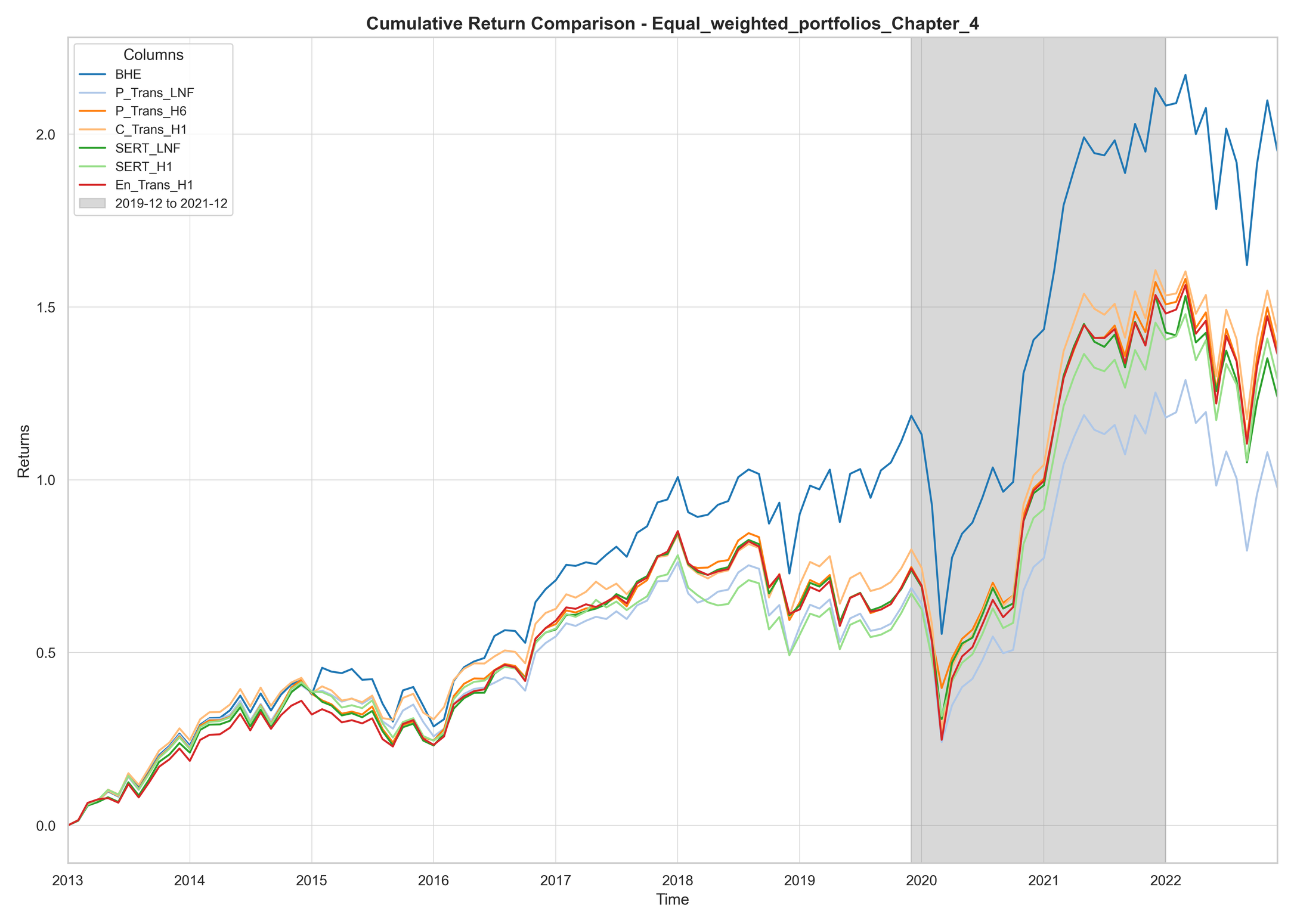}
\caption[Sign equal-weighted accumulative return plots considering dynamic transaction cost.]{Sign equal-weighted accumulative return plots considering dynamic transaction cost. The area covered by a grey shadow indicates the pandemic period. ‘BHE’ is the buy-and-hold strategy for the equal-weighted portfolio.}
\label{fig:cum_ew5322_ch3_turnover}
\end{sidewaysfigure}

\begin{figure}[htbp!]
\centering
\includegraphics[width=0.95\columnwidth, height=0.8\textheight, keepaspectratio]{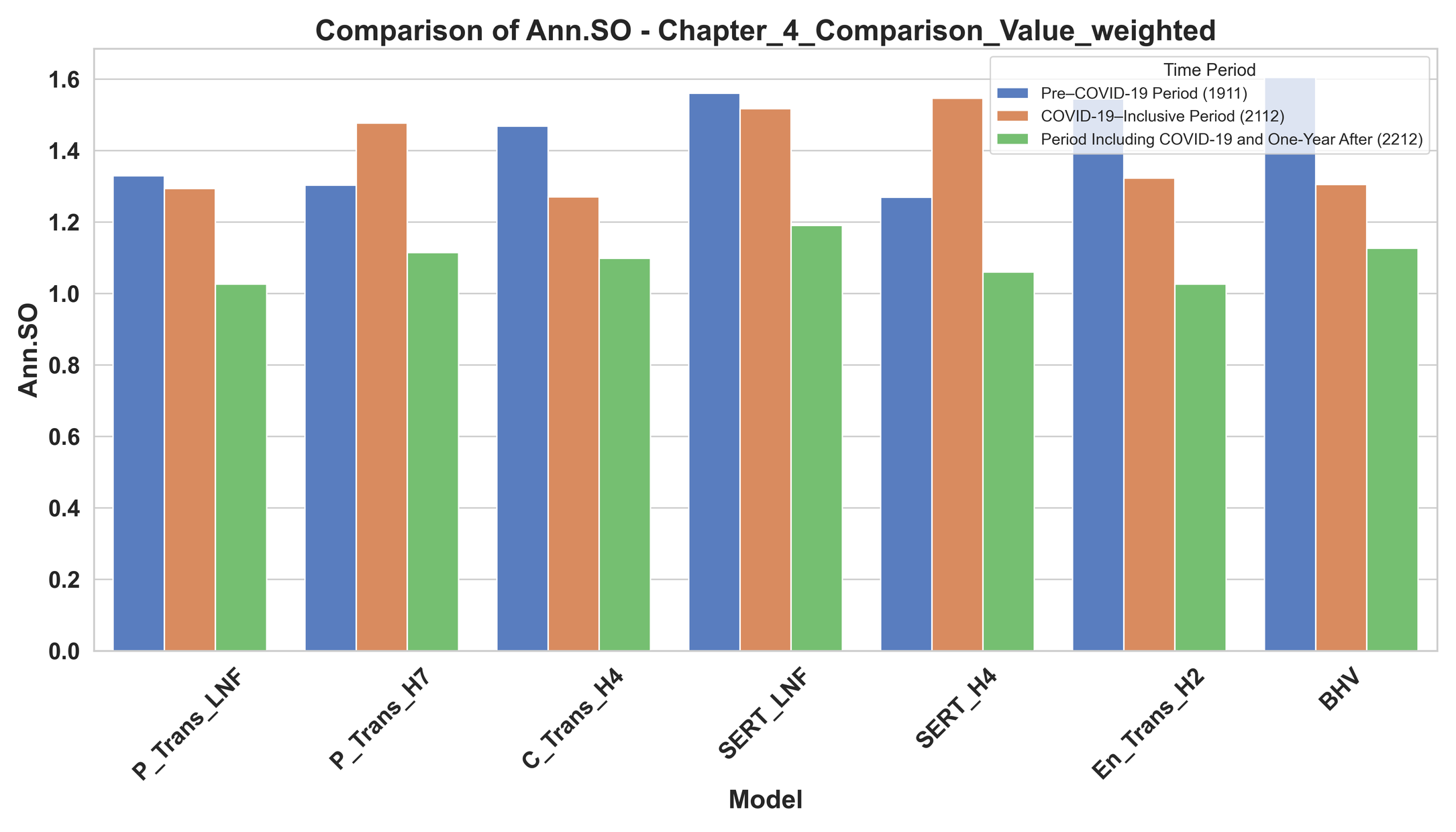}
\caption[Value-weighted accumulative return plots considering the dynamic transaction cost.]{Value-weighted accumulative return plots considering the dynamic transaction cost. ‘BHV’ is the buy-and-hold strategy for the value-weighted portfolio.}
\label{fig:sortino_vw5323_bar_ch3_turnover}
\end{figure}

\begin{sidewaysfigure}[htbp!]
\centering
\includegraphics[width=0.95\columnwidth, height=0.8\textheight, keepaspectratio]{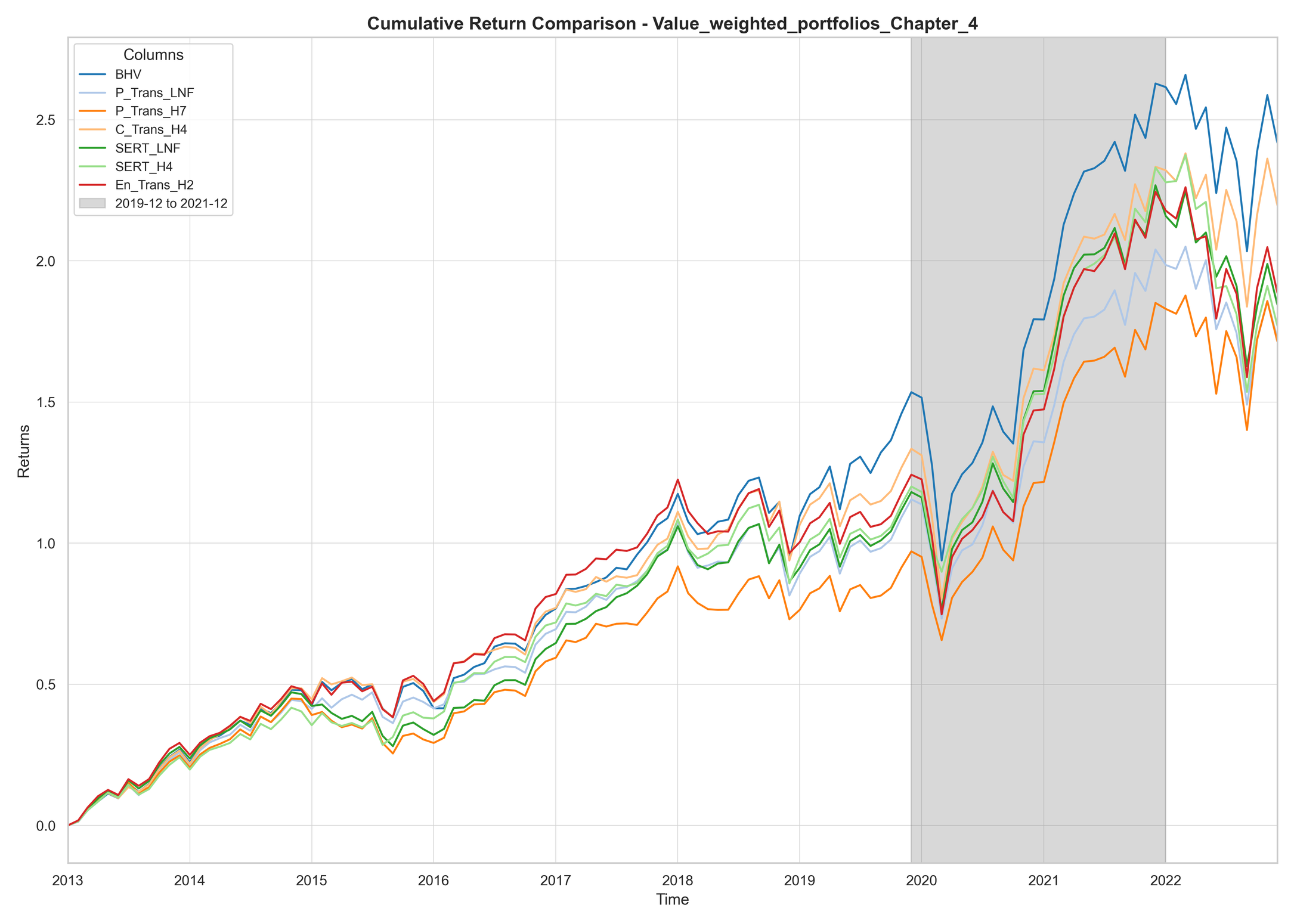}
\caption[Sign value-weighted accumulative return plots.]{Sign value-weighted accumulative return plots. The area covered by a grey shadow indicates the pandemic period. ‘BHE’ is the buy-and-hold strategy in the value-weighted method.}
\label{fig:cum_vw5324_ch3_turnover}
\end{sidewaysfigure}

The dynamic transaction cost robustness results comparison analysis is conducted to assess the overall impact of turnover rates on models' profitability in the equal-weighted method and the value-weighted method. In equal-weighted portfolio, the best model in each group of standard Transformer models is consistent with the ones in static transaction cost method, but the best models of encoder-only Transformers vary. Compared with the indicators in the static transaction cost method, it shows no evidence of static transaction cost (50bps) over-penalizes the models' profitability according to annualized returns, but it enlarges the maximum drawdown. In contrast, even if the dynamic transaction cost rate is lowering to 20bps, the Sharpe ratio and Sortino ratio are highly reduced due to the high turnover rate, and no model outperforms the buy-and-hold benchmark (BHE). However, in dynamic transaction cost scenario, the proposed pre-trained Transformer model (P\_Trans\_H6) and single-head SERT (SERT\_H1) outperform the benchmarks according to the Sortino ratio during the COVID-19 period, which shows in Figure~\ref{fig:sortino_ew5321_bar_ch3_turnover}. Non pre-training benchmark models, C\_Trans\_H1 and En\_Trans\_H1 present the resilience in the pre-COVID period, and P\_Trans\_H6 shows superiority lasting in Period Including COVID-19 and One-Year After (2212). In value-weighted portfolios, it is interesting to find that static transaction cost over-penalizes the profitability of the models again. The turnover adjusted transaction cost method significantly higher the annualized return, Sharpe ratio and Sortino ratio, and lower the maximum drawdown. According to Figure~\ref{fig:sortino_vw5323_bar_ch3_turnover}, the proposed pre-trained models slightly overweigh the SERT\_LNF benchmark in COVID-19–Inclusive Period (2112), are the most profitable models during that period. This  aligns with the value of turnover rate. The cumulative return plots, Figure~\ref{fig:cum_ew5322_ch3_turnover} and Figure~\ref{fig:cum_vw5324_ch3_turnover} verify the previous findings. Furthermore, the softmax signal filter narrows the difference between models. It significantly improves the models' profitability, and it optimizes the turnover rate, which leads the best proposed pre-trained models to outperform the buy-and-hold benchmarks.\\ 

\begin{table}[htbp]
    \centering
    \small
    \begin{tabular}{lcccccccc}
        \toprule
        \textbf{Model} & \textbf{AR} & \textbf{Ann.SR} & \textbf{SR} & \textbf{Ann.SO} & \textbf{SO} & \textbf{MDD} & \textbf{Turnover} \\
        \midrule
        \multicolumn{8}{c}{\textbf{Pre-COVID-19 Period (1911)}} \\
        \midrule
        P\_Trans\_LNF & 0.0733 & 0.6683 & 0.1929 & 0.8658 & 0.2499 & -0.1511 & 0.7229 \\
        P\_Trans\_H6  & 0.0787 & 0.7087 & 0.2046 & 0.9390 & 0.2711 & -0.1394 & 0.7589 \\
        C\_Trans\_H1  & 0.0837 & 0.7586 & 0.2190 & 1.0321 & 0.2979 & -0.1297 & 0.6654 \\
        SERT\_LNF     & 0.0782 & 0.7278 & 0.2101 & 1.0021 & 0.2893 & -0.1399 & 0.7301 \\
        SERT\_H1      & 0.0717 & 0.6578 & 0.1899 & 0.9092 & 0.2625 & -0.1625 & 0.8792 \\
        En\_Trans\_H1 & 0.0786 & 0.7247 & 0.2092 & 1.0519 & 0.3037 & -0.1483 & 0.6129 \\
        BHE           & 0.1141 & 0.8892 & 0.2567 & 1.2303 & 0.3552 & -0.1483 & 0.0120 \\
        \midrule
        \multicolumn{8}{c}{\textbf{COVID-19-Inclusive Period (2112)}} \\
        \midrule
        P\_Trans\_LNF & 0.0944 & 0.6968 & 0.2011 & 0.8052 & 0.2324 & -0.2956 & 0.6980 \\
        P\_Trans\_H6  & 0.1107 & 0.8454 & 0.2441 & 1.1946 & 0.3449 & -0.2428 & 0.6875 \\
        C\_Trans\_H1  & 0.1123 & 0.7503 & 0.2166 & 0.8926 & 0.2577 & -0.3105 & 0.5733 \\
        SERT\_LNF     & 0.1088 & 0.7804 & 0.2253 & 1.0253 & 0.2960 & -0.2921 & 0.6748 \\
        SERT\_H1      & 0.1049 & 0.7932 & 0.2290 & 1.1241 & 0.3245 & -0.2616 & 0.7834 \\
        En\_Trans\_H1 & 0.1089 & 0.7456 & 0.2152 & 0.9017 & 0.2603 & -0.3263 & 0.5202 \\
        BHE           & 0.1353 & 0.8512 & 0.2457 & 1.0181 & 0.2939 & -0.2890 & 0.0093 \\
        \midrule
        \multicolumn{8}{c}{\textbf{Period Including COVID-19 and One-Year After (2212)}} \\
        \midrule
        P\_Trans\_LNF & 0.0705 & 0.5036 & 0.1454 & 0.6125 & 0.1768 & -0.2956 & 0.7144 \\
        P\_Trans\_H6  & 0.0906 & 0.6479 & 0.1870 & 0.9366 & 0.2704 & -0.2428 & 0.6434 \\
        C\_Trans\_H1  & 0.0927 & 0.6036 & 0.1742 & 0.7622 & 0.2200 & -0.3105 & 0.5460 \\
        SERT\_LNF     & 0.0840 & 0.5875 & 0.1696 & 0.8044 & 0.2322 & -0.2921 & 0.6905 \\
        SERT\_H1      & 0.0865 & 0.6176 & 0.1783 & 0.8795 & 0.2539 & -0.2616 & 0.7367 \\
        En\_Trans\_H1 & 0.0898 & 0.5941 & 0.1715 & 0.7540 & 0.2177 & -0.3263 & 0.4956 \\
        BHE           & 0.1143 & 0.7017 & 0.2026 & 0.8883 & 0.2564 & -0.2890 & 0.0083 \\
        \bottomrule
    \end{tabular}
\caption[Back-testing performance of the equal-weighted portfolio between the best models in each group considering the dynamic transaction cost.]{Back-testing performance of the equal-weighted portfolio between the best models in each group considering the dynamic transaction cost. The notation of `Ann' means annualized. SR and SO denote the Sharpe Ratio and the Sortino Ratio, respectively.}
\label{tab:best_ew5321_ch3_turnover}
\end{table}

\clearpage
\begin{table}[htbp]
  \centering
  \small 
  \begin{tabular}{lcccccccc}
    \toprule
    \textbf{Model} & \textbf{AR} & \textbf{Ann.SR} & \textbf{SR} & \textbf{Ann.SO} & \textbf{SO} & \textbf{MDD} & \textbf{Turnover} \\
    \midrule
    \multicolumn{8}{c}{\textbf{Pre--COVID-19 Period (1911)}} \\
    \midrule
    P\_Trans\_LNF & 0.1125 & 1.0716 & 0.3094 & 1.3302 & 0.3840 & -0.1234 & 0.6952 \\
    P\_Trans\_H7  & 0.0983 & 0.9460 & 0.2731 & 1.3033 & 0.3762 & -0.1342 & 0.8495 \\
    C\_Trans\_H4  & 0.1256 & 1.0954 & 0.3162 & 1.4692 & 0.4241 & -0.1138 & 0.4638 \\
    SERT\_LNF     & 0.1145 & 1.1008 & 0.3178 & 1.5609 & 0.4506 & -0.1297 & 0.6941 \\
    SERT\_H4      & 0.1159 & 1.0813 & 0.3122 & 1.2697 & 0.3665 & -0.1305 & 0.7783 \\
    En\_Trans\_H2 & 0.1188 & 1.0585 & 0.3056 & 1.5451 & 0.4460 & -0.1169 & 0.6541 \\
    BHV           & 0.1388 & 1.1795 & 0.3405 & 1.6054 & 0.4634 & -0.1297 & 0.0120 \\
    \midrule
    \multicolumn{8}{c}{\textbf{COVID-19--Inclusive Period (2112)}} \\
    \midrule
    P\_Trans\_LNF & 0.1315 & 1.0401 & 0.3002 & 1.2939 & 0.3735 & -0.1959 & 0.6775 \\
    P\_Trans\_H7  & 0.1235 & 1.0156 & 0.2932 & 1.4776 & 0.4266 & -0.1596 & 0.7696 \\
    C\_Trans\_H4  & 0.1431 & 1.0245 & 0.2957 & 1.2709 & 0.3669 & -0.2279 & 0.4060 \\
    SERT\_LNF     & 0.1406 & 1.0701 & 0.3089 & 1.5182 & 0.4383 & -0.1914 & 0.6393 \\
    SERT\_H4      & 0.1430 & 1.1621 & 0.3355 & 1.5470 & 0.4466 & -0.1373 & 0.7028 \\
    En\_Trans\_H2 & 0.1397 & 1.0041 & 0.2899 & 1.3237 & 0.3821 & -0.2206 & 0.5801 \\
    BHV           & 0.1540 & 1.0688 & 0.3085 & 1.3057 & 0.3769 & -0.2352 & 0.0093 \\
    \midrule
    \multicolumn{8}{c}{\textbf{Period Including COVID-19 and One-Year After (2212)}} \\
    \midrule
    P\_Trans\_LNF & 0.1055 & 0.7924 & 0.2287 & 1.0260 & 0.2962 & -0.1959 & 0.6967 \\
    P\_Trans\_H7  & 0.1050 & 0.7752 & 0.2238 & 1.1152 & 0.3219 & -0.1656 & 0.7357 \\
    C\_Trans\_H4  & 0.1232 & 0.8376 & 0.2418 & 1.0988 & 0.3172 & -0.2279 & 0.3875 \\
    SERT\_LNF     & 0.1102 & 0.8095 & 0.2337 & 1.1912 & 0.3439 & -0.1963 & 0.6597 \\
    SERT\_H4      & 0.1073 & 0.8154 & 0.2354 & 1.0600 & 0.3060 & -0.2485 & 0.7390 \\
    En\_Trans\_H2 & 0.1118 & 0.7591 & 0.2191 & 1.0262 & 0.2963 & -0.2206 & 0.5654 \\
    BHV           & 0.1308 & 0.8654 & 0.2498 & 1.1270 & 0.3253 & -0.2352 & 0.0083 \\
    \bottomrule
  \end{tabular}
\caption[Back-testing performance of the value-weighted portfolio between the best models in each group considering the dynamic transaction cost.]{Back-testing performance of the value-weighted portfolio between the best models in each group considering the dynamic transaction cost. The notation of `Ann' means annualized. SR and SO are denoted as Sharpe Ratio and Sortino Ratio respectively.}
\label{tab:best_vw5322_ch3}
\end{table}

\clearpage
\begin{sidewaysfigure}[htbp!]
\centering
\includegraphics[width=0.95\columnwidth, height=0.8\textheight, keepaspectratio]{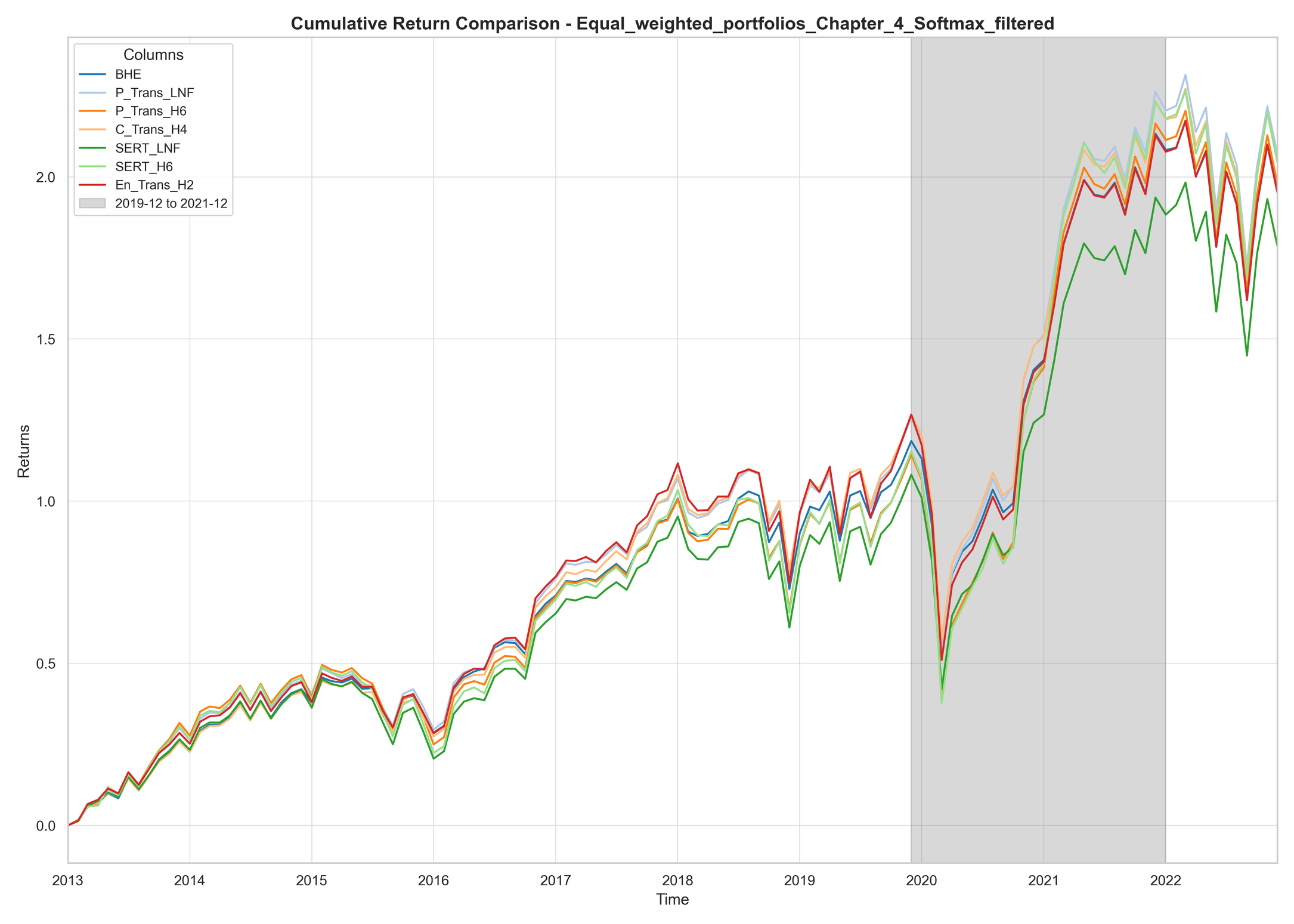}
\caption[Equal-weighted with Softmax trading signal filter accumulative return plots considering the dynamic transaction cost.]{Equal-weighted with Softmax trading signal filter accumulative return plots considering the dynamic transaction cost.}
\label{fig:cum_ew5325_soft_ch3_turnover}
\end{sidewaysfigure}

\clearpage
\begin{sidewaysfigure}[htbp!]
\centering
\includegraphics[width=0.95\columnwidth, height=0.8\textheight, keepaspectratio]{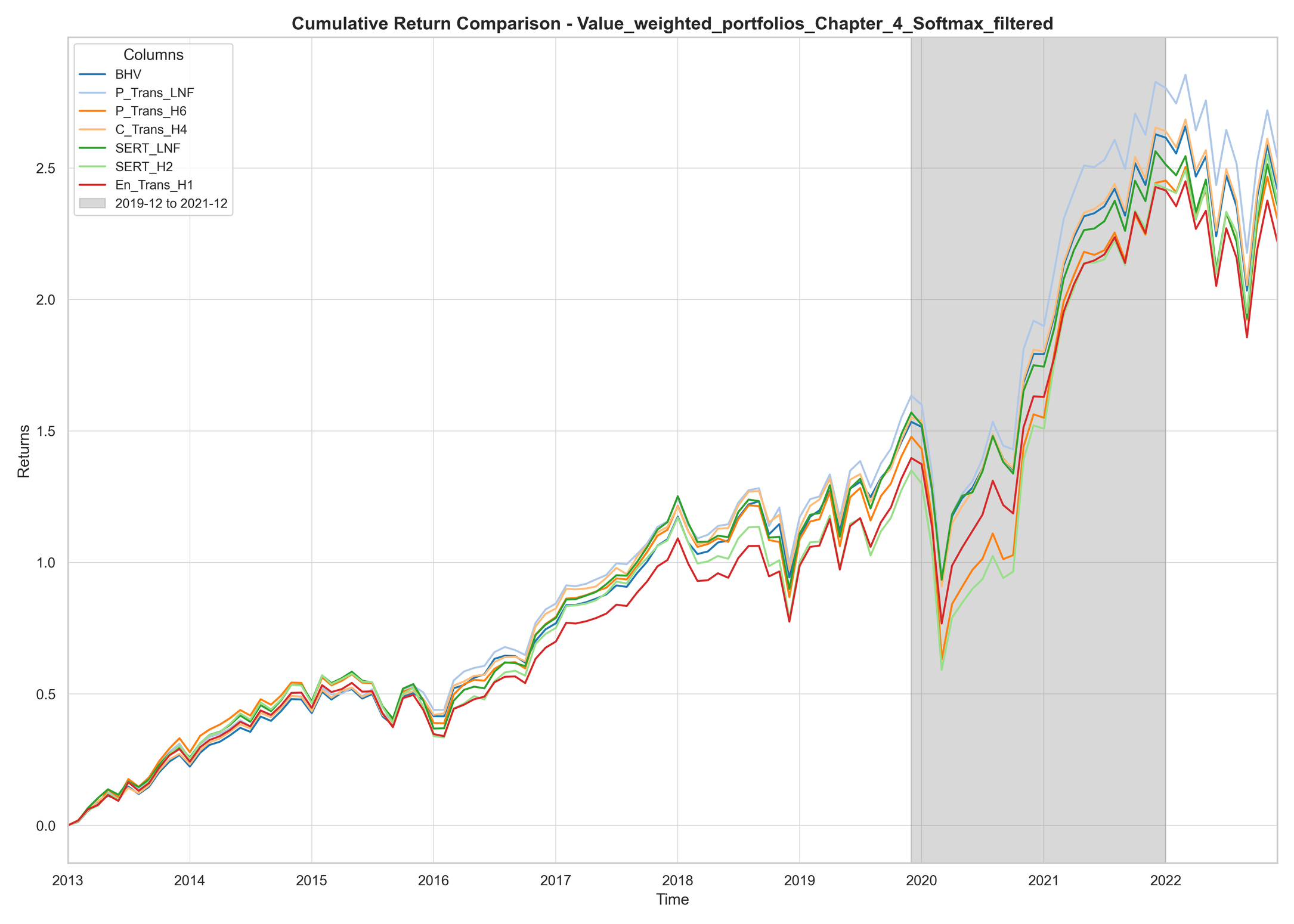}
\caption[Value-weighted with Softmax trading signal filter accumulative return plots considering the dynamic transaction cost.]{Value-weighted with Softmax trading signal filter accumulative return plots considering the dynamic transaction cost.}
\label{fig:cum_vw5326_soft_ch3_turnover}
\end{sidewaysfigure}

\section{Conclusion}\label{sec:Conclusion}
In this study, the innovative Transformer models, SERT and pre-trained Transformer, are developed specifically for financial and economic time series data. They are examined in stock pricing and factor investing assignments as the novel frameworks for the factor model. To the best of my knowledge, this is the first attempt for the pre-trained Transformer application on stock pricing and factor investing strategy building context. From the model engineering perspective, this article proves that the Transformer models, as a category of LLMs, are not only suitable for enormous datasets, such as datasets of NLP tasks, speech or image recognition and video generation, but also minuscule datasets such as financial and economic numerical time series data. They capture the variation of the economic data with high temporal sparsity characteristics and significantly improve the prediction accuracy under extreme market conditions, such as the price collapses and lift caused by the pandemic. In addition, the pre-training process verifies its necessity for improving model performance when the data contains the aforementioned characteristics and has missing values and noise. Moreover, the encoder-only Transformer is the simplest and most commonly used LLM in Fintech; it shows advantages in model fitness as well as prediction performance compared to the pre-trained Transformer models. Single-block SERT presents instability and inconsistency to some extent. They show sensitivity to the hyperparameters during the training process. These break the illusion that model complexity is positively linked to the model overfitting issue. This study also finds that attention head numbers affect the prediction performance at an unnoticeable level. Pre-trained Transformer models are relatively more stable than SERTs, which show a positive relationship with attention head numbers, and there is an optimal number of attention heads for each configuration. In addition, the layer normalization first setting does not perform excellent in the predictive power of this case. It could possibly work better in a deeper structure and a noisier dataset. However, they perform well in backtesting under a dynamic transaction costs configuration. \\

The models are examined for extreme market fluctuation adaptivity via three separate periods: the period before COVID-19, the period including COVID-19 and the period including one year after COVID-19. During these periods, the U.S. stock market experienced a moderate uptrend, an extreme uptrend and an extreme downtrend. The proposed models show magnificent adaptivity to extreme market fluctuations. In other words, they achieve excellent performance in capturing patterns in highly varied stock prices. Moreover, in the static transaction cost scenario, the factor investing strategy based on the proposed models shows high profitability, and some of them show even higher performance than the ‘buy-and-hold’ benchmarks under the equal-weighted portfolio method. They also show a fantastic capability for hedging the downside risks. It is worth noting that the softmax signal filter only removes the difference between models, making them approach the buy-and-hold benchmarks, but does not facilitate strategy-wise performance in static transaction cost scenarios. However, in dynamic transaction-cost scenarios, which are included in the turnover rate and the adjusted large-cap-specific rate (20bps), the high turnover rate significantly erodes the profitability of these models in most cases and reduces backtesting performance to levels below buy-and-hold benchmarks. Whereas, even with the high conservative static transaction cost restrictions and high turnover rates, the best proposed models exhibit strong resilience during the COVID-19 period, which proves the profitability of the best proposed models in terms of the downside risks. Additionally, in the static transaction scenario, the absolute capital gain of the value-weighted portfolio is highly reduced, which is caused by over-penalization of the high static transaction cost. However, when considering the dynamic transaction cost with lower rates (20bps), the value-weighted portfolios show higher profitability, particularly during the pandemic period (2112). The softmax signal filter, which restricts the turnover rates, significantly improves the profitability of the models in the dynamic transaction cost scenarios, especially for value-weighted methods.\\

Some open questions remain for future researchers. For example, when employing the observable factors for the factor model, it may have omitted factor issues, which cause omitted variable bias or missing factor problems, although Transformer models already mitigate these issues via autoencoder structure and self-attention mechanisms to some extent. Namely, the latent factors employed in this chapter are completely derived from the observable original factors. Future researchers can attempt alternative models from the Transformer model family for extracting factors directly from the returns, for example, generative Transformer models such as generative pre-trained Transformer (GPT). Furthermore, although we study Transformer models on large capital stocks, we find that the proposed models are excellent at coping with extreme fluctuations, and higher volatility appears to enhance their performance.\\ 

This implies that Transformer models are extra sensitive to higher variations, namely, models perform better with highly volatile assets. In that sense, it is possibly worth trying the proposed models on small capital stocks, emerging markets or cryptocurrencies, which have high volatility. Also, the firm characteristic-sorted portfolios as factors increase the economic meanings to a degree, but they sacrifice some of the information originally contained in the firm characteristics. Researchers may explore innovative factor-organizing methods which balance information completeness and economic explanations. Finally, and no less importantly, researchers can further explore the LLM-based portfolio optimization or asset allocation technique as employed in \citet{Ma2023AttentionApproach}.\\

\section*{Acknowledgements}
First, I am profoundly thankful to my supervisor, Professor Peter N Smith (University of York), for his guidance and support throughout this research. I am grateful to Dr Penn Rainford (University of York), Dr Mark Stevenson (University of Oxford) and Dr Mark Hallam (University of York) for their insightful comments. Special thanks to Qiran Lai (University of York) and Jingbo Yang (University of York) for their technical support. I also appreciate the valuable ideas of the Transformer model from Dr Frank Soboczenski (University of York) and model structure suggestions from Dr Dimitar Kazakov (University of York), Professor Michael Thornton (University of York) and Professor Jia Chen (University of Macau). Finally, I would like to thank Dr Grega Smrkolj (Newcastle University) for his advice on LaTeX editing and diagram creation, as well as my family and friends for their encouragement during this study.\\

Constructive feedback is greatly appreciated.







\clearpage
\bibliography{references}
\clearpage
\begin{appendices}
\section{Transformer model Estimation}\label{sec:Transformer_Estimation_ch4}
Assume the input is given by
\begin{equation}
X\in R^{n\times d}
\end{equation}
where $n$ is the length of the features and $d$ is the dimension of features. The MSE loss function form can be presented as:
\begin{equation}
L = \frac{1}{N} \sum_{i=1}^{N} (y_i - \hat{y}_i)^2
\end{equation}
\subsection*{Gradients of output linear layer}\label{subsec:Gradients of output linear layer}
\begin{equation}
\hat{y}_i = W_{\text{out}} h_{L,i} + b_{\text{out}}
\end{equation}
Where $h_{L,i}$ is the outputs from previous layer, $W_{\text{out}}$  and $b_{\text{out}}$  are learnable parameters. For the gradient $g$, we have,
\begin{equation}
\begin{aligned}
g_{W_{\text{out}}} &= \frac{\partial L}{\partial W_{\text{out}}} = \frac{\partial}{\partial W_{\text{out}}} \left[ \frac{1}{N} \sum_{i=1}^{N} (y_i - \hat{y}_i)^2 \right] \\
&= \frac{1}{N} \sum_{i=1}^{N} 2 (y_i - \hat{y}_i) \cdot \frac{\partial (y_i - \hat{y}_i)}{\partial W_{\text{out}}} \\
&= \frac{2}{N} \sum_{i=1}^{N} (\hat{y}_i - y_i) h_{L,i}^{T}
\end{aligned}
\end{equation}
\begin{equation}
\begin{aligned}
g_{b_{\text{out}}} &= \frac{\partial L}{\partial b_{\text{out}}} = \frac{1}{N} \sum_{i=1}^{N} 2 (y_i - \hat{y}_i) \cdot \frac{\partial (y_i - \hat{y}_i)}{\partial b_{\text{out}}} \\
&= \frac{2}{N} \sum_{i=1}^{N} (\hat{y}_i - y_i)
\end{aligned}
\end{equation}
\subsection*{Gradients of Attention layer}\label{subsec:Gradients of Attention layer}
\begin{equation}
\begin{aligned}
Q &= h W_Q, \quad K = h W_K, \quad V = h W_V \\
S &= \frac{Q K^{T}}{\sqrt{d_k}} \\
A &= \text{softmax}(S) \\
O &= A V \\
Z &= \text{Concat}(O_1, O_2, \dots, O_H) W_O
\end{aligned}
\end{equation}
\begin{itemize}
    \item \( A = \text{softmax}\left( \frac{Q K^T}{\sqrt{d_k}} \right) \in \mathbb{R}^{N \times N} \),
    \item \( h \in \mathbb{R}^{N \times d} \), \( W_V \in \mathbb{R}^{d \times d_k} \),
    \item \( V \in \mathbb{R}^{N \times d_k} \), \( O \in \mathbb{R}^{N \times d_k} \).
\end{itemize}
For $O=AV$,
\begin{equation}
\frac{\partial L}{\partial W_V} = \frac{\partial L}{\partial O} \cdot \frac{\partial O}{\partial V} \cdot \frac{\partial V}{\partial W_V}.
\end{equation}
\begin{align}
O_{ij} = \sum_k A_{ik} V_{kj},
\frac{\partial O_{ij}}{\partial V_{mn}} = A_{im} \cdot \delta_{jn},
\frac{\partial O}{\partial V} = A,
\frac{\partial L}{\partial V} = \frac{\partial L}{\partial O} A^T.
\end{align}
\begin{align}
V_{kj} = \sum_m h_{km} W_{V,mj},
\frac{\partial V_{kj}}{\partial W_{V,mn}} = h_{km} \cdot \delta_{jn},\\
\frac{\partial L}{\partial W_{V,mn}} = \sum_{k,j} \frac{\partial L}{\partial V_{kj}} \cdot h_{km} \cdot \delta_{jn} = \sum_k \frac{\partial L}{\partial V_{kn}} \cdot h_{km},\\
\frac{\partial L}{\partial W_V} = h^T \cdot \frac{\partial L}{\partial V} = h^T \cdot \left( \frac{\partial L}{\partial O} A^T \right).
\end{align}
\begin{equation}
g_{W_V} = \frac{\partial L}{\partial W_V} = h^{T} \left( \frac{\partial L}{\partial O} \right)^{T} A
\end{equation}
For $W_Q$,
\begin{equation}
g_{W_Q} = \frac{\partial L}{\partial W_Q} = \frac{\partial L}{\partial O} \cdot \frac{\partial O}{\partial A} \cdot \frac{\partial A}{\partial S} \cdot \frac{\partial S}{\partial Q} \cdot \frac{\partial Q}{\partial W_Q}
\end{equation}
\begin{align}
\frac{\partial O}{\partial A} &= V \\
\frac{\partial L}{\partial A} &= \frac{\partial L}{\partial O} V^{T} \\
\frac{\partial A_i}{\partial S_j} &= 
\begin{cases}
A_i (1 - A_i), & i = j \\
- A_i A_j, & i \ne j
\end{cases} \\
\frac{\partial A}{\partial S} &= \text{diag}(A) - A A^{T} \\
\frac{\partial L}{\partial S} &= \frac{\partial L}{\partial A} \cdot \left[ \text{diag}(A) - A A^{T} \right] \\
S_{ij} &= \frac{\sum_{m} Q_{im} K_{jm}}{\sqrt{d_k}} \\
\frac{\partial S_{ij}}{\partial Q_{mn}} &= \frac{K_{jn}}{\sqrt{d_k}} \cdot \delta_{im} = \frac{K^{T}}{\sqrt{d_k}}\\
\frac{\partial L}{\partial Q} &= \frac{\partial L}{\partial S} \cdot \frac{K}{\sqrt{d_k}} \\
Q_{ij} &= \sum_{m} h_{im} W_{Q,mj} \\
\frac{\partial Q_{ij}}{\partial W_{Q,mn}} &= h_{im} \cdot \delta_{jn} \\
\frac{\partial Q}{\partial W_Q} &= h^{T} \\
g_{W_Q} &= \frac{\partial L}{\partial W_Q} = h^{T} \left( \frac{\partial L}{\partial S} \cdot \frac{K}{\sqrt{d_k}} \right)^{T}
\end{align}
For $W_K$,
\begin{align}
\frac{\partial L}{\partial W_K} = \frac{\partial L}{\partial O} \cdot \frac{\partial O}{\partial A} \cdot \frac{\partial A}{\partial S} \cdot \frac{\partial S}{\partial K} \cdot \frac{\partial K}{\partial W_K}\\
O_{ij} = \sum_k A_{ik} V_{kj}\\
\frac{\partial O_{ij}}{\partial A_{mn}} = V_{nj} \cdot \delta_{im}\\
\frac{\partial L}{\partial A} = \frac{\partial L}{\partial O} V^T\\
\frac{\partial A_i}{\partial S_j} = 
\begin{cases} 
A_i (1 - A_i), & i = j, \\
-A_i A_j, & i \neq j,
\end{cases}\\
\frac{\partial A}{\partial S} = \text{diag}(A) - A A^T\\
\frac{\partial L}{\partial S} = \left( \frac{\partial L}{\partial O} V^T \right) \cdot \left( \text{diag}(A) - A A^T \right)\\
S_{ij} = \frac{1}{\sqrt{d_k}} \sum_m Q_{im} K_{jm}\\
\frac{\partial S_{ij}}{\partial K_{mn}} = \frac{1}{\sqrt{d_k}} Q_{im} \cdot \delta_{jn}\\
\frac{\partial S}{\partial K} = \frac{Q^T}{\sqrt{d_k}}\\
\frac{\partial L}{\partial K} = \frac{\partial L}{\partial S} \cdot \frac{Q}{\sqrt{d_k}}\\
K_{kj} = \sum_m h_{km} W_{K,mj}\\
\frac{\partial K_{kj}}{\partial W_{K,mn}} = h_{km} \cdot \delta_{jn}\\
\frac{\partial L}{\partial W_{K,mn}} = \sum_k \frac{\partial L}{\partial K_{kn}} \cdot h_{km}\\
\frac{\partial L}{\partial W_K} = h^T \cdot \frac{\partial L}{\partial K} = h^T \cdot \left( \frac{\partial L}{\partial S} \cdot \frac{Q}{\sqrt{d_k}} \right)\\
g_{W_K} = \frac{\partial L}{\partial W_K} = h^T \left( \frac{\partial L}{\partial S} \cdot \frac{Q}{\sqrt{d_k}} \right)^T.
\end{align}
In the case of multi-head attention,
\begin{align}
Z = \text{Concat}(O_1, O_2, \dots, O_H) W_O\\
\frac{\partial L}{\partial W_O} = \frac{\partial L}{\partial Z} \cdot \frac{\partial Z}{\partial W_O}\\
\end{align}
Assume $C = \text{Concat}(O_1, O_2, \dots, O_H) \in \mathbb{R}^{N \times (H d_k)}$,
\begin{align}
Z = C W_O\\
Z_{ij} = \sum_k C_{ik} W_{O,kj}\\
\frac{\partial Z_{ij}}{\partial W_{O,mn}} = C_{im} \cdot \delta_{jn}
\frac{\partial L}{\partial W_{O,mn}} = \sum_{i,j} \frac{\partial L}{\partial Z_{ij}} \cdot C_{im} \cdot \delta_{jn} = \sum_i \frac{\partial L}{\partial Z_{in}} \cdot C_{im}\\
\frac{\partial L}{\partial W_O} = C^T \cdot \frac{\partial L}{\partial Z}\\
g_{W_{O}}=\frac{\partial L}{\partial W_O} = \text{Concat}(O_1, O_2, \dots, O_H)^T \cdot \frac{\partial L}{\partial Z}\\
\end{align}
\subsection*{Gradients of MLP autoencoder feedforward NN (FFN)}\label{subsec:Gradients of MLP autoencoder feedforward NN (FFN)}
\begin{equation}
\text{FFN}(h) = W_2 \cdot \text{ReLU}(W_1 h + b_1) + b_2
\end{equation}
Let $z=W_1 h+b_1$, $a=\text ReLU (z)$,
For $W_2$,
\begin{equation}
g_{W_2} = \frac{\partial L}{\partial W_2} = \frac{\partial L}{\partial \text{FFN}} \cdot a^{T}
\end{equation}
For $b_2$,
\begin{equation}
g_{b_2} = \frac{\partial L}{\partial b_2} = \frac{\partial L}{\partial \text{FFN}}
\end{equation}
For $W_1$,
\begin{equation}
g_{W_1} = \frac{\partial L}{\partial W_1} = \frac{\partial L}{\partial a} \cdot \frac{\partial a}{\partial z} \cdot h^{T}
\end{equation}
Where $\frac{\partial a}{\partial z}=1$, if $z>0$, $0$ for otherwise.
\subsection*{Gradients of Embedding layer}\label{subsec:Gradients of Embedding layer}
In the embedding layer, $x$ is reflected as $h_0=W_{em} x$,
\begin{equation}
g_{W_{\text{em}}} = \frac{\partial L}{\partial W_{\text{em}}} = \frac{\partial L}{\partial h_0} \cdot x^{T}
\end{equation}
\subsection*{Adam optimizer}\label{subsec:Adam optimizer}
Adam updates parameters by maintaining the first and second moments of the gradients.
\begin{algorithm}[H]
\caption{Adam Optimization}
\begin{algorithmic}[1]
\State \textbf{Initialize:} $m_0 = 0$, $v_0 = 0$, $l = 0$
\State \textbf{Set hyperparameters:} learning rate $\eta = 0.001$, $\beta_1 = 0.9$, $\beta_2 = 0.999$, $\epsilon = 10^{-8}$
\While{$\theta_l$ not converged}
    \State $l \gets l + 1$
    \State Compute gradient: $g_l = \nabla_\theta L(\theta_{l-1})$
    \State $m_l \gets \beta_1 m_{l-1} + (1 - \beta_1) g_l$
    \State $v_l \gets \beta_2 v_{l-1} + (1 - \beta_2) g_l \odot g_l$ \Comment{$\odot$: element-wise multiplication}
    \State $\hat{m}_l \gets \frac{m_l}{1 - \beta_1^l}$ \Comment{Bias correction}
    \State $\hat{v}_l \gets \frac{v_l}{1 - \beta_2^l}$
    \State $\theta_l \gets \theta_{l-1} - \eta \cdot \frac{\hat{m}_l}{\sqrt{\hat{v}_l} + \epsilon}$
\EndWhile
\State \textbf{Return:} $\theta_l$
\end{algorithmic}
\begin{flushleft}
\textit{Source: Adapted from \citet{Gu2020EmpiricalLearning}}
\end{flushleft}
\end{algorithm}
where $g_t$ and $\theta_t$ represent the gradients and parameters computed from previous sections respectively.
\subsection*{Early stopping}\label{subsec:Early stopping}
\begin{algorithm}[H]
\caption{Early Stopping}
\begin{algorithmic}[1]
\State \textbf{Initialize:} $j = 0$, $\epsilon = \infty$, select patience parameter $p$.
\While{$j < p$}
    \State Update $\theta$ using the training algorithm (e.g., for $h$ steps).
    \State Calculate the prediction error from the validation sample, denoted as $\epsilon'$.
    \If{$\epsilon' < \epsilon$}
        \State $j \gets 0$
        \State $\epsilon \gets \epsilon'$
        \State $\theta' \gets \theta$
    \Else
        \State $j \gets j + 1$
    \EndIf
\EndWhile
\State \textbf{Return:} $\theta'$
\end{algorithmic}
\begin{flushleft}
\textit{Source: Adapted from \citet{Gu2020EmpiricalLearning}}
\end{flushleft}
\end{algorithm}
\subsection*{Layer Normalization}\label{subsec:Layer Normalization}
\begin{algorithm}[H]
\caption{Layer Normalization}
\begin{algorithmic}[1]
\State \textbf{Input:} Input vector $x \in \mathbb{R}^d$, learnable parameters $\gamma$, $\beta$
\State Compute the mean: $\mu = \frac{1}{d} \sum_{i=1}^{d} x_i$
\State Compute the variance: $\sigma^2 = \frac{1}{d} \sum_{i=1}^{d} (x_i - \mu)^2$
\State Normalize: $\hat{x}_i = \frac{x_i - \mu}{\sqrt{\sigma^2 + \epsilon}}$ for $i = 1, \dots, d$
\State Apply scale and shift: $y_i = \gamma_i \hat{x}_i + \beta_i$ for $i = 1, \dots, d$
\State \textbf{Return:} Output vector $y$
\end{algorithmic}
\end{algorithm}

\section{Extended Positional Encoding Explanation}\label{sec:positional_encoding_ch4}
\textbf{Self-attention}
In addition, while the self-attention mechanism can measure the similarity between any two vectors from different time steps is not explained by the Transformer model’s literature. For this part, we can dig a bit deeper into the dot product-based attention score calculation. From Equation~\eqref{eq:sattention2_ch3}, the attention score is calculated from the scaled dot product of Q and K matrix. Thus, we denote $S \in \mathbb{R}^{d_{\text{model}} \times d_{\text{model}}}$, then we have:
\begin{equation}
S = QK^T
\end{equation}
For each $s_{i,j}$ in S,
\begin{equation}
s_{i,j} = q_i \cdot k_j = \left\lVert q_i \right\rVert \cdot \left\lVert k_j \right\rVert \cdot \cos\theta
\label{eq:s_derivative_ch3}
\end{equation}
Where $q_i, k_j \in \mathbb{R}^{d_{\text{model}}}$ are vectors from Q and K. From Equation~\eqref{eq:s_derivative_ch3}, the dot product can measure the similarity of two vectors via directional similarity. In other words, when vectors with similar directions, the dot product value is high, which means the two vectors in different time steps have higher similarity, while when the vectors with opposite directions, the dot product value is low, which means the two vectors in different time steps are irrelevant. Two orthometric vectors dot product value is zero. $\sqrt{d_{\text{model}}}$ is the scale for the dot product value to moderate the anomalies. Since the attention weights are computed by the softmax function, higher weights are distributed to vectors in time steps that have higher similarities. \\

However, the dot product measures the similarity between vectors in different time steps, but without positional codes, the self-attention mechanism cannot identify sequential information. For example, it recognizes ‘I love cats’ and ‘cats love me’ as the same sentence, because it cannot detect different sequential meanings. Thus, when positional encodings are added in, the attention score is altered as:
\begin{align}
Q &= W^Q (X + PE), \quad K = W^K (X + PE) \\
s_{ij} &= \left(W^Q X_i + W^Q PE_i\right) \cdot \left(W^K X_j + W^K PE_j\right)^\mathrm{T} \\
&= (W^Q X_i)(W^K X_j)^\mathrm{T}
+ (W^Q X_i)(W^K PE_j)^\mathrm{T}\\
&+ (W^Q PE_i)(W^K X_j)^\mathrm{T}
+ (W^Q PE_i)(W^K PE_j)^\mathrm{T}
\end{align}
Where $(W^Q X_i ) (W^K X_j )^T$ conveys the original features' interaction information, $(W^Q X_i ) (W^K PE_j )^T$ and $(W^Q PE_i ) (W^K X_j )^T$ conveys features and positional interaction information which represent how the Query features at one time step are influenced by the positional information of the Key at another time step, $(W^Q PE_i ) (W^K PE_j )^T$ contains pure positional interaction information. Therefore, it is more advanced and efficient than traditional statistical models since this attention mechanism simultaneously captures differences of lagged inputs, and features in this case. \\

\textbf{Cross-attention}
Analogous to the self-attention mechanism introduced in Section~\ref{subsubsec:Multi-head self-attention4142}, cross-attention in the Transformer is more powerful than self-attention since the Query from the decoder enables the cross-attention to capture the spatial-temporal information from a fuller angle. Concretely,
\begin{align}
Q_{de} &= W^{Q_{de}} (Y + PE_{de}), \quad K = W^K (X + PE_{en}) \\
s_{ij} &= \left( W^{Q_{de}} Y_i + W^{Q_{de}} PE_{de,i} \right) \cdot \left( W^K X_j + W^K PE_{en,j} \right)^\mathrm{T} \\
&= (W^{Q} Y_i)(W^{K} X_j)^\mathrm{T}
+ (W^{Q} Y_i)(W^{K} PE_{en,j})^\mathrm{T} \notag \\
&\quad + (W^{Q} PE_{de,i})(W^{K} X_j)^\mathrm{T}
+ (W^{Q} PE_{de,i})(W^{K} PE_{en,j})^\mathrm{T}
\end{align}
Where $(W^Q Y_i) (W^K X_j)^T$ demonstrate the spatial-temporal information of input $X$ and output $Y$. $(W^Q Y_i ) (W^K PE_{en,j})^T$ demonstrates the interaction of stock returns and the positional information of the encoder, while $(W^Q PE_{de,i}) (W^K X_j)^T$ demonstrates the interaction of decoder positional information and input $X$. $(W^Q PE_{de,i}) (W^K PE_{en,j})^T$ logs the similarity of positional information from the encoder and decoder. \\

\section{Extended Introduction of MLP Autoencoder}\label{sec:app_G_MLP_auto_ch2}
\begin{figure}[htbp!]
\centering
\includegraphics[width=0.95\columnwidth, height=0.4\textheight, keepaspectratio]{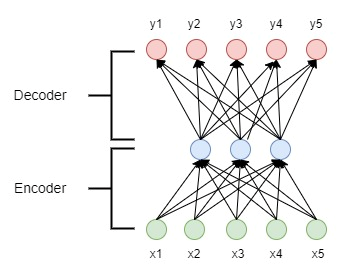}
\caption[An example of a single-layer MLP autoencoder NN structure.]{An example of a single-layer MLP autoencoder NN structure. The green dots represent the input vectors, while the red dots represent the output vectors. The blue dots indicate the neurons of the latent layer. $x_1,\cdots,x_5$ are the inputs and $y_1,\cdots,y_5$ are the outputs. The entire structure is a feedforward neural network.}
\label{fig:mlp_autoencoder1_4151_ch3}
\end{figure}

The feed-forward neural network module in the Transformer encoder block or decoder block is a single-layer MLP autoencoder NN structure. Also, the MLP autoencoder is the simplest autoencoder structure among the autoencoder family. Figure~\ref{fig:mlp_autoencoder1_4151_ch3} shows an example of the single-layer MLP autoencoder module. It contains one encoder block and one decoder block as well. All nodes in the input layer or output layer are connected with the nodes of the latent layer or bottleneck layer (blue dots layer in Figure~\ref{fig:mlp_autoencoder1_4151_ch3}, which attaches encoder and decoder blocks). If $x_1,\cdots,x_5$ are the input vectors, while $y_1,\cdots,y_5$ are output vectors, the matrix of the input could be denoted as $X$, where $X=[x_1,\cdots,x_5]$, and the output could be denoted as $Y$, where $Y=[y_1,\cdots,y_5]$. Further, the input layer to the latent layer can be presented as:
\begin{equation}
Z = g(WX + b)
\end{equation}
If $Z$ is assumed to be the output of the latent layer and $Z=[z_1,\cdots,z_3]$. Here, $W\in R^{3\times5}$ is the weight matrix and $b \in R^{3\times1}$ is the bias vector. $g(\cdot)$ is the activation function for the encoding process such as ReLU.\\

From the output of the latent layer to the decoder, it can be presented as:
\begin{equation}
\hat{Y} = g'\left( W' Z + b' \right)
\end{equation}
Where $W'\in R^{3\times5}$,$b'\in R^{5\times1}$ are weight matrix and bias vector for the decoder, while $g'(\cdot)$ is the activation function for the decoding process. $g'(\cdot)$ and $g(\cdot)$ can be identical or distinct. \\
Generally, the encoder is a method to project $X\in R^n$ to latent space (latent layer) output $Z\in R^m$, hence a generalized encoder with multiple hidden layers can be simplified as:
\begin{equation}
Z = f_{\text{en}}(X; \theta_{\text{en}})
\end{equation}
Where $f_{en}(\cdot)$ is the projection form of the encoder, $\theta_{en}$ presents the parameters of the encoder which contain the weight matrix and bias vectors. If the projection form is MLP, the multi-hidden-layer encoder can be extended as:
\begin{align}
h^{(1)} &= \sigma^{(1)} \left( W^{(1)} X + b^{(1)} \right) \\
h^{(2)} &= \sigma^{(2)} \left( W^{(2)} h^{(1)} + b^{(2)} \right) \\
&\ \vdots \notag \\
Z = h^{(L)} &= \sigma^{(L)} \left( W^{(L)} h^{(L-1)} + b^{(L)} \right)
\end{align}
Where $h^{(L)}\in R^{d_L\times1}$ is the L$th$ hidden layer output, and $\sigma^{(L)}(\cdot)$ is the L$th$ hidden layer’s activation function. $W^{(L)}\in R^{d_L×d_{L-1}}$ and $b^{(L)}\in 
R^{d_L\times1}$ are the weight matrix and bias vector in the Lth hidden layer. \\

Similarly, the decoder projects the encoder’s output $Z\in R^m$ to $\hat Y \in R^n$:
\begin{equation}
\hat{Y} = f_{\text{de}}(Z; \theta_{\text{de}})
\end{equation}
The $f_{de}(\cdot)$ is assumed MLP projection, $\theta_{de}$ presents the parameters of the decoder which contain the weight matrix and bias vectors. Furthermore, the decoder with hidden layers can be generalized as:
\begin{align}
h^{(L+1)} &= \sigma^{(L+1)} \left( W^{(L+1)} Z + b^{(L+1)} \right) \\
h^{(L+2)} &= \sigma^{(L+2)} \left( W^{(L+2)} h^{(L+1)} + b^{(L+2)} \right) \\
&\ \vdots \notag \\
\hat{Y} = h^{(2L)} &= \sigma^{(2L)} \left( W^{(2L)} h^{(2L-1)} + b^{(2L)} \right)
\end{align}
$W^{(2L)}\in R^{d_{2L}\times d_{2L-1}}$ and $b^{(2L)}\in R^{d_{2L}\times1}$ are the weight matrix and bias vector in the L$th$ hidden layer. All activation function $\theta$ here employs the ReLU function.\\

Corresponding to naïve MLP models, the autoencoder can be estimated by minimising the loss function such as mean square error (MSE) via the method of Stochastic Gradient Descent (SGD) with the Adaptive Moment Estimation optimizer (Adam) as well. The full presentation of the simplified autoencoder can be written as:\\
\begin{equation}
\hat{Y} = f_{\text{de}} \left( f_{\text{en}}(X; \theta_{\text{en}}); \theta_{\text{de}} \right)
\end{equation}
And the parameter $\theta_{en}$, $\theta_{de}$ can be estimated via minimizing the loss function:
\begin{equation}
\min_{\theta_{\text{en}}, \theta_{\text{de}}} L(Y, \hat{Y})
\end{equation}

\section{Mapping of Model Names (Original vs. New)}\label{tab:name_mapping}
\begin{table}[htbp]
  \centering
  \begin{tabular}{llll}
    \toprule
    Original Name & New Name & Original Name & New Name \\
    \midrule
    SERT1 & SERT\_LNF & Trans1 & P\_Trans\_LNF \\
    SERT2 & SERT\_H1 & Trans2 & P\_Trans\_H1 \\
    SERT3 & SERT\_H2 & Trans3 & P\_Trans\_H2 \\
    SERT4 & SERT\_H3 & Trans4 & P\_Trans\_H3 \\
    SERT5 & SERT\_H4 & Trans5 & P\_Trans\_H4 \\
    SERT6 & SERT\_H6 & Trans6 & P\_Trans\_H6 \\
    SERT7 & SERT\_H7 & Trans7 & P\_Trans\_H7 \\
    T-En1 & En\_Trans\_H1 & Trans8 & C\_Trans\_H1 \\
    T-En2 & En\_Trans\_H2 & Trans9 & C\_Trans\_H2 \\
    T-En3 & En\_Trans\_H4 & Trans10 & C\_Trans\_H4 \\
    \bottomrule
  \end{tabular}
\end{table}

\clearpage
\section{Revised Python code of OOS \texorpdfstring{$R^2$}{R2} and MSE calculation}\label{sec:app_python_r2_mse_ch2}
Please note that the input actual return form and predicted return form, the row labels are the stock permno, the column labels are the time steps.\\

\begin{lstlisting}
import pandas as pd
import numpy as np
import os
import glob

path_history = r"x:\xxx\xxxx\xx.csv"
r_history_df = pd.read_csv(path_history, index_col=0)

stock_benchmarks = r_history_df.iloc[:, :-120].mean(axis=1)

path_actual_oos = r"x:\xxx\xxxx\xx.csv"
actual_all = pd.read_csv(path_actual_oos, index_col=0)

actual_sub_df = actual_all.iloc[:, -120:]

actual_values = actual_sub_df.values 

stock_benchmarks = stock_benchmarks.reindex(actual_sub_df.index)

benchmark_vector = stock_benchmarks.values.reshape(-1, 1)

input_folder = r"x:\xxx\xxxx\xx"

if not os.path.exists(details_folder):
    os.makedirs(details_folder)

summary_list = []

for file_name in os.listdir(input_folder):
    if file_name.endswith('.csv'):
        file_path = os.path.join(input_folder, file_name)
        
        pred_df = pd.read_csv(file_path, index_col=0)
        pred_df = pred_df.reindex(actual_sub_df.index)
        
        pred_df_sub = pred_df
        pred_values = pred_df_sub.values
        
        errors = actual_values - pred_values
        numerator = np.sum(errors**2, axis=1)
        
        benchmark_errors = actual_values - benchmark_vector
        
        denominator = np.sum(benchmark_errors**2, axis=1)

        stock_mse = np.mean(errors**2, axis=1)
        
        with np.errstate(divide='ignore', invalid='ignore'):
            stock_r2 = 1 - (numerator / denominator)
            stock_r2[~np.isfinite(stock_r2)] = np.nan
        
        stock_details = pd.DataFrame({
            'permno': actual_sub_df.index,
            'Benchmark_Used': stock_benchmarks.values,
            'OOS_R2': stock_r2,
            'OOS_MSE': stock_mse
        })
        
        detail_file_name = f"Detail_{file_name}"
        stock_details.to_csv(os.path.join(details_folder, detail_file_name), index=False)
        
        avg_r2 = np.nanmean(stock_r2)
        avg_mse = np.nanmean(stock_mse)
        
        summary_list.append({
            'Model': file_name,
            'Avg_OOS_R2': avg_r2,
            'Avg_OOS_MSE': avg_mse
        })
\end{lstlisting}

\end{appendices}
\end{document}